# 150 years of ground-based solar instrumentation at Meudon observatory (1876-2026)

*Jean-Marie Malherbe, emeritus astronomer (Jean-marie.Malherbe@obspm.fr)*

*LIRA – Paris Observatory - University of Paris Sciences et lettres (OP-PSL)*

*24 January 2026*

## Abstract

The Sun has been observed through a telescope for four centuries. However, its study made a prodigious leap at the end of the nineteenth century with the appearance of photography and spectroscopy, then at the beginning of the following century with the invention of the coronagraph and monochromatic filters, and finally in the second half of the twentieth century with the advent of large ground-based telescopes and space exploration. This article retraces the main stages of solar instrumental developments in Meudon, from its foundation by Jules Janssen in 1876 to the present day, limited to ground-based or balloon instrumentation, designed in Meudon and installed there or in other places (Nançay, Pic du Midi, Canary Islands). The Meudon astronomers played a pioneering role in the history of solar physics through the experimentation of innovative techniques. After the golden age of inventions, came the time of large instruments, studied in Meudon but often installed in more favourable sites, and that of space, in a framework of international collaboration, but this is not discussed here.

**Keywords** : Sun, solar physics, instrumentation, observation, imaging, spectroscopy, radio, visible, telescope, refractor, spectrograph, balloon, history of science

## Introduction

Section 1 of this article presents the essentials to know about the Sun, allowing us to understand the progression of ideas and knowledge in terms of technical and instrumental developments. Section 2 describes the major advances in the period prior to Janssen's settlement in Meudon in 1876; section 3 provides a brief history of the royal domain of Meudon and section 4 relates the Janssen period up to 1907. Section 5 is devoted to the work of Deslandres and d'Azambuja in spectroscopic imaging, an activity that is still alive today. Section 6 summarizes the Lyot period with the introduction of coronography and polarimetry, later developed by Dollfus, Charvin and Leroy in the 1960s (section 7). Deep coronagraphy led to the foundation of the Saint Véran station in 1974 by Felenbok (section 8). It was also the subject of totally innovative balloon flights by Dollfus in 1956-1957 (Section 9). Magnetography (measurement of the solar magnetic field) started after 1960 and is presented in section 10; the acquired expertise led to the construction of the large THEMIS telescope in the Canary Islands in 1999, which is still active, under the direction of Rayrole. High spatial resolution spectroscopy has motivated several achievements, including the 9 m spectrograph in 1959 (Section 11) under the auspices of Michard, then the turret spectrograph at the Pic du Midi in 1980 by Mouradian (section 12). The construction of the Meudon solar tower in 1970 was the starting point of a great adventure in 2D imaging spectroscopy, the MSDP technique successively applied by Mein to 5 telescopes (Section 13). Finally, we discuss the birth of solar radio astronomy in Nançay (section 14) with Denisse, Steinberg and Pick after 1953.

It is not possible to exhaustively list all the instrumental achievements carried out during these 150 years in Meudon because they are too numerous. During this period, the observatory, which was in the meantime attached to the Paris Observatory in 1926, played a leading role, innovative and world-renowned in solar physics, with the introduction of spectroscopy, the invention of the coronagraph and birefringent filters, polarimetry, magnetography techniques, the start of radio astronomy, balloon exploration, and the beginnings of the space age well established at the observatory today. Mein & Mein (2020) focused on Michard's period.

**1 - What is the Sun ?**

The Sun is an ordinary star in our galaxy. It is a sphere of hot gas whose diameter is 110 times that of the Earth, and the mass 300000 times that of our planet. The Sun is composed of 90% Hydrogen, the rest being mostly Helium. The majority of terrestrial elements, such as metals, exist in trace amounts in the Sun, but in a gaseous state due to the surface temperature of 6000 degrees. The internal structure of the Sun is opaque, and therefore unobservable. The solar core, at high temperature and pressure, is a "nuclear reactor" that fuses hydrogen into helium, the corresponding mass loss (4 million tons/s) producing the power radiated by the star throughout space (3.86 $10^{26}$ W). The Earth, above the atmosphere, receives 1.75 $10^{17}$ W, i.e. a flux of 1365 W/m². The age of the Sun is estimated at 5 billion years, it would be mid-life. Only the surface, called the photosphere, and its atmosphere can be observed through a telescope. The photosphere, at 5800 degrees, reveals spots (Figure 1) which are regions of concentrated magnetic field (0.1 to 0.25 T). The chromosphere, just above, is partially ionized and more tenuous. It is only visible in spectroscopy, we see dark filaments and bright areas, the faculae (Figure 1). Its temperature is 8000 degrees. The Sun is surrounded by a vast corona, a faint glory that only appears during total eclipses. It is made up of charged particles (electrons, protons, ions) that are constantly escaping from the Sun, making up the solar wind. It bathes the interplanetary medium and causes another loss of mass of a few million tons/s. The corona is very hot (more than a million degrees), so totally ionized, but very tenuous. The electrons flow out because their thermal velocity exceeds the gravitational escape speed (600 km/s); as they move out, they create an electric field that accelerates the protons (1860 times heavier) and together they form the solar wind that travels at 500 km/s.

The Sun is a variable star, in the sense that its photon flux varies by one thousandth with a periodicity of 11 years. This variation in luminosity is linked to the cyclical appearance of magnetized spots and faculae (Figure 2), with weak or strong cycles. The spotted area is small or absent during the minima, but it can reach a few thousandths of the disk surface (Figure 3) during the maxima. Active regions (spots and faculae) are often unstable at maxima, releasing the surplus of magnetic energy they store, in the form of violent eruptions and ejections of matter, which can spread out in the interplanetary medium and cross the Earth. It is from the Sun-Earth interaction that the aurorae borealis originate, which also exist on some planets (Jupiter, Saturn, Uranus) with a magnetosphere.

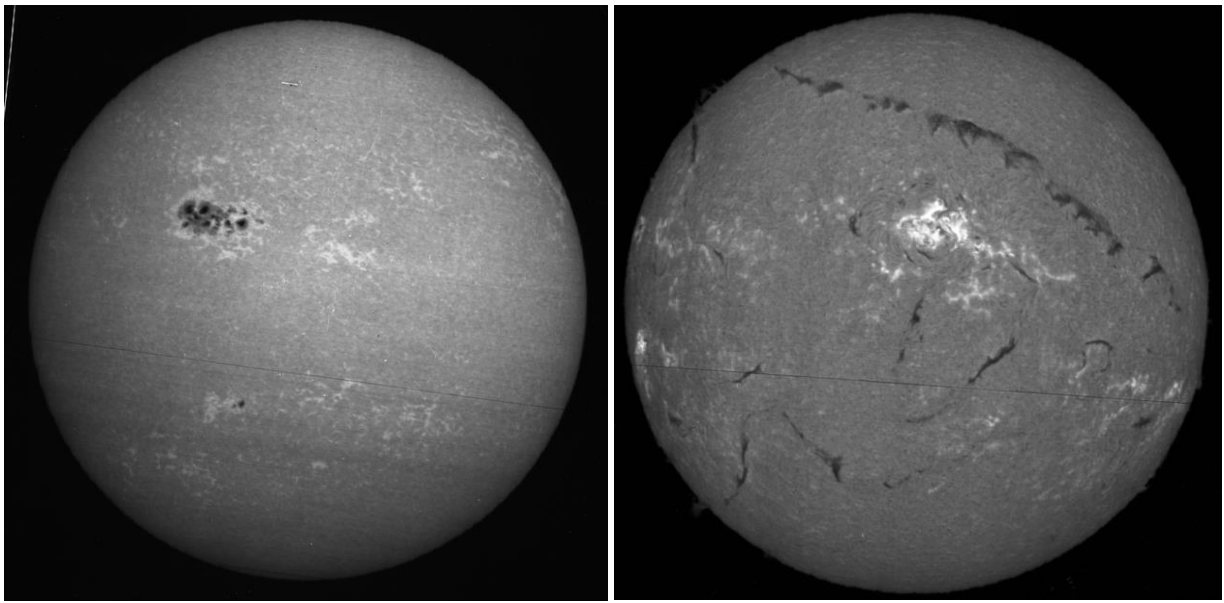

*Figure 1 : The surface of the Sun or photosphere with its magnetic spots (left), and the chromosphere, the layer just above the photosphere, visible only in spectroscopy, with its dark filaments and bright facular areas (right). All these structures are more or less magnetized. Credit: OP.*

The observation of the Sun, from Galileo Galilei until the middle of the nineteenth century, essentially consisted of studying the movements of the spots. Their survey became very regular from 1750 onwards, which allowed the German Heinrich Schwabe to discover the existence of the 11-year cycle around 1850. However, the more sporadic surveys of the seventeenth century showed the existence of an abnormally prolonged minimum, known as the Maunder minimum (its discoverer) between 1650 and 1700; other deep, but less durable minima, were noticed by Gleissberg with a coarse periodicity of 100 years (Figure 2). The 11-year cycle is therefore modulated. The magnetic cycle lasts 22 years (reversal of the polarity of the poles every 11 years during maxima), it was discovered by George Hale in the USA, at the beginning of the twentieth century, about ten years after he understood the magnetic nature of the spots (1908) by noticing the Zeeman effect in their spectrum (this effect splits the spectral lines; the splitting depends on the magnetic field).

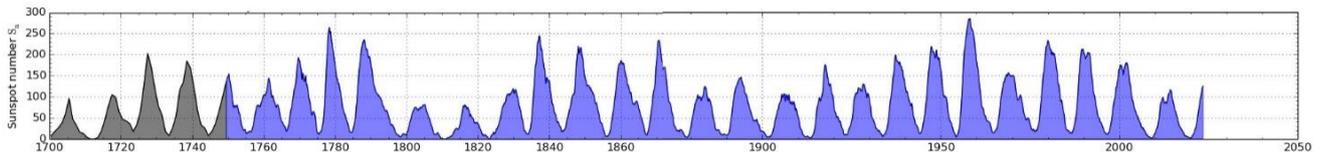

*Figure 2 : counting of sunspots since 1700. The 11-year cycle is modulated by a longer period cycle, still poorly understood, around 100 years. Credit SIDC, Royal Observatory of Brussels.*

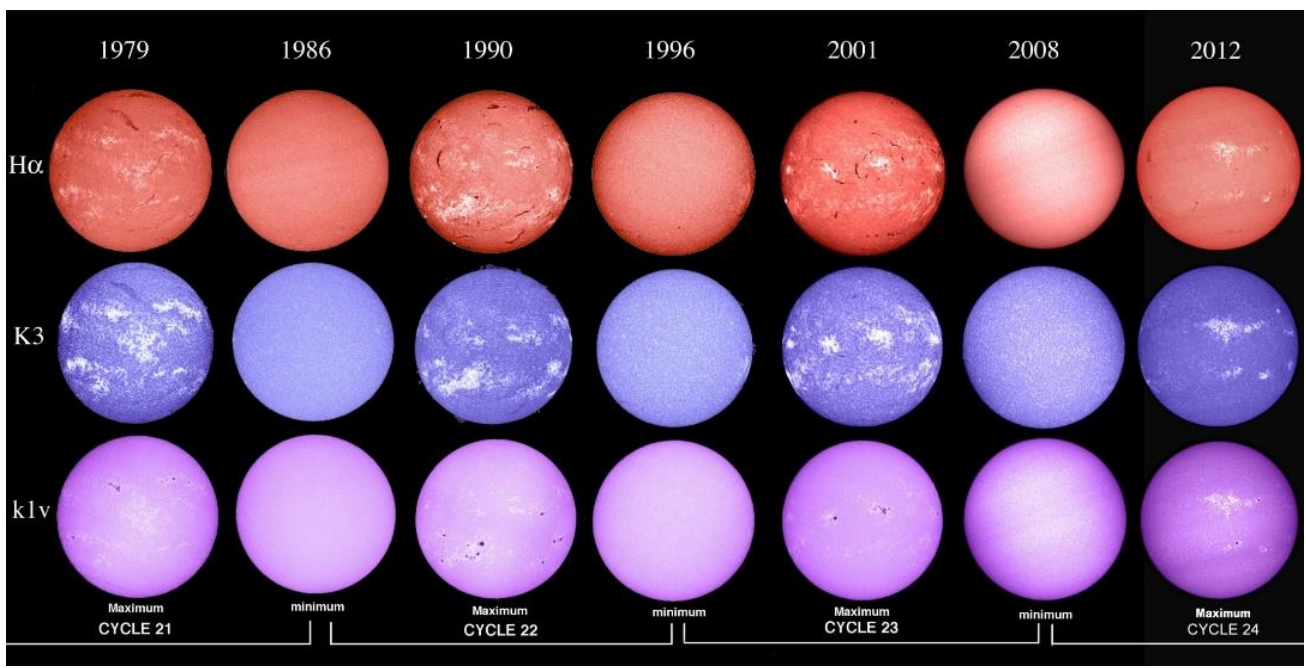

*Figure 3 : extract from the collection of monochromatic images of Meudon in the lines of Hydrogen and Calcium, providing views at several altitudes (in the chromosphere with dark filaments and bright faculae, in the photosphere with magnetized spots), over three solar cycles. OP Credit.*

**2 – Jules Janssen enters the scene (1868)**

Until the middle of the nineteenth century, drawing was systematic in solar observation. The first white light photograph dates back to 1845, it is attributed to Léon Foucault and Hippolyte Fizeau in Paris with the daguerreotype process (Figure 4). The appearance of photography was a first revolution in the history of solar physics, because it allowed an objective conservation of observations, unlike drawings in which a personal interpretation of the observer is present. Also, the Englishman Warren de la Rue took the first photograph of a solar eclipse (1860) in Spain (Figure 5), using his photoheliograph, an instrument specially designed for imaging the Sun in white light on a photographic plate.

We owe to Jules Janssen (Figure 6) the introduction of physical astronomy in France, a new science now called astrophysics, and whose aim is the physicochemical study of the cosmos (constitution, chemical and physical nature). Indeed, the Paris Observatory was essentially devoted to celestial mechanics (description and study of the movement of the celestial bodies). Janssen, inspired by the work of early spectroscopists such as Kirschoff and Bunsen, applied the technique to astronomy, with the aim of isolating light from the dark lines of the solar spectrum, and thus studying the plasma that forms them, i.e. the atmospheric layers that are inaccessible to observation in white light. Janssen thus created a second revolution during the eclipse of August 18, 1868 in India, in Guntoor (Launay, 2021), when he demonstrated, with Norman Lockyer, the possibility of observing prominences at any time, i.e. outside the eclipse, thanks to spectroscopy (Janssen, 1969a, 1969b).

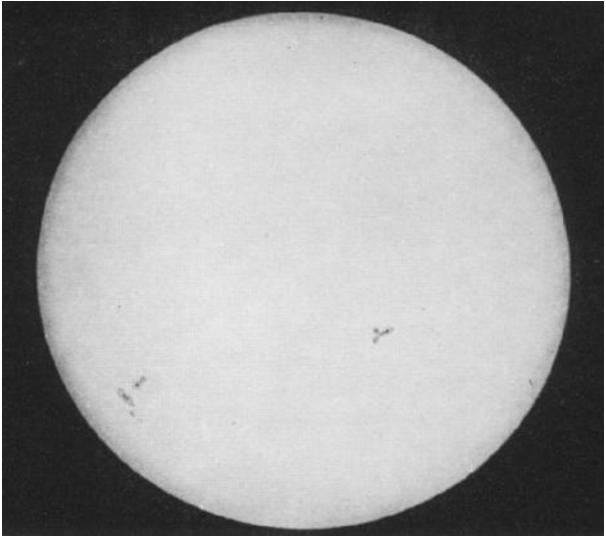

*Figure 4 : First photograph of the Sun (in white light) by Léon Foucault and Hippolyte Fizeau (1845), at the Paris Observatory (daguerreotype process). Photography then became systematic in astronomy, allowing an archiving of observations freed from the personal interpretation of each observer.*

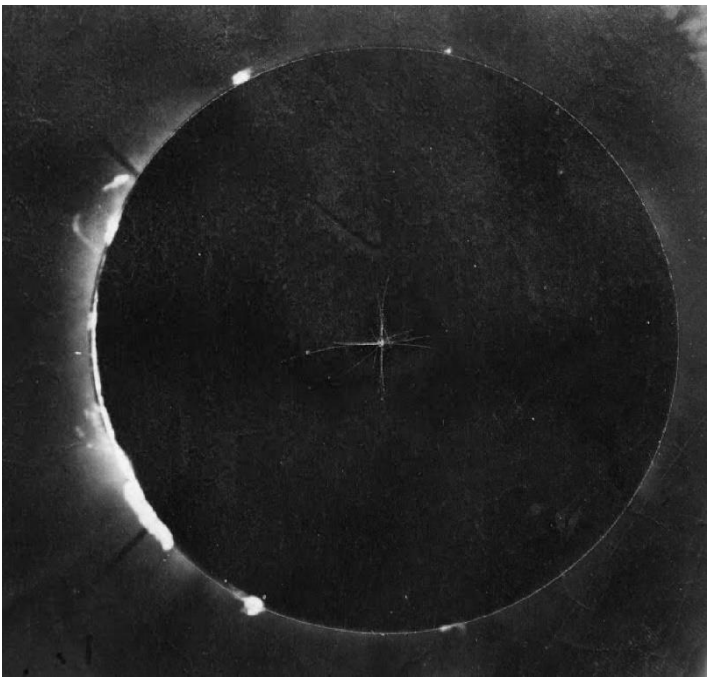

*Figure 5 : first photograph of a solar eclipse (1860) by the British Warren de la Rue, in Spain, with his photoheliograph, specially designed for the photography of the Sun. We can see the prominences around the limb, i.e. structures of Hydrogen whose natural color is red, because they emit in the Hα line. They remained invisible outside of eclipses, except in spectroscopy, which Janssen discovered in 1868.*

Thus, on the day of the eclipse, Janssen (Figure 7) visually recognized a prominence at the limb of the Sun. These structures appear naturally to the eye in red colour, during a total eclipse, because they are made of Hydrogen emitting in the Hα line. The corona, 1000 times less dense, is also revealed by the scattering of white light issued from the disk (Thomson scattering by electrons). But outside of eclipses, these structures are no longer visible, because their luminosity is much fainter than that of the disk (a million times less for the nearby corona). The day after the eclipse, Janssen pointed his spectroscope at the prominence location that was noticed during the eclipse, and it appeared in the intense red line of the spectrum, the Hα line of Hydrogen. A great discovery! Janssen has found a method to use spectroscopy to observe the structures of the solar atmosphere that are hidden outside of eclipses. This advance had a worldwide impact, as it was at the origin of many techniques developed later for the study of spectral lines and monochromatic imaging, allowing to reveal atmospheric layers at various altitudes. Spectroscopy will therefore make it possible to do more physics. Janssen was admitted to the french Academy of Sciences in 1873.

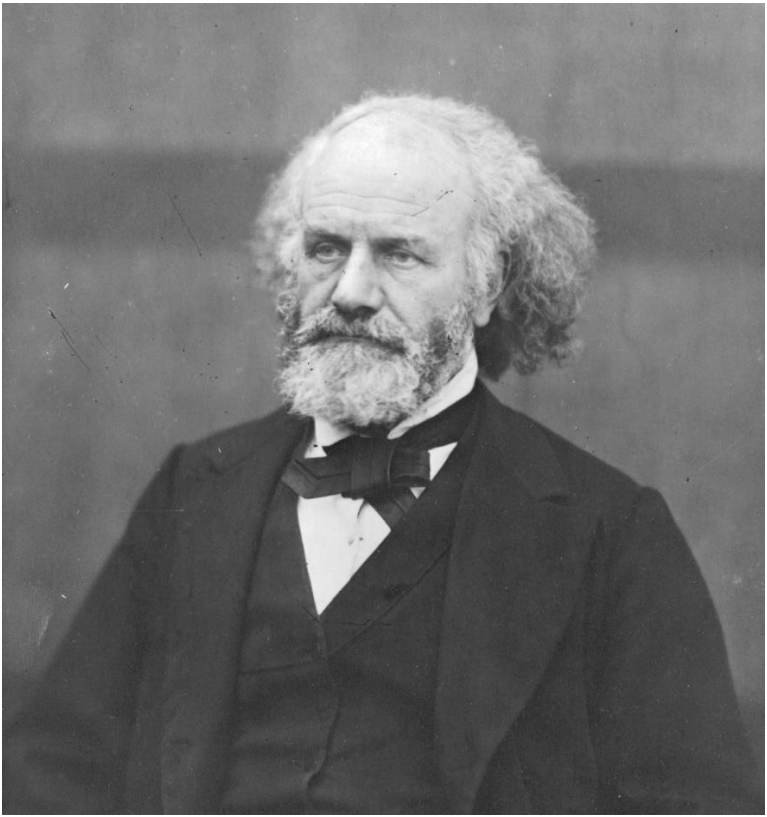

*Figure 6 : Jules Janssen (1824-1907), credit OP. Janssen founded the Meudon Observatory in 1876. A true "adventurer" or "globetrotter" of celestial physics, as Françoise Launay explains in her book published in 2008 by Vuibert and which traces his work and life (Launay, 2008). Janssen's numerous publications and conferences were compiled by Henri Dehérain and published in a two-volume book (Dehérain, 1929, 1930).*

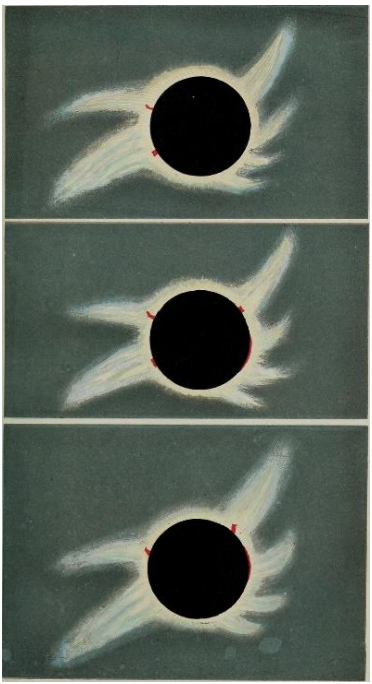
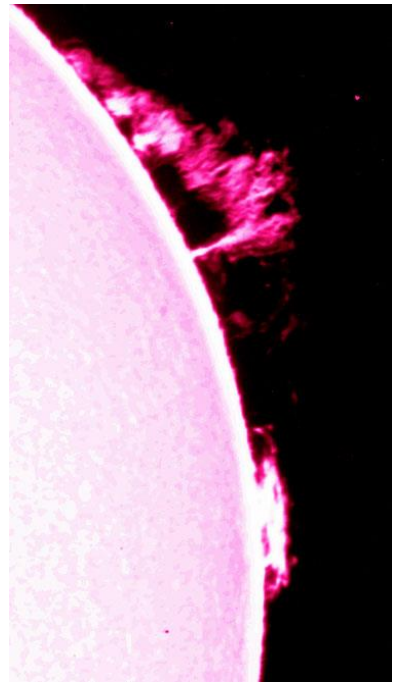

*Figure 7*

*On the left, drawings of the 1868 eclipse, showing the white corona and the red prominences of the solar limb. The corona and prominences are visible to the eye only during eclipses, without spectroscopic means.*

*On the right, a recent example of a prominence seen in the light of the red line (Hα) of Hydrogen, outside any eclipse, thanks to the scans of the spectroheliograph in daily operation in Meudon.*

*Credit OP.*

During the 1968 eclipse, an unknown yellow line (He D3) appeared in the spectrum of the solar limb. Janssen noticed it but paid little attention to it, it was Lockyer who found there a new element called Helium, still unknown on Earth; this is the second major discovery of solar spectroscopy!

Janssen then developed a telescope for observing the Sun in white light (Figures 8 and 9) and designed a special apparatus (Figure 9, left) for the observation of the transit of Venus in front of the Sun, in December 1874, visible from Japan, a nice event to improve the calculation of the distance from the Sun to Earth. It was the first chronophotography camera, which can be considered as one of the ancestors of the cinematograph that would come later. This device, called the "photographic revolver", had a sensitive surface in the form of a rotating circular wheel with 48 sectors, thus being able to take 48 successive views of a rapid phenomenon such as the points of contact between Venus and the solar limb.

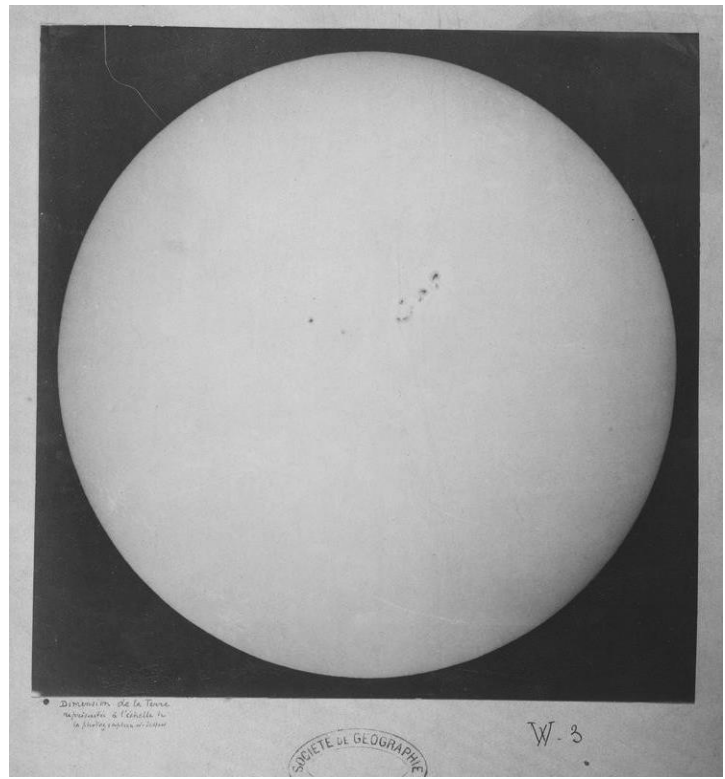

*Figure 8 : image taken by Janssen on July 20, 1874, well before his arrival at Meudon. Credit Gallica/BNF.*

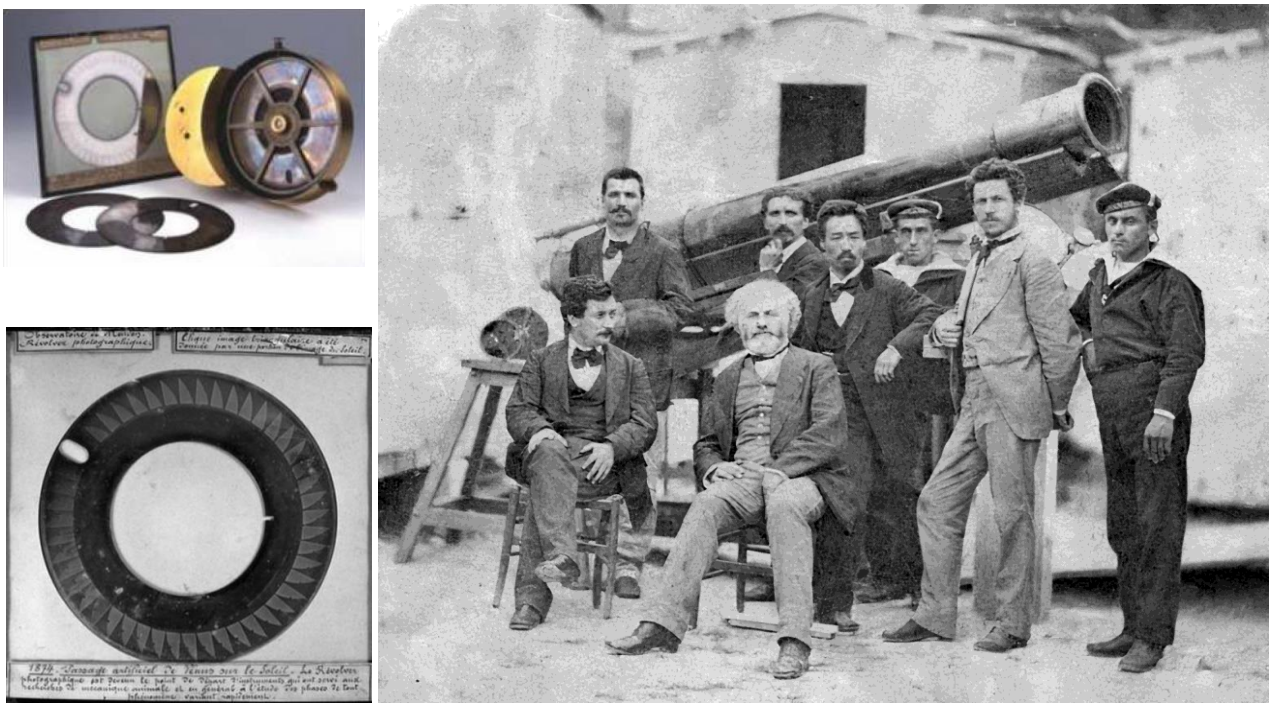

*Figure 9 : Janssen, the photographic revolver (credit OP) and the solar telescope for observing the transit of Venus, at the end of 1874. It should be noted that the wooden support was recovered to carry a refractor of much better optical quality that was used after 1876 in Meudon.*

On the strength of these great successes, in spectroscopy and imaging, Janssen considered that the need for a new observatory entirely dedicated to physical astronomy was becoming imperative. For that purpose, he obtained from the French government the allocation of the former royal domain of Meudon to astronomy, at the end of 1875. At this nice place overlooking Paris, in the countryside, still occupied by the army, Janssen found the new castle of the Grand Dauphin devastated in 1870 by a fire during a bombardment by the Prussians. He undertook its transformation into an observatory, a project that lasted nearly 20 years.

### 3 – Brief history of the royal estate of Meudon

In 1693, Monseigneur or the Grand Dauphin, the eldest son of King Louis XIV, inherited the castle of Choisy. However, the castle of Meudon, near Versailles and Paris, was more prestigious for a future king. Hence, Louis XIV bought it and exchanged it with the widow of its owner, the Marquis de Louvois, superintendent of the King's buildings. The castle, with its classic U-shape (figures 10 to 14), was magnificent with huge gardens. Built during the Renaissance, it was modernised around 1650 by Le Vau.

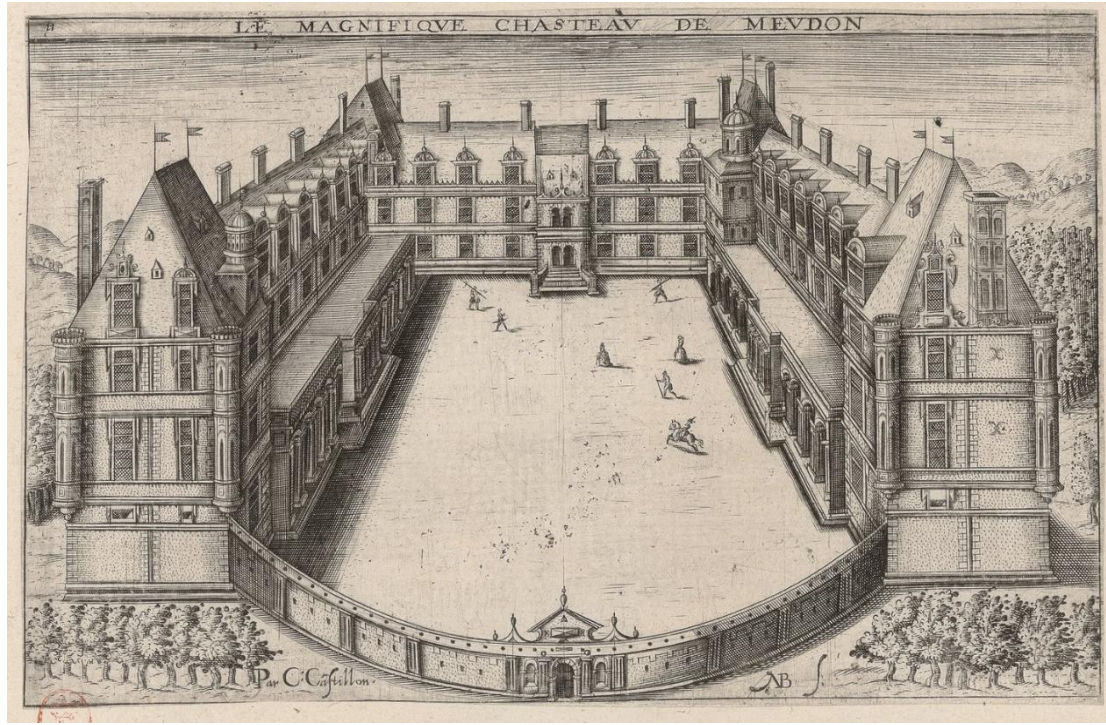

*Figure 10 : the castle of Meudon and its courtyard of honour. Credit Gallica/BNF.*

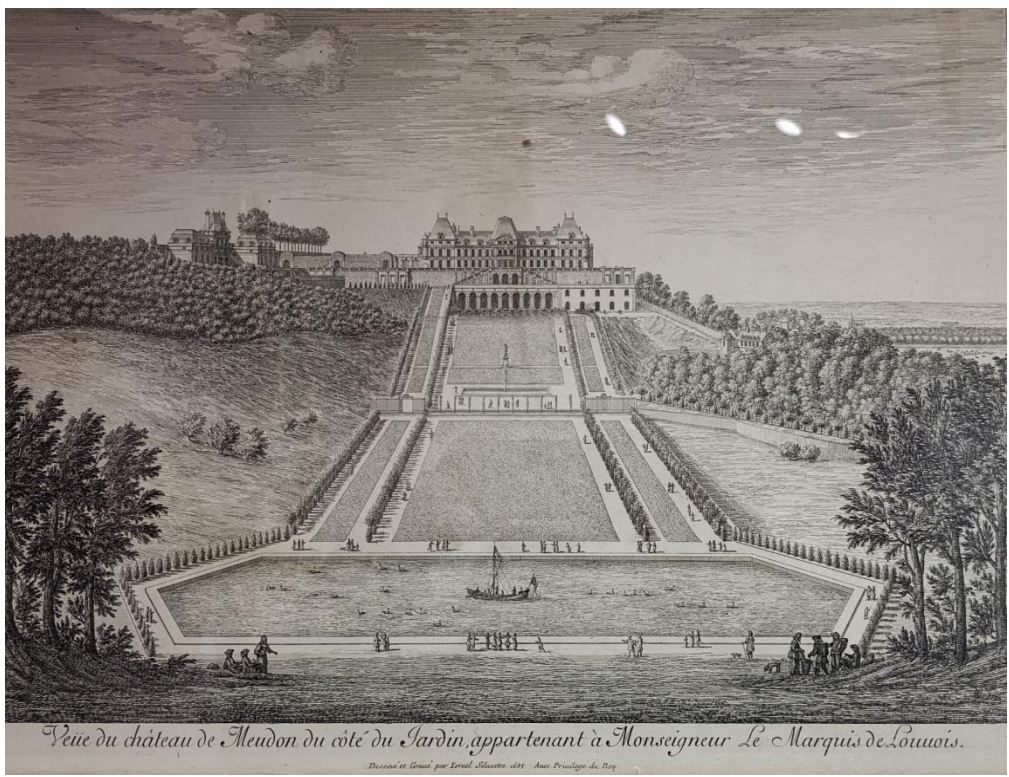

*Figure 11 : the castle of Meudon and the orangery in the background; in the foreground, the gardens. To the left, on the hillside, a small Renaissance castle (the Grotte). Credit Palace of Versailles.*

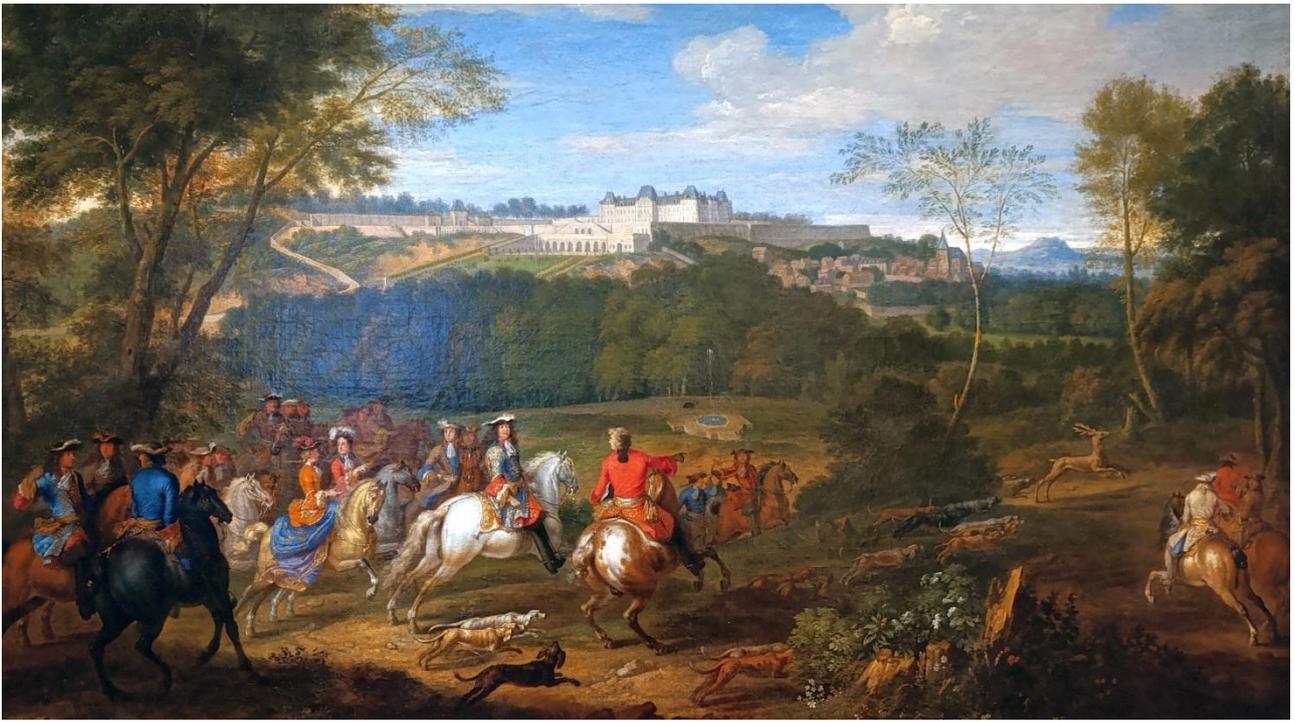

*Figure 12 : the castle of Meudon (at centre), the Renaissance Grotte (at left), the orangery and the village of Meudon on the right. Credit Palace of Versailles.*

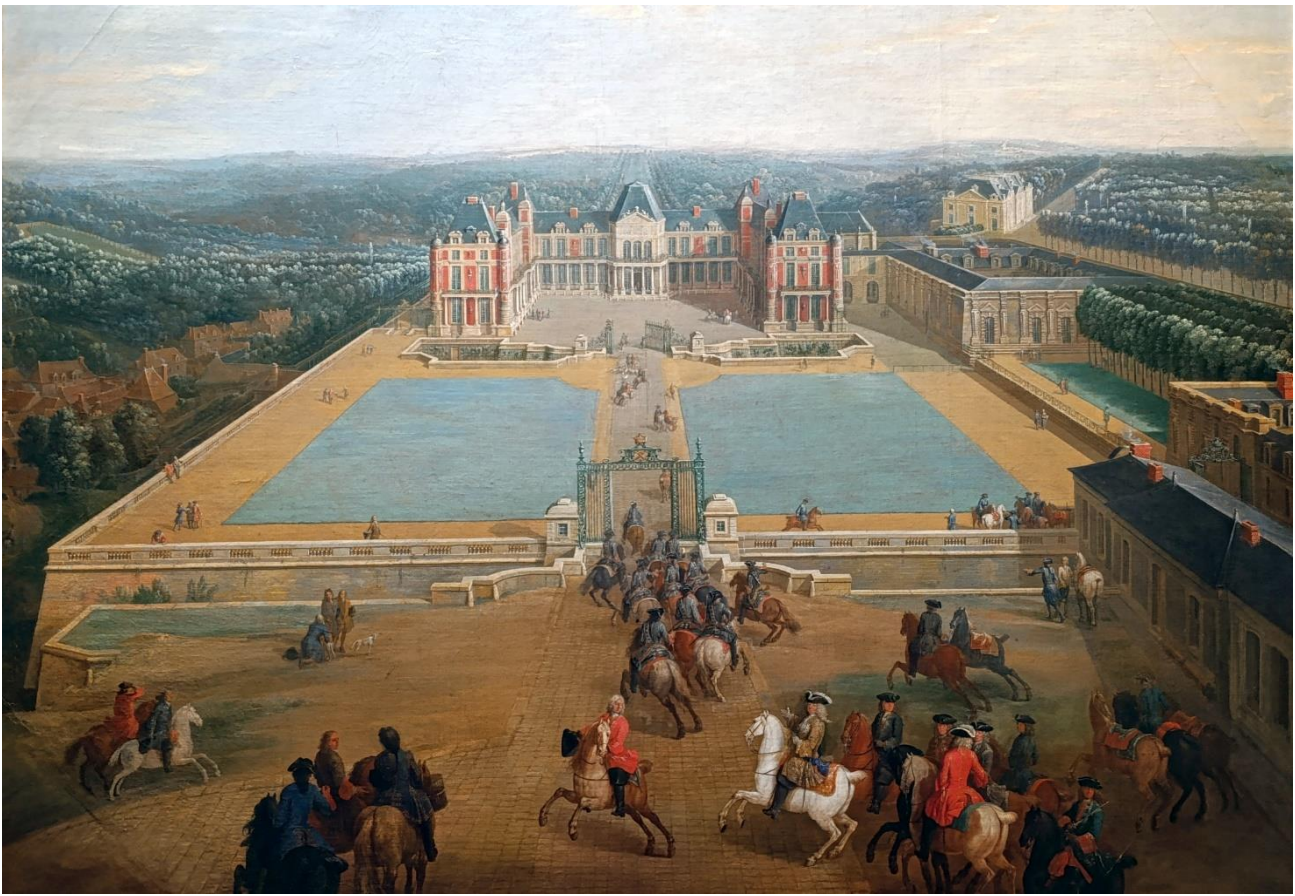

*Figure 13 : the castle of Meudon and its main courtyard, known as the old castle, and the new castle in the background on the right, built on the site of the Grotte which was demolished. Credit Palace of Versailles.*

Monseigneur, sacrificing the Renaissance Grotte, made the new Castle built between 1706 and 1709 by Hardouin Mansart on the same site, in order to accommodate his courtiers and guests. The gardens were organized by Le Nôtre who completed the Grande Perspective, which can be seen in figure 13 behind the castle, and in figure 11. It was the golden age of the "Château de Meudon". Unfortunately, Monseigneur caught smallpox in 1711 and died quickly. Never a king, he was forgotten... The old castle was ravaged by fire in 1795 and demolished in 1804 under Napoleon 1st. As for the new castle, it was bombed in 1870 (Figure 15).

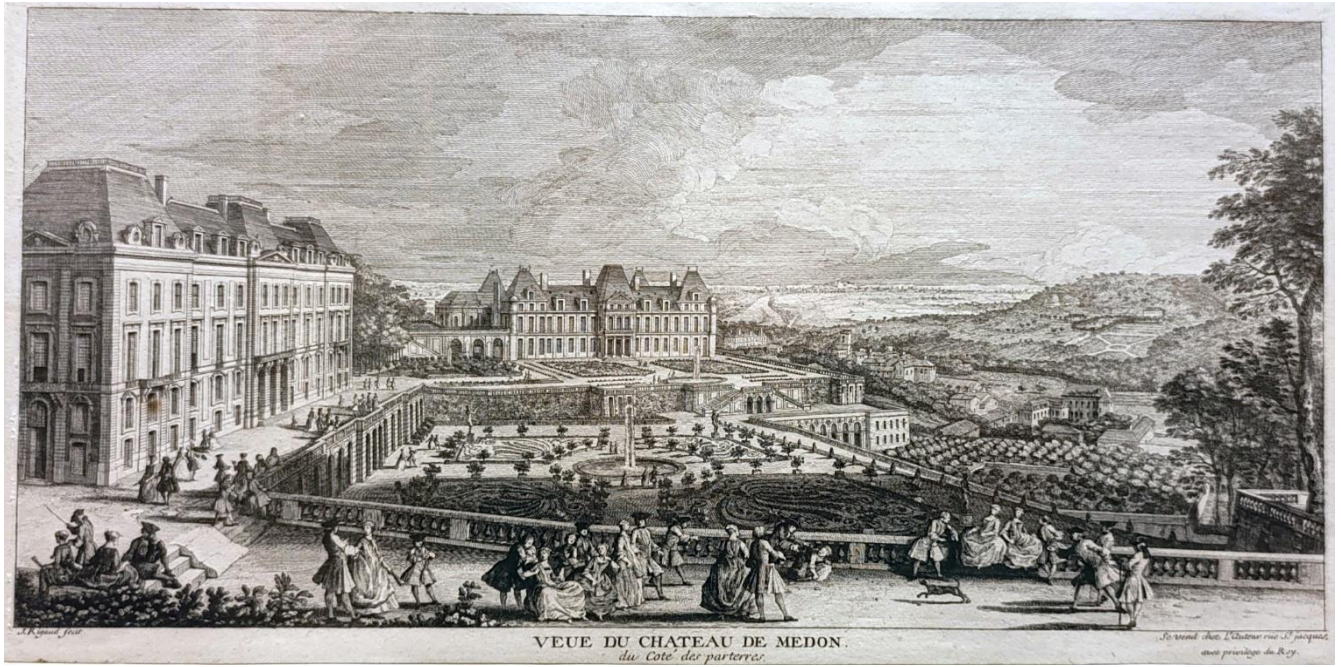

*Figure 14 : the new castle of Meudon (on the left) and the old castle (in the background). Credit Versailles.*

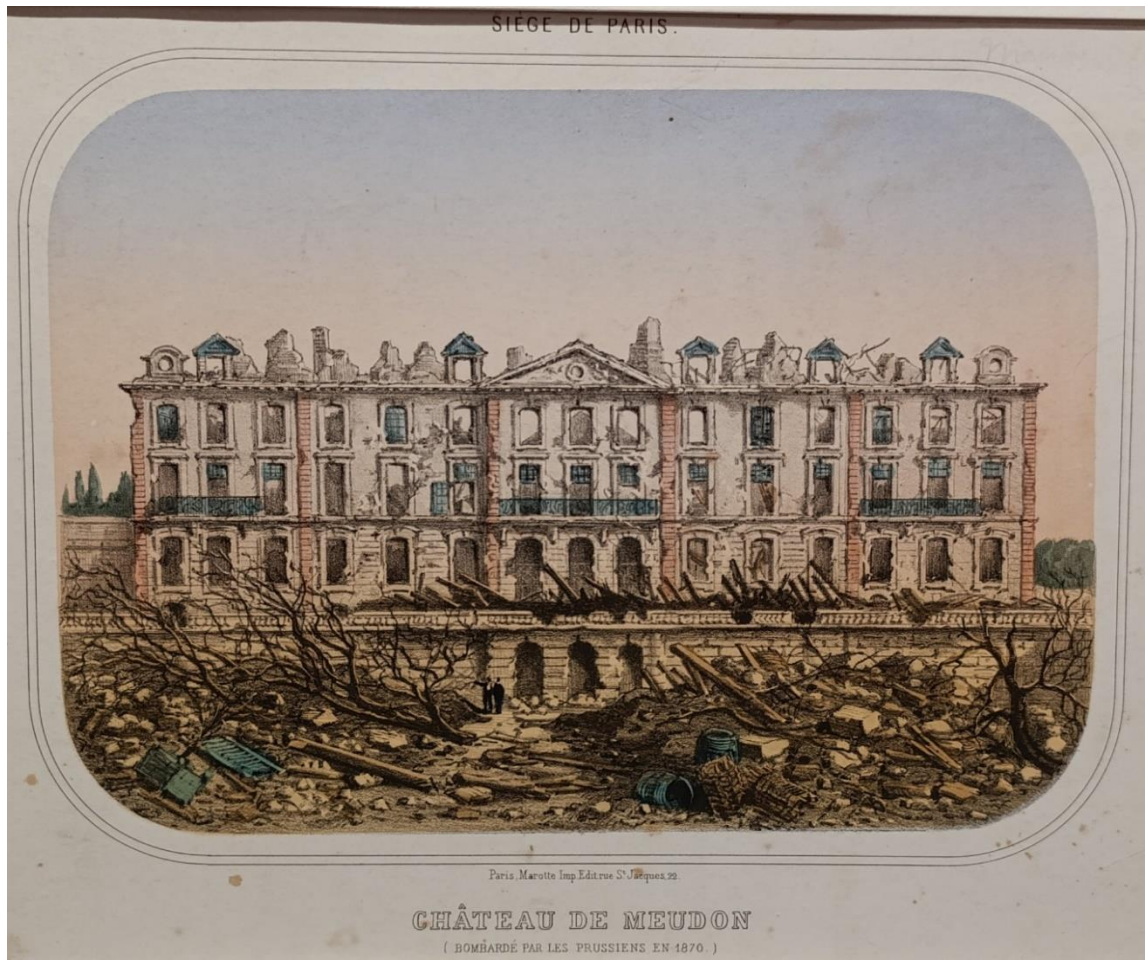

*Figure 15 : the new castle of Meudon ravaged by fire in 1870. Credit Palace of Versailles.*

Janssen obtained the allocation of the Meudon estate by the government to transform it into an observatory at the end of 1875 and moved there in 1876, although the official decree dates back to 1879. The wings, which were irretrievable, were destroyed (Figure 16) and the central part was preserved and surmounted by a large dome, 18 metres in diameter (Figure 17). Since 1893, it has housed the largest refractor in Europe, composed of two superimposed telescopes of 67 and 83 cm, 16 m in focal length. The whole set of instruments and laboratories is described in the volume 1 of the annals by Janssen (1896).

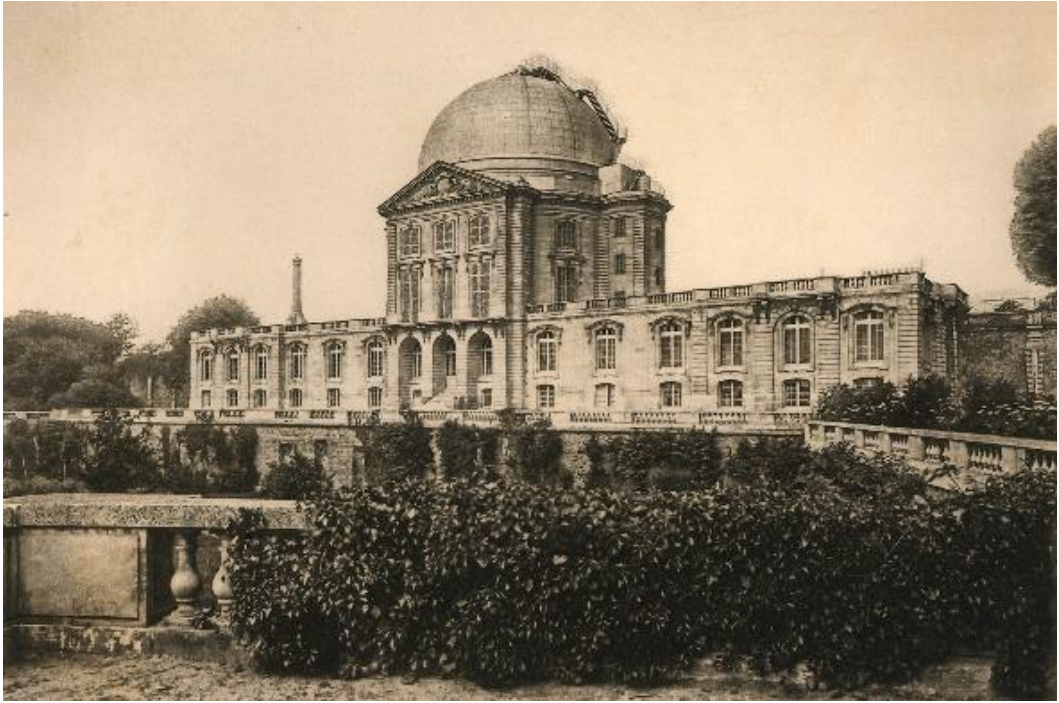

*Figure 16 : the New Castle restructured and then transformed into an observatory by Janssen. Credit OP.*

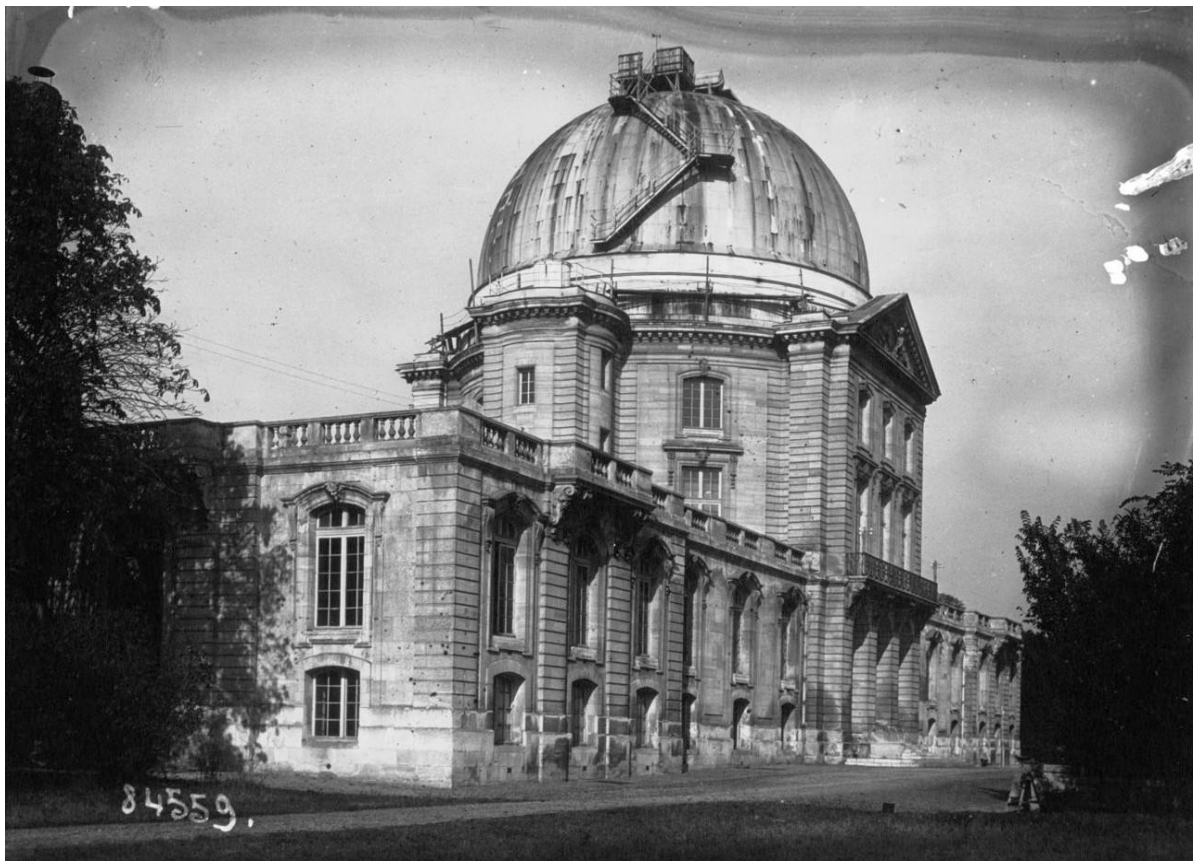

*Figure 17 : The great dome of Meudon in 1920 was built on the ruins of the New Castle. Credit Gallica/BNF.*

## 4 – Janssen in Meudon and the installation of solar instrumentation

During the restoration work on the "château neuf" and the construction of the large dome, which lasted 18 years, Janssen was setting up and operating a new refractor specially dedicated to the photography of the solar surface in white light (Figures 18 to 20). The optics was optimized by Adam Prazmowski (Lecocguen & Launay, 2005), with a focal length of 2.2 m. The primary image of the Sun (only 2 cm) was magnified by a projection lens for 30 x 30 cm² glass photographic plates. Janssen developed an associated photographic laboratory, with a wet collodion process, and invented a curtain shutter to minimize exposure times (1/2000 s), and freeze atmospheric turbulence as well as possible (camera made by the Gautier company).

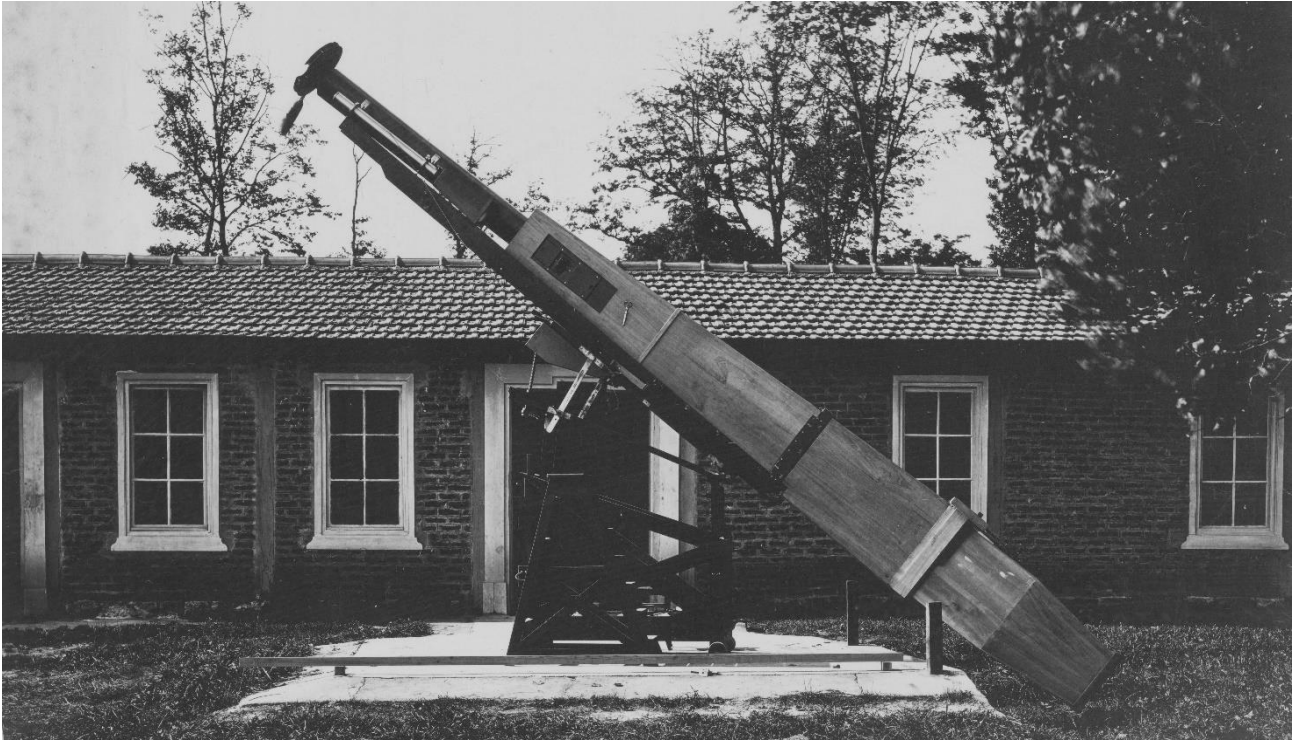

*Figure 18 : Janssen's 135 mm/2.2 m refractor in the garden of Meudon, in 1880. OP Credit.*

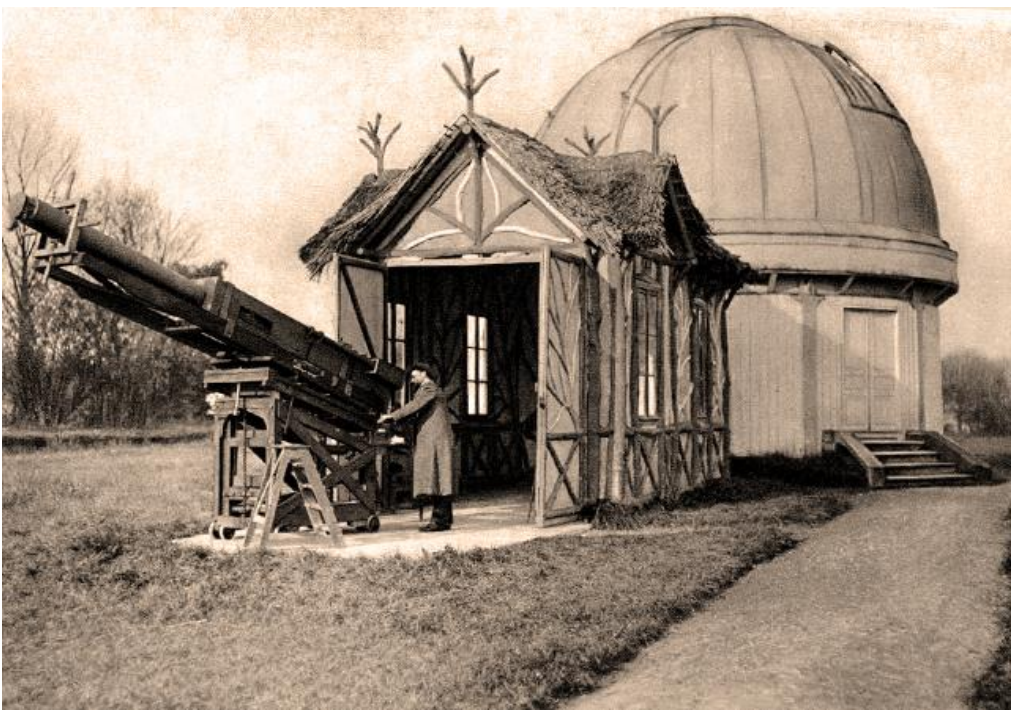

*Figure 19 : Janssen refractor in the park of Meudon at the end of the nineteenth century. At the beginning, it was installed on a wooden rolling support. The entrance lens had a diameter of 135 mm and a focal length of 2.2 m, but a magnifying lens considerably enlarged the images to use 30 x 30 cm² plates at the focus. Credit OP.*

In practice, the photographic plates were mostly sensitive in the blue, so despite the white light observation, the result gave images from the blue part of the spectrum, around the G-band of the Fraunhofer spectrum, centred at 430 nm wavelength. For this reason, Prazmowski had optimized the lens of Janssen's refractor for the blue part of the spectrum, a band roughly from 400 nm to 460 nm.

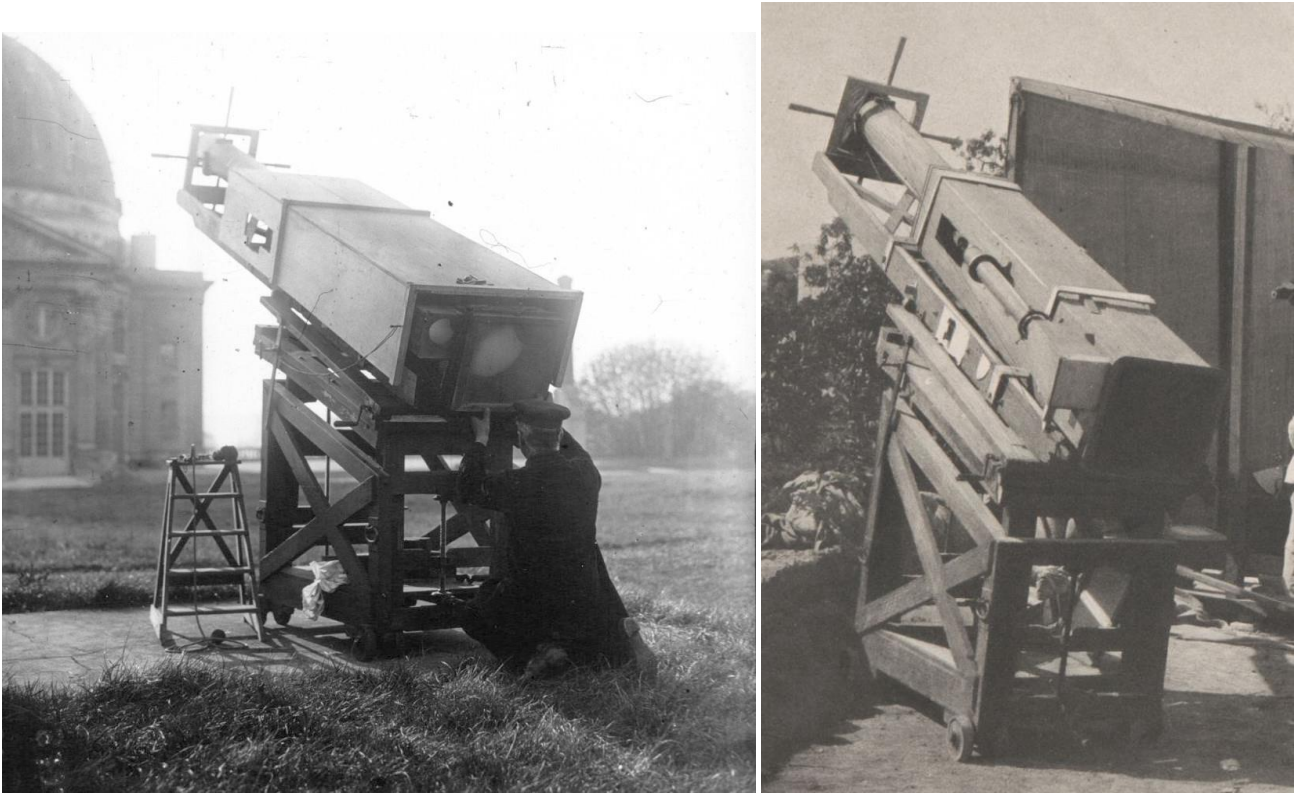

*Figure 20 : Janssen's 135 mm/2.2 m refractor in the park of Meudon on the left and transported to Spain (right) to observe the eclipse of 1905. Credit OP.*

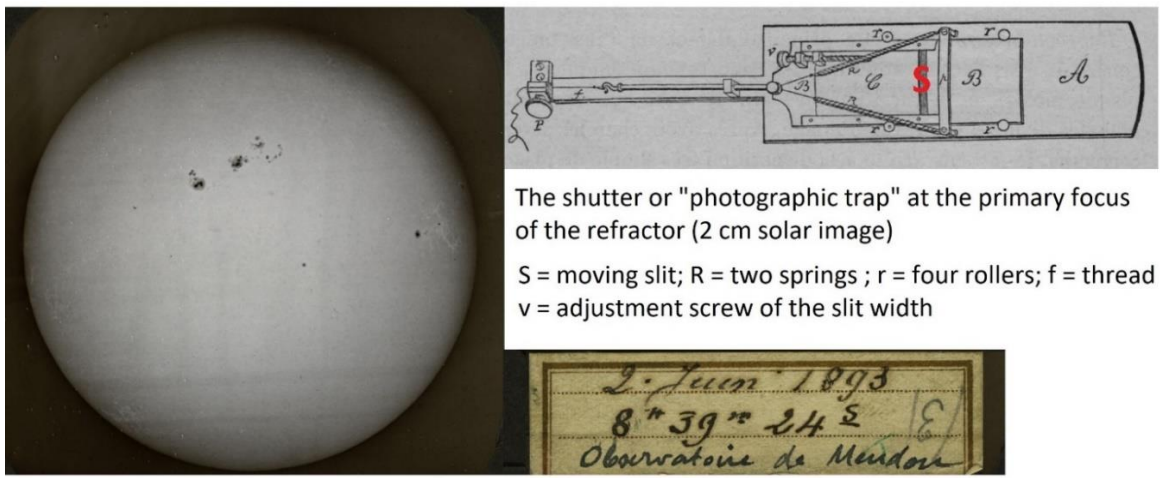

*Figure 21 : ultra-fast shutter designed by Janssen for solar imaging. OP Credit.*

Thanks to the fast shutter in Figure 21, Janssen was able to freeze the atmospheric turbulence and took 6000 photographs with his assistants, of which unfortunately only 1% have come down to us, the rest having been lost or destroyed. These are the first very detailed, good-resolution photographs of the solar surface, in particular of the granulation, a signature of the underlying convection, which is difficult to observe at sea level sites because of the small size of its structures (about 1"). Janssen published a sample of 50 plates in his atlas of solar photographs, which are archived in the observatory library, and online at Gallica (Janssen, 1903). Among the oldest photographs of granulation, the one in Figure 22, almost 150 years old, is exceptional for the epoch and for the Meudon site, which is now very degraded by the growth of Paris city.

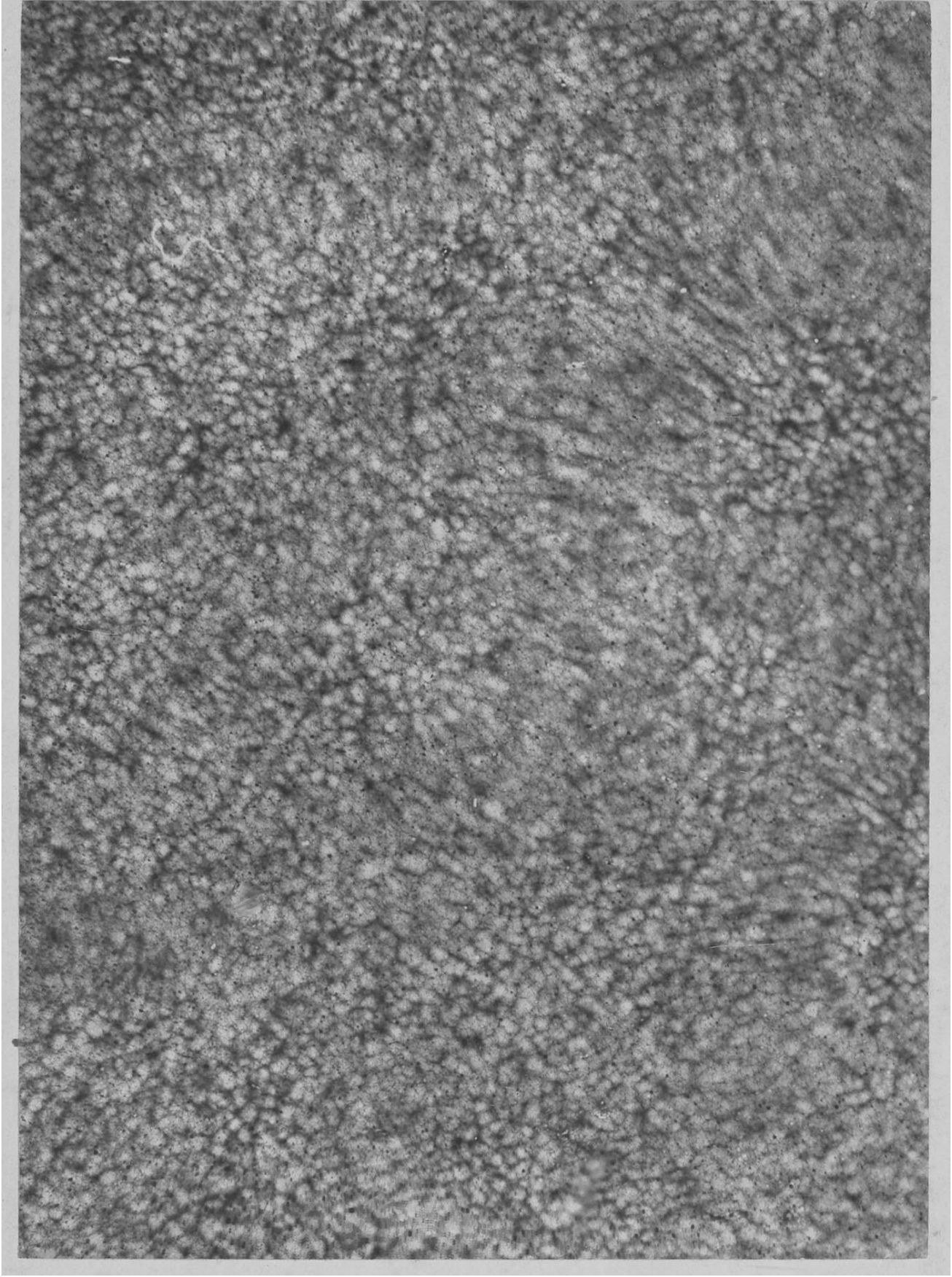

*Figure 22 : A fine example of a photograph of solar granulation, Janssen, October 10, 1877. OP Credit.*

After this pioneering work in high-resolution imaging of the sun, Janssen became interested in the presence (or not) of oxygen in the solar spectrum. It is known that the Earth's atmosphere, which contains water vapour, di-oxygen and di-nitrogen, contaminates the solar spectrum through the lines of these molecules, which are often found in the red (bands A, B). The question was whether oxygen lines could also be of solar origin. The first idea, that one can have to study such a question, is to go to high altitude sites, in order to reduce the thickness of the Earth's atmosphere above the observer, and thus decrease the intensity of the telluric lines as one rises (those of solar origin do not vary). Janssen therefore undertook to transport his spectroscope to Mont Blanc, on whose slopes Joseph Vallot was already doing glaciology with his observatory located at 4350 m. A first incursion into the massif (1888) brought Janssen to the "Grands Mulets" refuge at 3050 m. In 1890, he made his first ascent, under epic conditions. His difficulties in locomotion and the adventures of the ascent through complex and crevassed glaciers did not make him back down: he considered travelling seated in a ladder chair or in a sleigh with the assistance of 12 guides (Figure 23). Stopping at the Vallot observatory, Janssen reached the summit; he was struck by its vast horizon, and decided to build an observatory, which could have been permanent if it had benefit from rocky foundations, which was not the case.

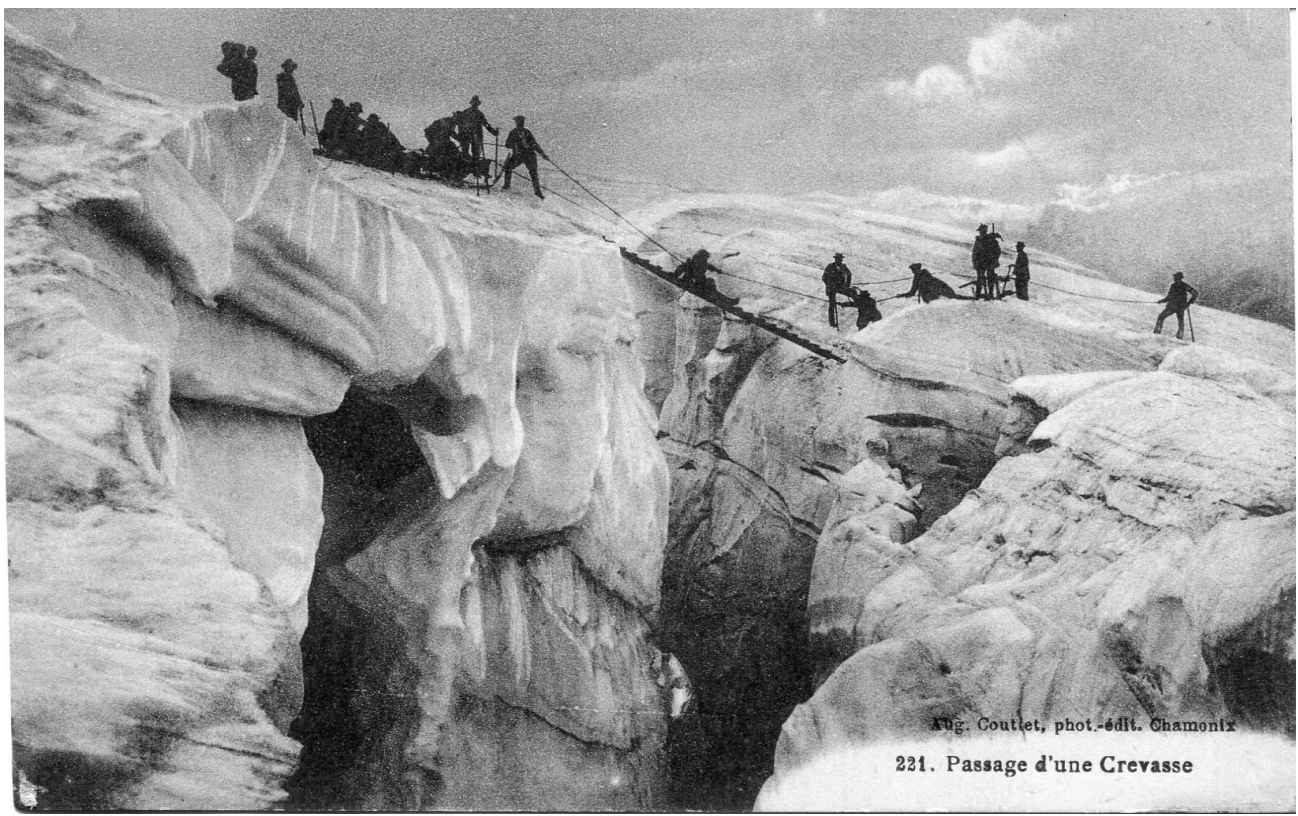

*Figure 23 : Janssen's epic ascent by sleigh, in the glaciers to Mont Blanc. Credit J.-M. Malherbe.*

Faced with the impossibility of a solid foundation, Janssen decided to place his observatory directly in the ice at the summit. The wooden building (Figure 24) was designed and erected in Meudon before it was transported to Mont Blanc in hundreds of small parts for carriers. There were two levels, with the lower level having to be buried in the snow (Figures 26 and 27). Actuators were used to adjust the building's foundation and prevent ice movement. The turret was dedicated to weather observations (but a long-running meteorograph made by Jules Richard was permanently put in the observatory). The large refractor (Figure 25, Henry brothers' objective) with a diameter of 30 cm and a focal length of 5 m was supplied with light by a polar siderostat (60 cm plane mirror) fixed in front of the lens, the optical axis remaining parallel to the Earth's axis of rotation. The telescope protruded on one side; observers worked in the lower level. Janssen came to inaugurate it in 1893 and remounted in 1895. However, he did not use the large refractor himself for his observations of the solar spectrum, but a Duboscq spectroscope that he always carried with him. He noticed a decrease in the intensity of the di-oxygen lines with altitude. We now know that there is oxygen in the solar atmosphere, but only in the form of atomes or ions. The telescope was equipped with a 1 m grating spectrograph in 1904. Milan Stefanik, co-founder of the Czechoslovak Republic and astronomer, visited the observatory several times.

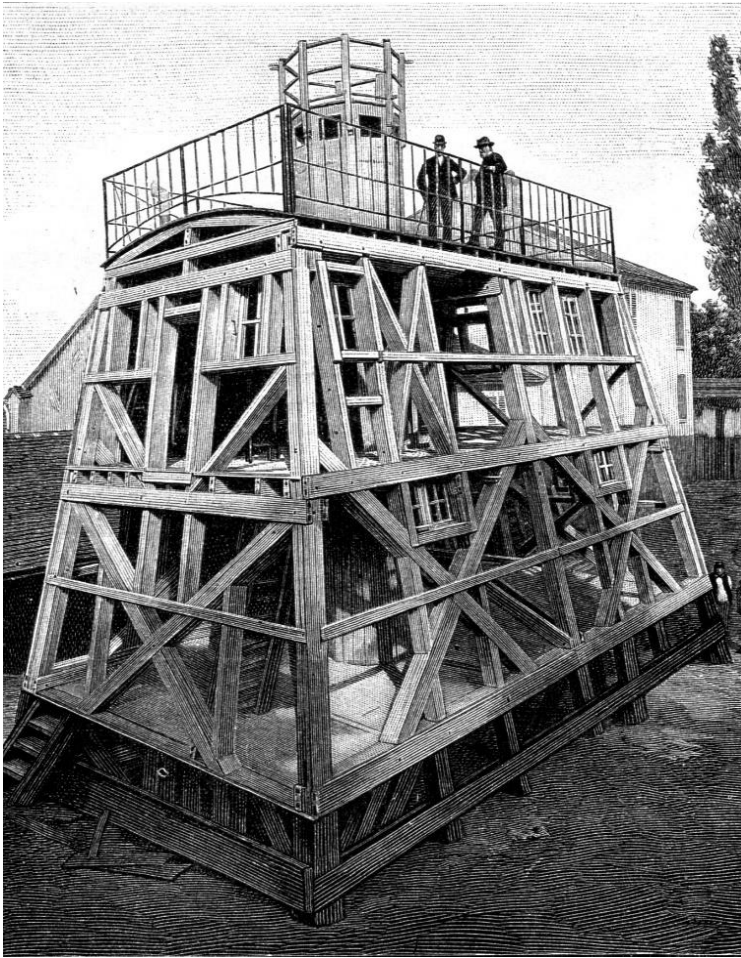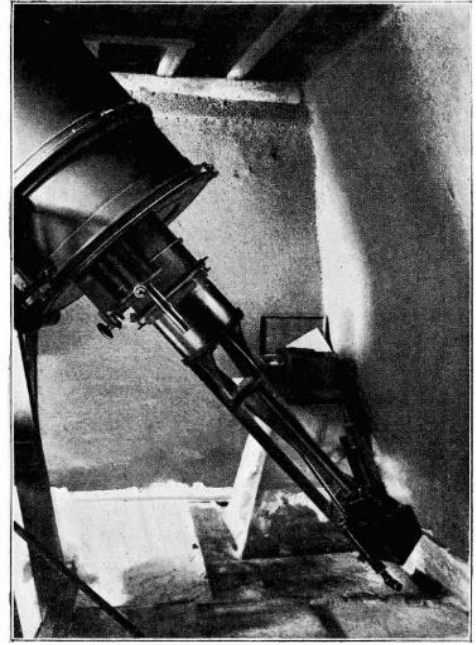

*Figure 24 : the Mont Blanc observatory built and tested in Meudon before being transported on the back of carriers or by sleigh to the summit. On the right, the small spectrograph of 1 m. Credit OP.*

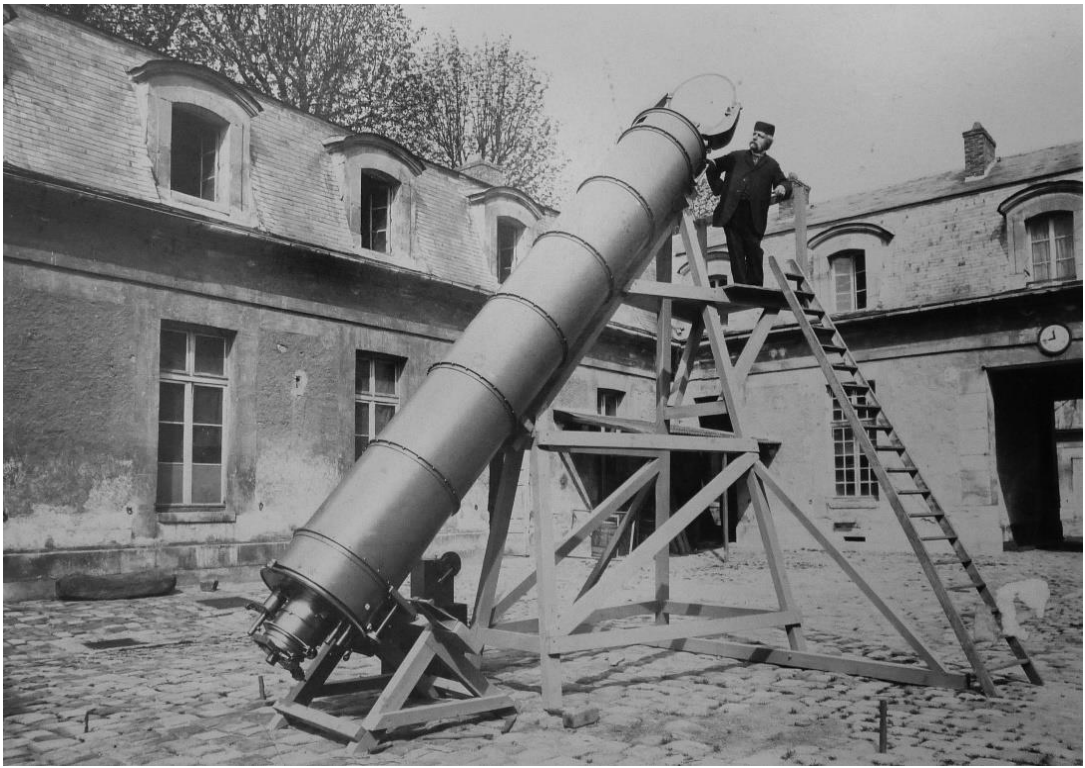

*Figure 25 : the large polar refractor of 30 cm in diameter, 5 m in focal length (1893) built in the courtyard of Meudon observatory. There was a double tube, one rotating around its axis. OP Credit.*

The Mont Blanc observatory operated until 1909 (two years after Janssen's death), with numerous expeditions, often astronomical, but sometimes multidisciplinary. As it was submitted to the pressure and movement of the ice, it was dislocated at the end of its life, and was finally abandoned in 1909.

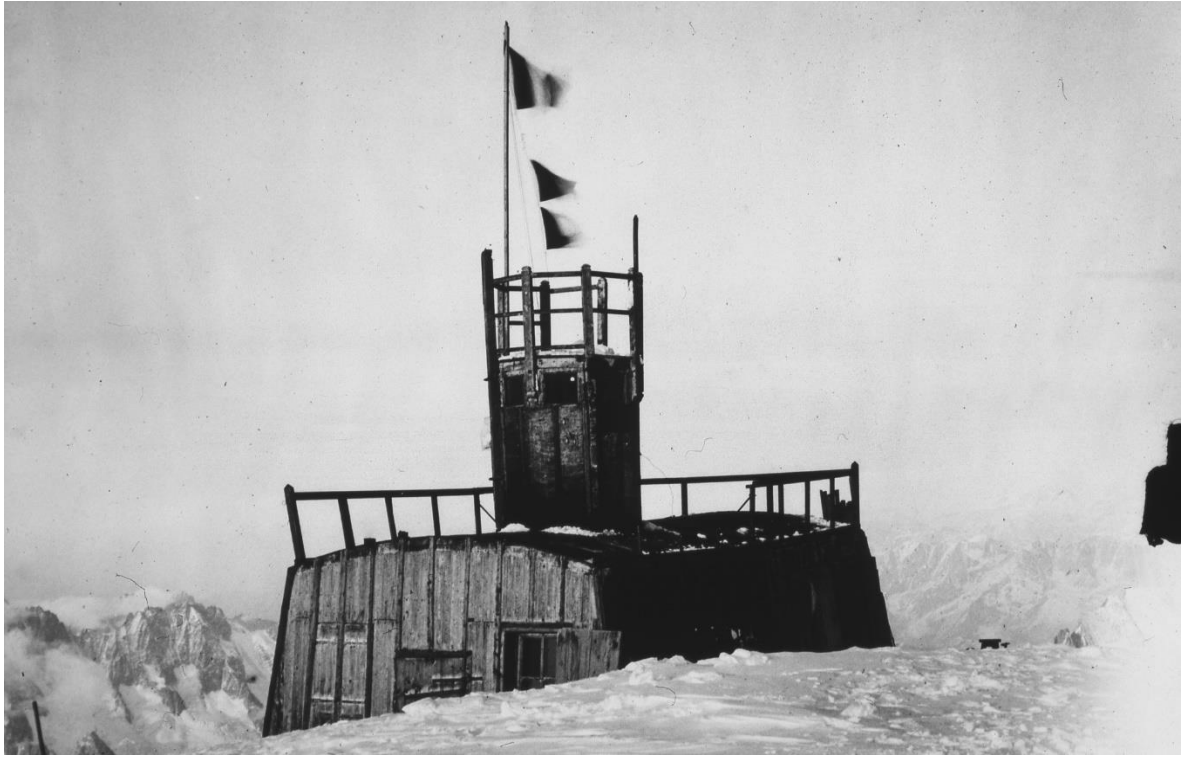

*Figure 26 : the Mont Blanc observatory set on the summit ice. The turret is used for meteorological measurements. West and North sides. Credit OP.*

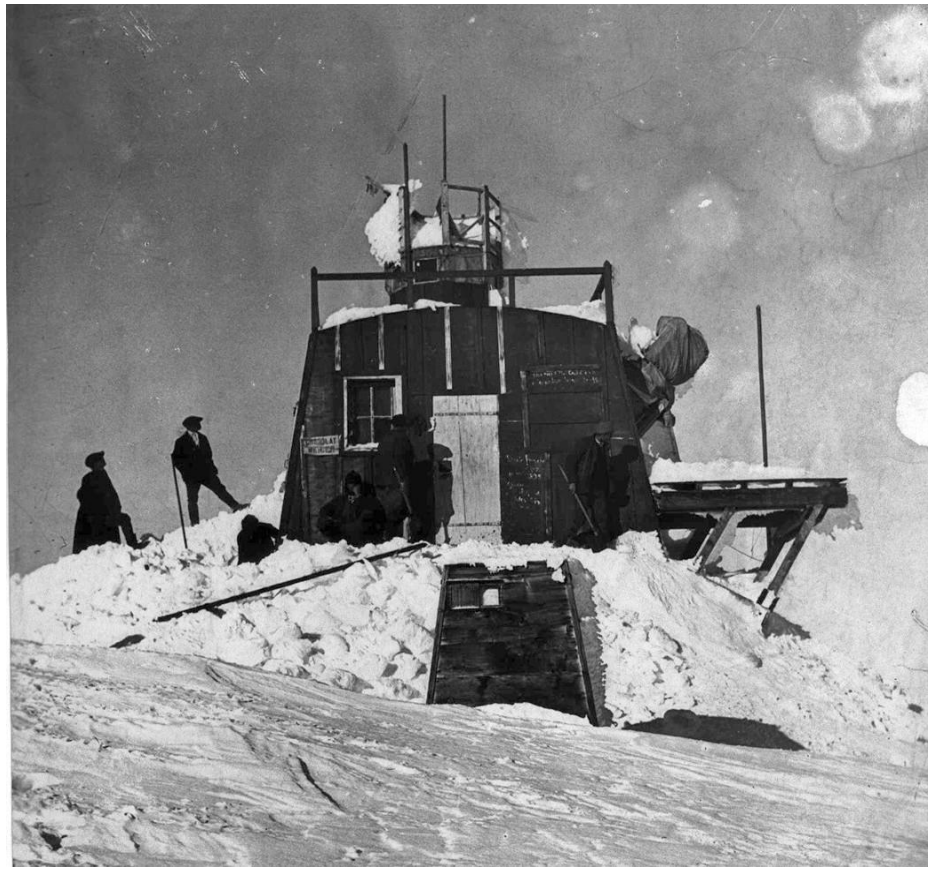

*Figure 27 : the Mont Blanc Observatory. The large fixed refractor protrudes to the right. Credit OP.*

## 5 – Deslandres, d'Azambuja and spectroscopic imaging of the Sun

While Janssen was busy at Mont Blanc, Henri Deslandres (future academician, figure 28) was hired in 1888 at the Paris Observatory by Admiral Ernest Mouchez, the director, to found a spectroscopy laboratory. It was quite natural that he turned to solar spectroscopy, where everything had to be developed. He remembered that Janssen, after the eclipse of 1868, had stated the principle of the spectrohelioscope, a scanning imaging spectrograph (Janssen 1869c): "by rotating the spectrograph on its axis, the input slit scans the solar surface, and by placing in the spectrum a second slit isolating a well-chosen spectral line, a monochromatic image is formed on the retina through an eyepiece." To do this, it is necessary to benefit from retinal persistence, so it is necessary to rotate the spectroscope fast enough (Janssen's instrument had a crank). If the eye is replaced by a photosensitive plate, a recording spectroheliograph has been made. This was Janssen's idea, enunciated in 1869, taken up and improved by Deslandres in Paris, but also by George Hale in the USA, independently. This is how the spectroheliograph was born in Paris, providing the first monochromatic images of the Sun in 1893 (called spectroheliograms), and the first recordings of prominences in 1894 (Figure 29).

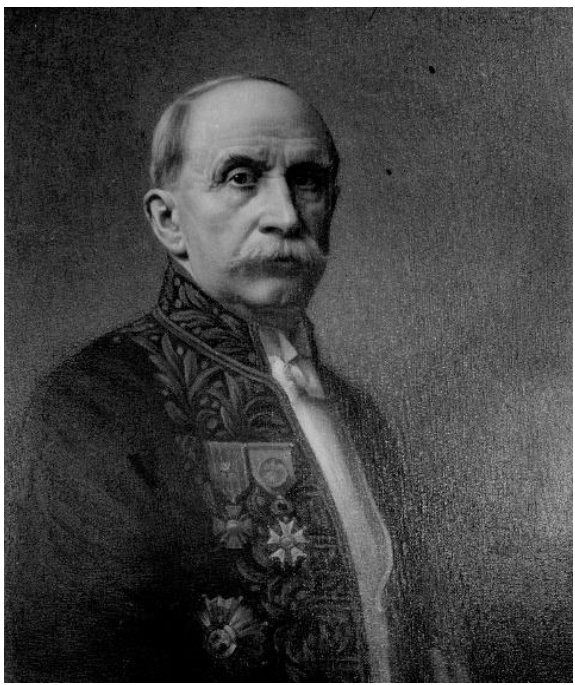

*Figure 28 : Henri Deslandres (1853-1948). Credit OP. Inventor of the spectroheliograph in 1893 in Paris, he moved to Meudon in 1898 where he installed his instruments and then started a new work on a large quadruple spectroheliograph (1907) that he made with his student Lucien d'Azambuja. The latter was responsible for organizing a service of daily observations of the Sun in 1908, which he developed and which continues today. Deslandres returned to Paris in 1926 to be appointed director of the new establishment formed by the administrative merging of Paris and Meudon observatories. The story of Deslandres' spectroheliograph is described in detail by Dollfus (2003a, 2003b, 2005).*

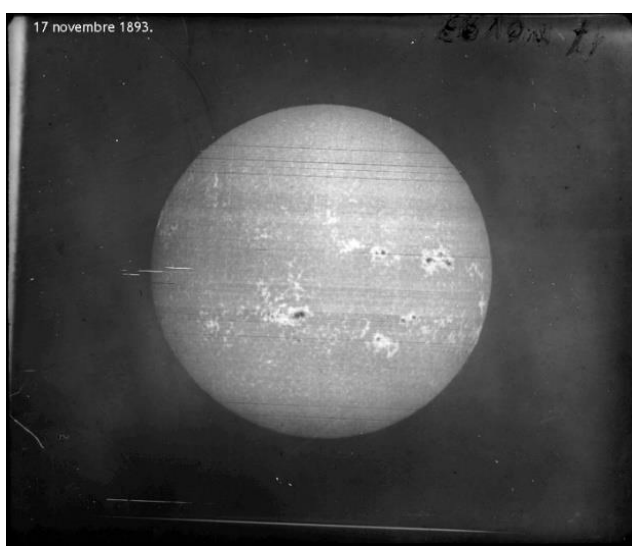 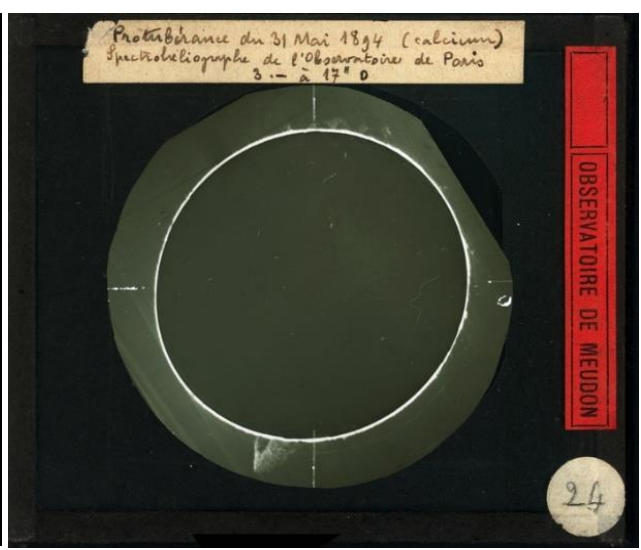

*Figure 29 : First spectroheliograms or monochromatic images made in Paris in 1893 and 1894 by Henri Deslandres. The chosen spectral line is the purple "K" line of ionized calcium. The image on the left shows the active regions surrounded by bright areas, the one on the right (with a simple mask upon the solar disk to avoid overexposure) reveals the prominences at the limb. Credit OP.*

Thinking that he would find more space and resources in Meudon than in Paris, Deslandres moved there in 1898 with his instruments, which took their place in the building of the small siderostat (figure 30). In a fixed position in the laboratory, they were supplied with sunlight by a polar siderostat (Figure 31).

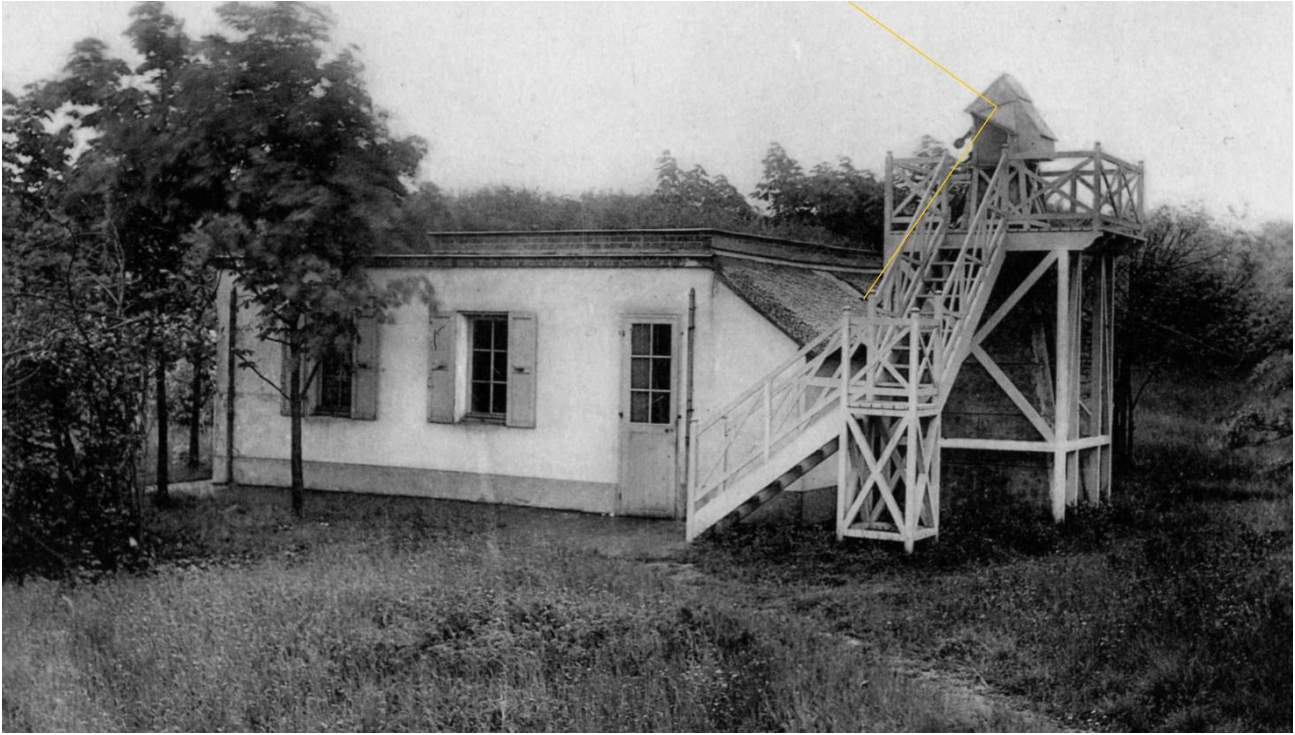

*Figure 30 : building of the laboratory where Deslandres installed his instruments in 1898. Credit OP.*

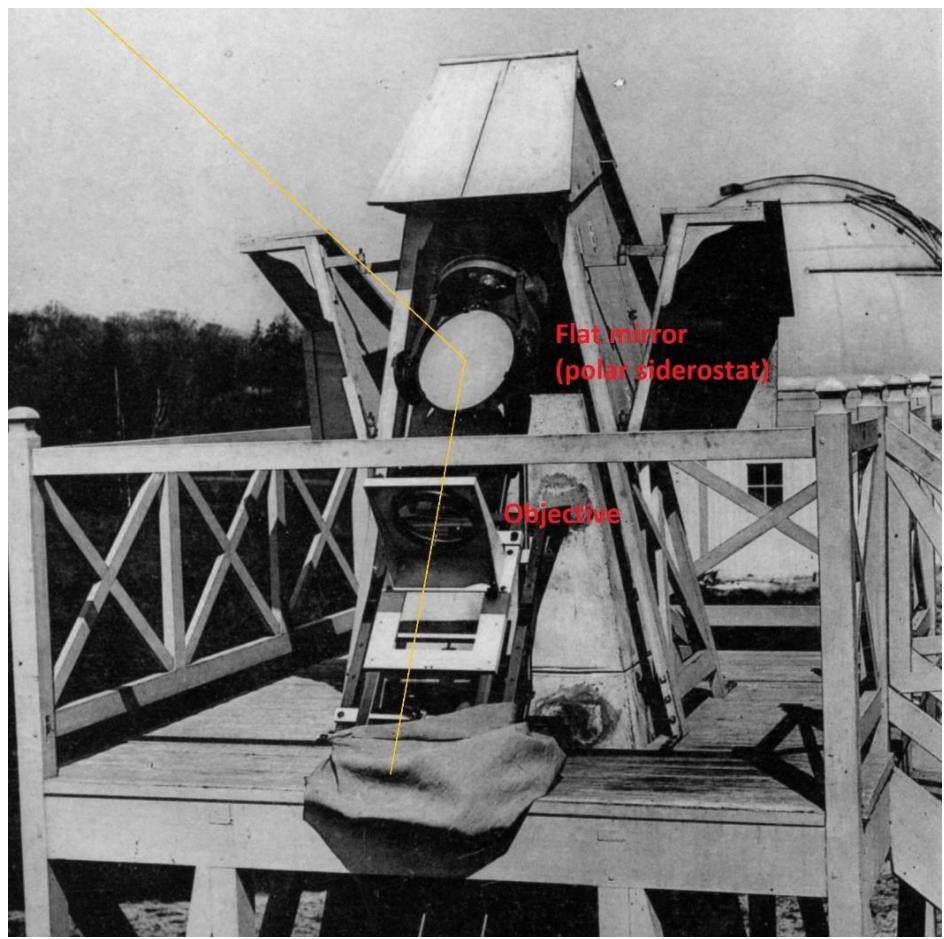

*Figure 31 : the small polar siderostat feeding the fixed solar instruments in 1898. OP Credit.*

At that time (Figure 32) there were two spectroheliographs described by Deslandres (1910):

- the shape spectroheliograph recording the morphology of chromospheric structures in the core of a spectral line such as CaII K, using a thin output slit. The solar surface was swept by a continuous movement of the spectrograph on rails.

- the velocity spectroheliograph, with a wide output slit recording the full spectrum of a line such as CaII K, in order to detect mass motions by Doppler effect on the line profiles. The solar surface was then covered in sections, with a discontinuous displacement of the spectrograph on its rails. About 80 sections were recorded on the photo plate with 30" steps on the sun.

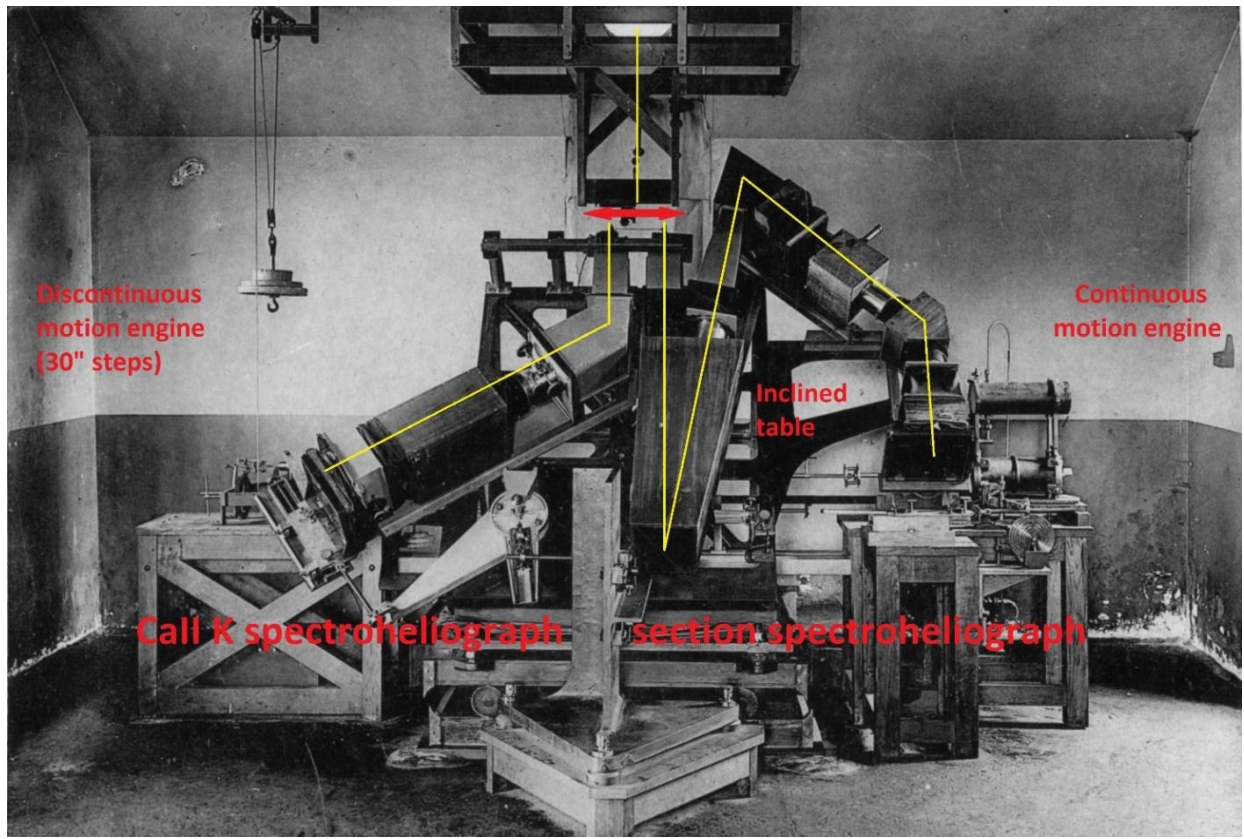

*Figure 32 : the velocity spectroheliograph (left) and the shape spectroheliograph (right) in 1898. The entrance slit scanned the sun by translating the spectrographs onto rails. Credit OP.*

Deslandres hired an assistant in 1899, Lucien d'Azambuja (Figure 33), who was only 15 years old, and started working with him on a large quadruple spectroheliograph (Figures 34 to 37, Malherbe, 2023a, 2023b, 2024), which was completed in 1908, and which was intended for both daily observations and scientific research. D'Azambuja (1020a, 1920b) described this powerful instrument in detail.

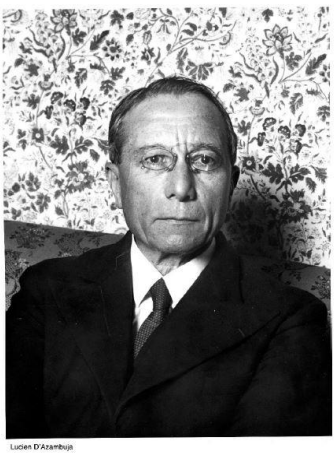

*Figure 33 : Lucien d'Azambuja (1884-1970), "hired by Deslandres at the age of 15, was a man who was not very tall, always dressed to the nines, in a three-piece suit, a shirt with a hard collar, and a twinkle in his eyeglasses. An excellent instrumentalist, he was uncompromising on the service... ». D'Azambuja did his thesis on the observations of the quadruple spectroheliograph that he presented at the same time (D'Azambuja, 1930). He remained in Meudon for 60 years, until his wife's retirement in 1959. He was 75 years old at that time. His work and the one of his wife Marguerite is told by Malherbe (2024). An incredibly long career today! Credit OP.*

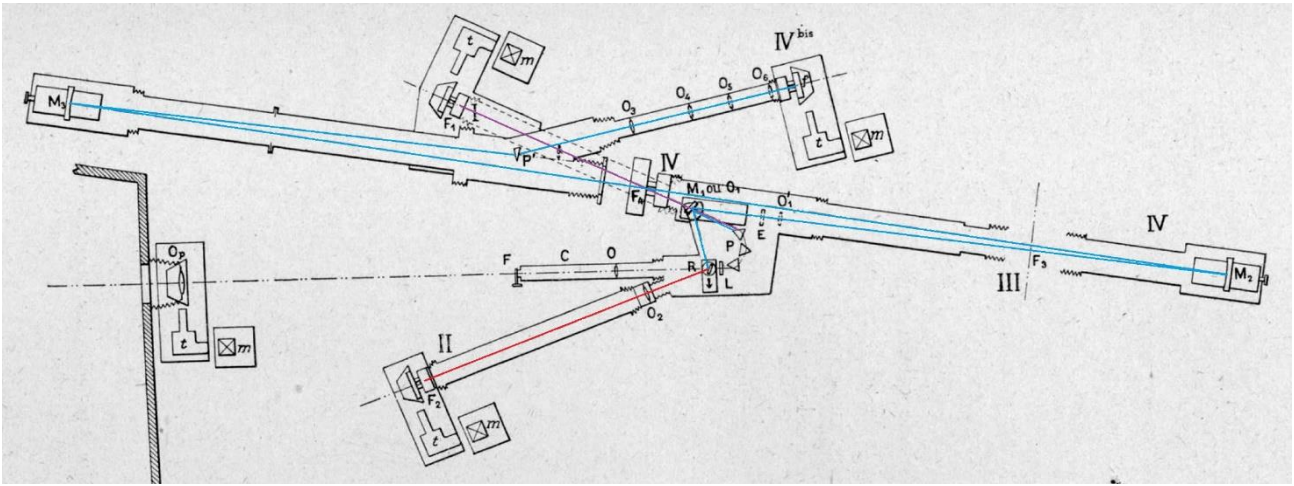

*Figure 34* : the large quadruple spectrograph of Deslandres and Azambuja (1908). Spectrographs I and II: systematic observations of the Hα and CaII K lines (3 m focal length). Spectrographs III and IV: research observations, the spherical mirror of spectrograph IV has a focal length of 7 m. Its total length is 14 m. The collimator is common to all combinations (1.3 m focal length). The imaging lens has a focal length of 4 m. The whole is fed by the coelostat with two mirrors of 40 and 50 cm of *Figure 35*. Credit OP.

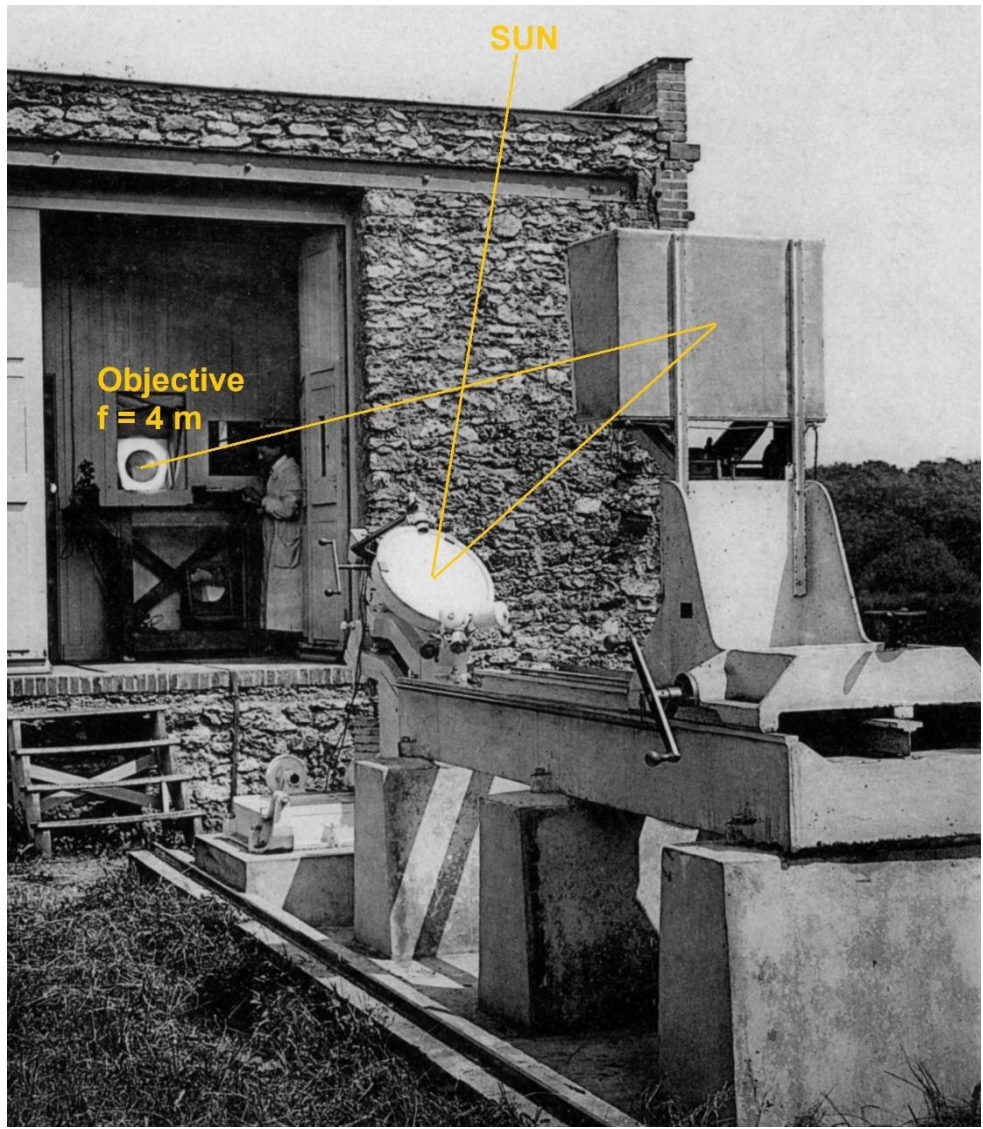

*Figure 35* : coelostat with two mirrors feeding the 4 m objective of the large quadruple spectrograph of Deslandres and d'Azambuja (1908). OP Credit.

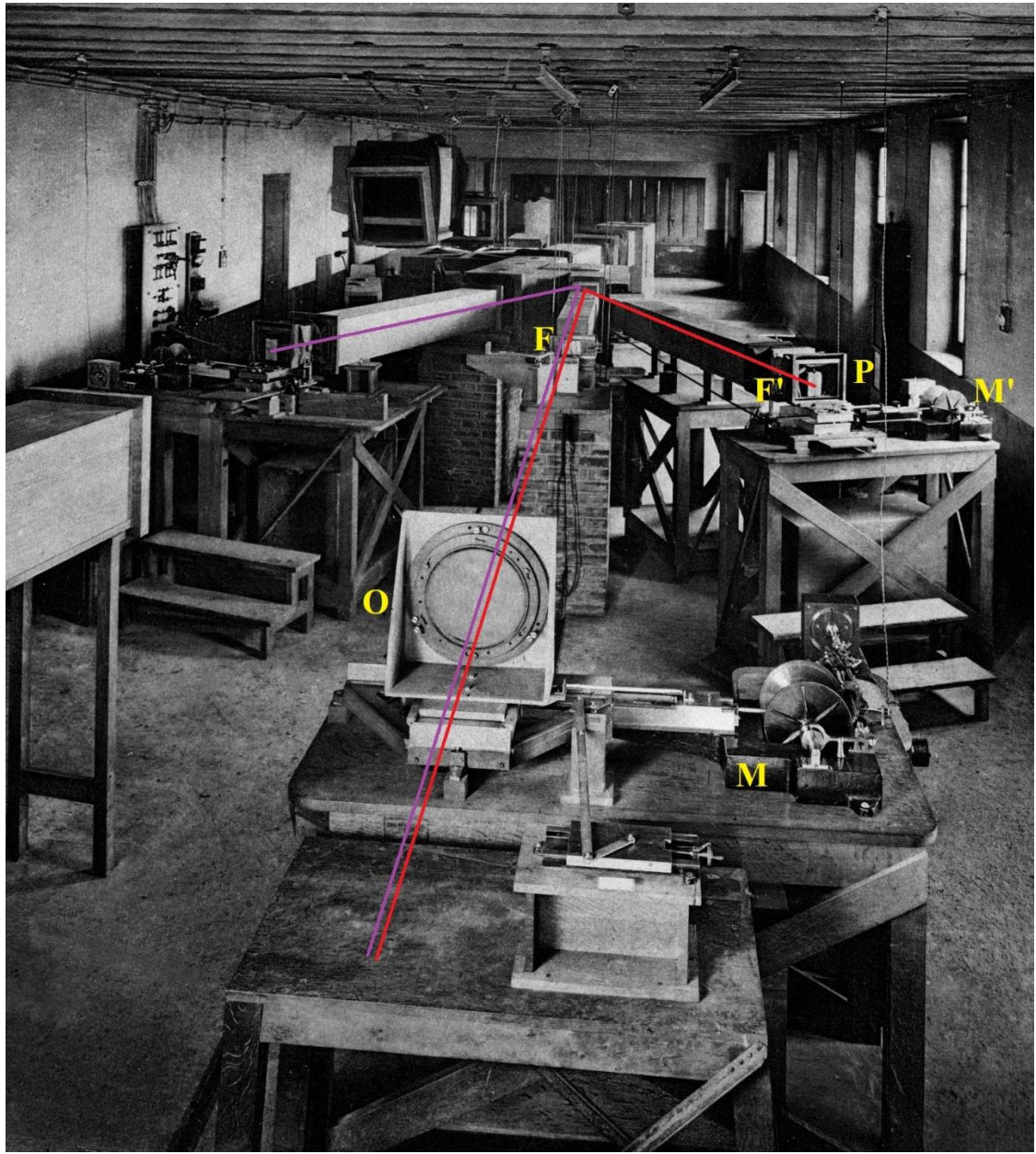

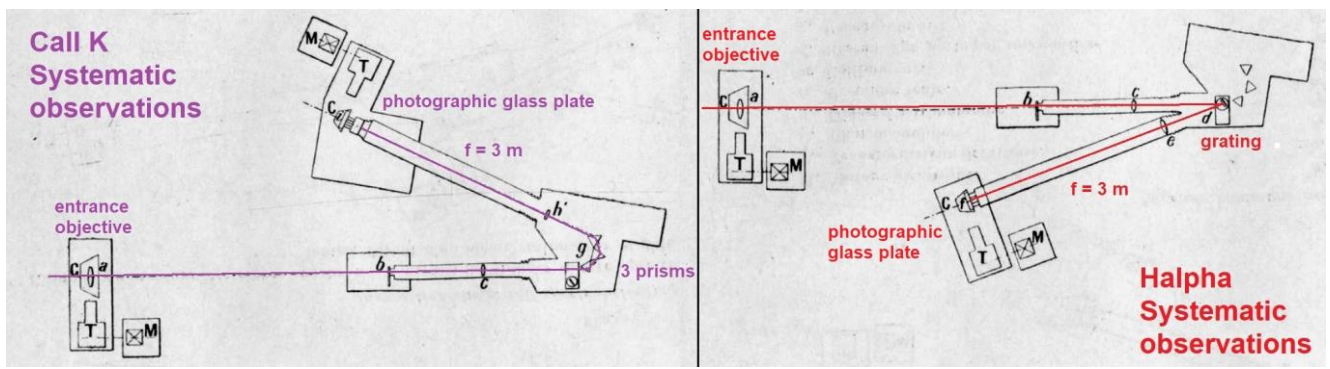

*Figure 36 : the large quadruple spectroheliograph of 1908 in its version dedicated to systematic observations. Two 3 m chambers for Hydrogen and Calcium. There were two dispersive elements, a train of 3 prisms (CaII K) and a plane Rowland grating (Hα). In front of us we see the imaging lens O of the Sun, common to all the chambers, with a focal length of 4 m. It was in right/left translation to move the Sun at a constant speed (a few minutes) on the slit of the spectrograph (M engine). We see the 13 x 18 cm² plate holder (P and motor M') for Hα and the associated output slit F' in the spectrum. F is the common entrance slit. The current instrument retains these basic principles but has been renovated several times. Credit OP.*

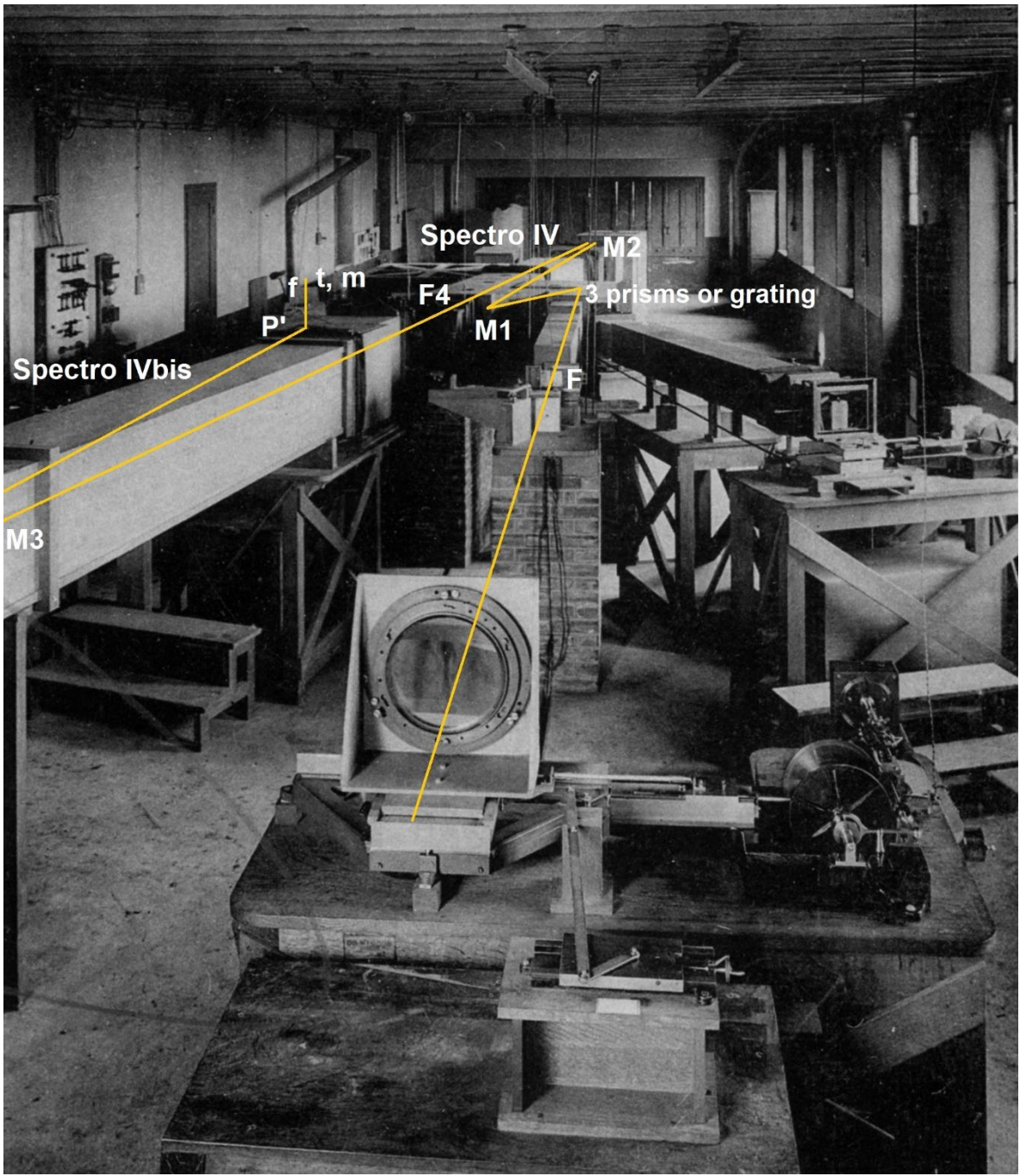
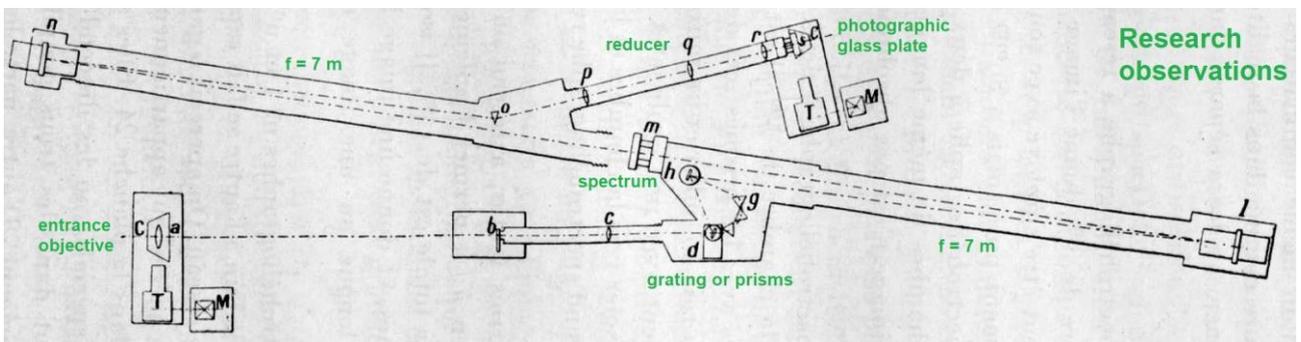

*Figure 37 : the large quadruple spectroheliograph of 1908 in its version dedicated to research observations. A concave chamber with a focal length of 7 m and two dispersive elements, a train of 3 prisms and a Rowland plane grating. F is the entrance slit. M1 is a plane mirror; M2 a concave mirror with a focal length of 7 m. The spectrum is formed in F4. The assembly consisting of M3 (second concave mirror of 7 m) and an objective (p, q, r or s) is an afocal reducer. P' is a prism to select the order of interference. OP Credit.*

D'azambuja was responsible for organizing a service of daily and systematic observations as early as 1908 (Figures 38 and 39). Images of the solar atmosphere were recorded in two lines, Calcium and Hydrogen, which give a synthetic view of the solar events in progress. In 2026, this service is still active, but it has been digitized since the year 2000. The instrument recorded more than a hundred thousand monochromatic pictures. The old glass plates and 13 x 18 cm² film images have been scanned. Marguerite d'Azambuja succeeded Lucien, her husband, at the head of the service, when he retired in 1954; she herself left in 1959, and it was Marie-Josèphe Martres who continued, with Gualtiero Olivieri until 1990; then Elisabeth Nesme-Ribes until 1996 followed by Jean-Marie Malherbe until 2023; currently Guillaume Aulanier is in charge of the service.

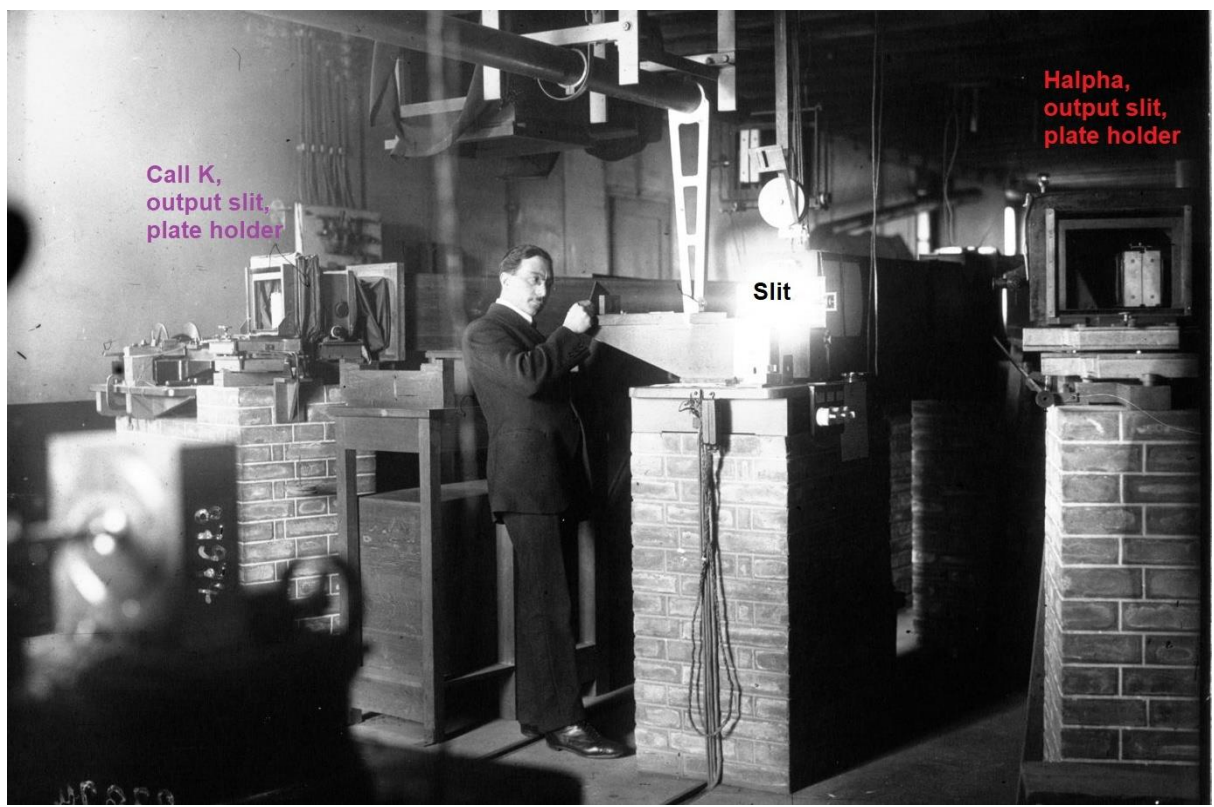

*Figure 38 : Lucien d'Azambuja observing in 1921. On the left, the Calcium chamber, on the right the Hydrogen Hα chamber, we see the associated slits in the spectrum and the plate holders. Credit Gallica/BNF.*

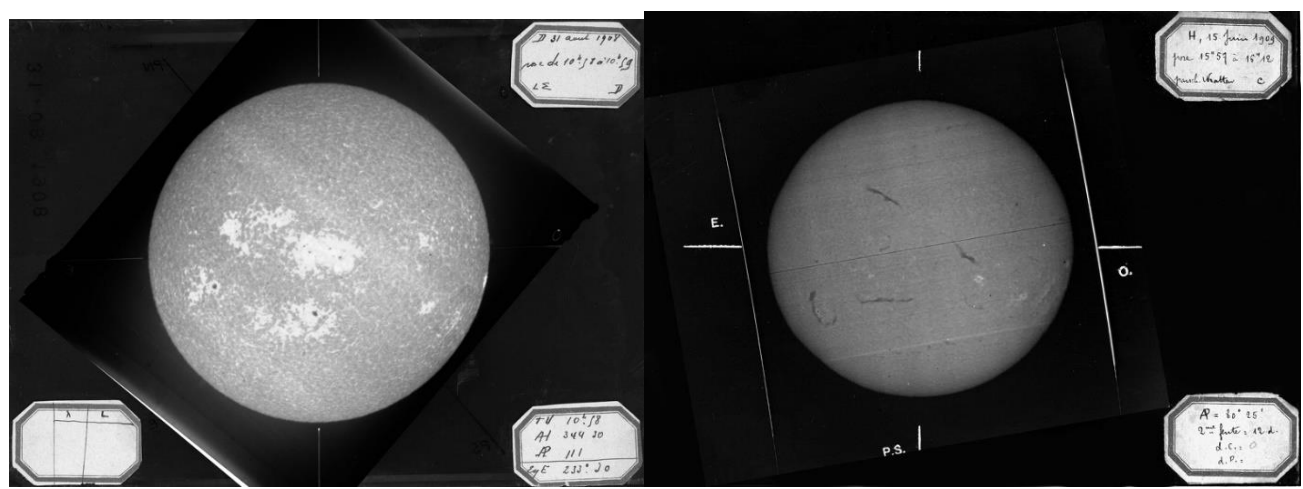

*Figure 39 : The first spectroheliograms from the huge collection of the Meudon spectroheliograph (1908-2026), on the left in Calcium (1908) and on the right in Hydrogen Hα (1909). Credit: OP.*

Until 1939, the systematic observations were those of Figure 40. Daily images were taken in CaII K1v (purple wing of the line), CaII K3 (center line) and Hα, as well as a spectroheliogram by sections (80 full spectra by 30" steps) for the measurement of radial velocities. The latter, requiring far too much analysis work, was abandoned, due to lack of personnel, during the Second World War in favour of a CaII K3 long pose to make the prominences (10 times less intense than the disk) appear clearly at the limb.

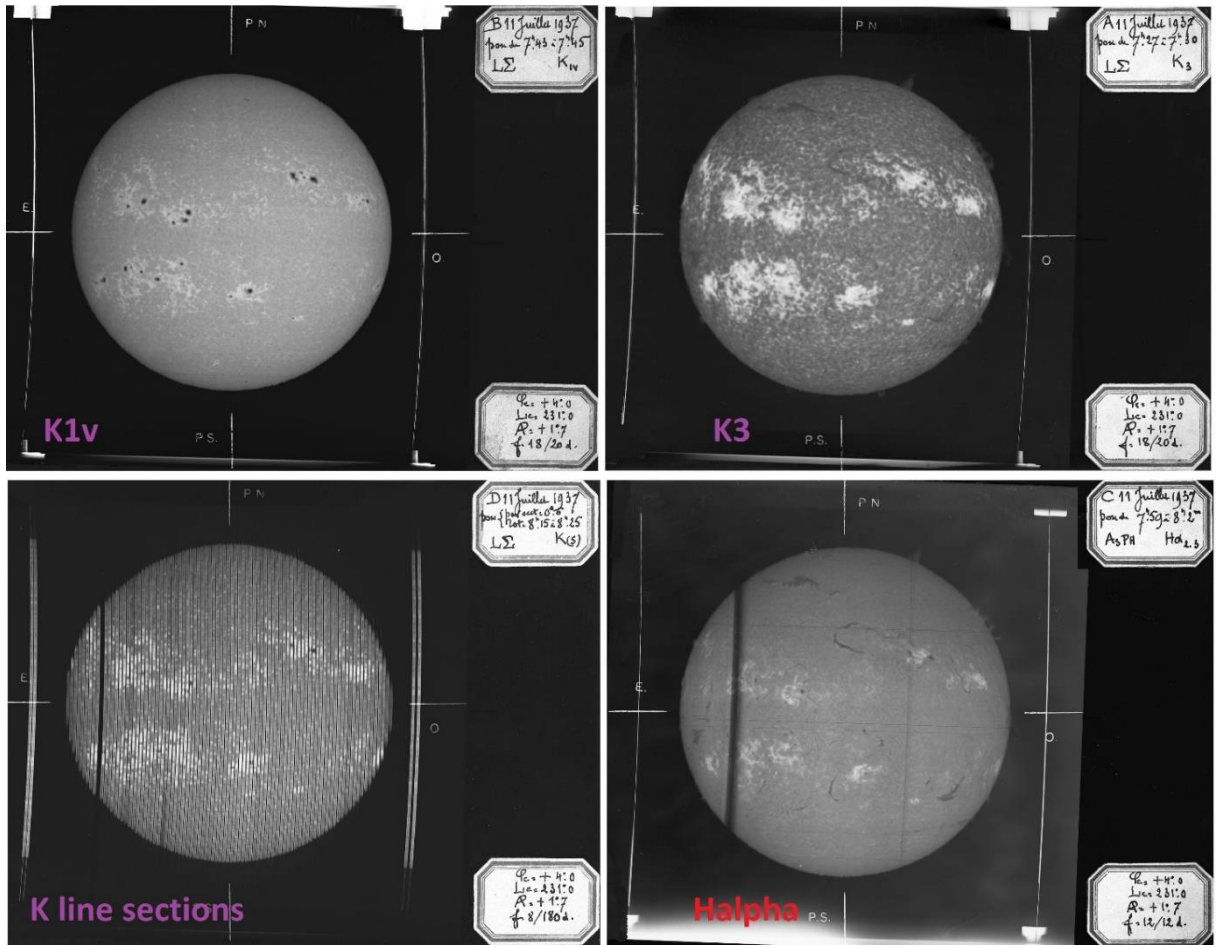

*Figure 40 : Spectroheliograms of the shapes, according to the observed lines, reveal the layers of the solar atmosphere. Bottom left, the spectroheliogram by sections for radial velocities. OP Credit.*

**6 – Lyot, the coronagraphy and monochromatic imaging**

Between 1930 and 1950, Bernard Lyot (1899-1952, figure 41) distinguished with two remarkable inventions that toured the astronomical world and that still equip many instruments on the ground and in space today: the coronagraph and the birefringent monochromatic filter.

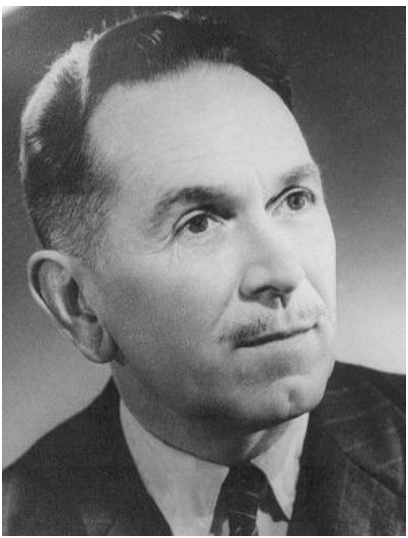 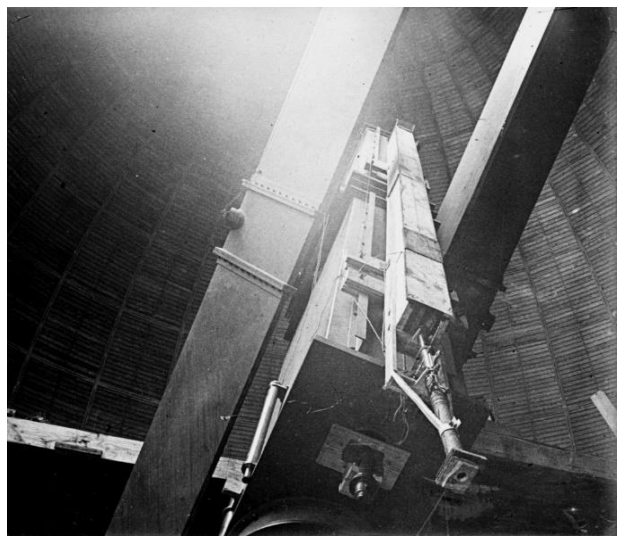

*Figure 41*

*Bernard Lyot (photograph by Harcourt) and his first coronagraph, fixed here in 1930 on the mount of the "carte du ciel" telescope in Paris. Credit OP.*

It is said of Lyot that he was "a charming and shy man despite his celebrity (he was an academician). His distraction was proverbial, if he borrowed something from you, it was better to follow him and get the object back before he had made the trip to the Pic du Midi...". The coronagraph (Figure 42) was created in 1930 in Meudon (Lyot, 1932). We know that the solar corona was only observable during total eclipses, because it is very faint. The coronagraph is an instrument whose optical formula is specially designed and optimized to observe the low solar corona (a million times less bright than the disk), outside eclipses, provided that it benefits from a clear and pure sky (this is the case in high mountains). This is the reason why the device was designed in Meudon, but tested and used at the Pic du Midi at 2870 m. However, in 1950, Lyot completed the coronagraph with a coronameter capable of isolating the linearly polarized light component of the corona, allowing it to be observed even at sea level, as in Meudon.

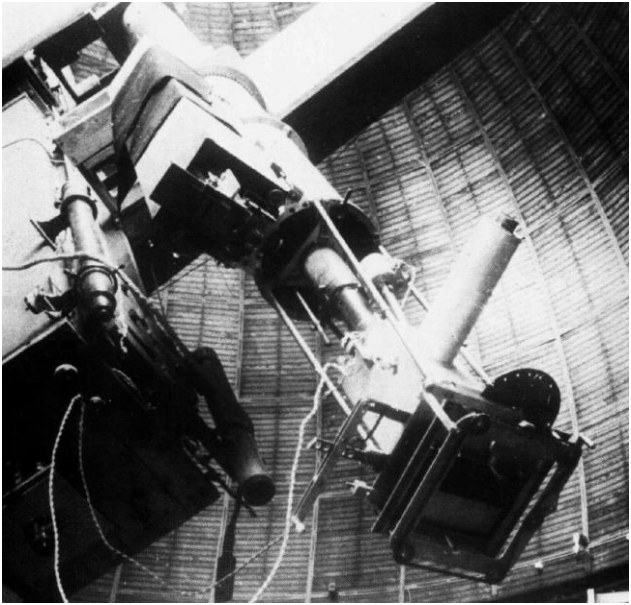

*Figure 42*

*Lyot's coronagraph at the Pic du Midi in 1936 and schematic diagram. A is the special glass input lens. JB is a rejection mask (or cone) with a diameter equal to that of the solar image. The objective C forms an image of the pupil in AA' which is diaphragmed (to remove diffraction rings) and whose centre is masked by the Lyot stop E (elimination of defects). F is an imaging objective forming the coronal image in BB'. Credit OP.*

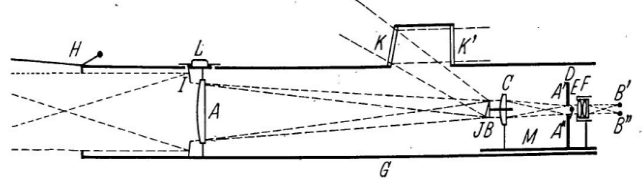

The second invention dates back to 1933, it is the birefringent monochromatic filter (Figures 43 and 44) forming a grooved spectrum, with thin equidistant peaks when many optical stages are present between polarizers. It is based on the interference of the ordinary ray and the extraordinary ray in linearly polarized light (Lyot, 1944). If the coronagraph allows the lower corona to be observed in white light under good conditions, it must be supplemented with a filter to isolate the spectral lines emitted by the solar structures visible in the limb, resulting from cold prominences (Hydrogen lines, 8000 degrees) or hot magnetic loops (highly ionized iron lines, 1 million degrees, Figure 45), without which it is not possible to discern them. The filter is also able to isolate the absorption lines of the solar disk that reveal the dark filaments and active regions, but remains less selective than the spectroscopy-based spectroheliograph of Deslandres and d'Azambuja. These two inventions by Lyot revolutionized solar physics, they have been adopted by the majority of observatories in the world, and they can even be found on modern satellites or space instruments (coronagraphs of SOHO, 1996, and of STEREO A and B, 2006; adjustable Lyot filter of HINODE, 2006).

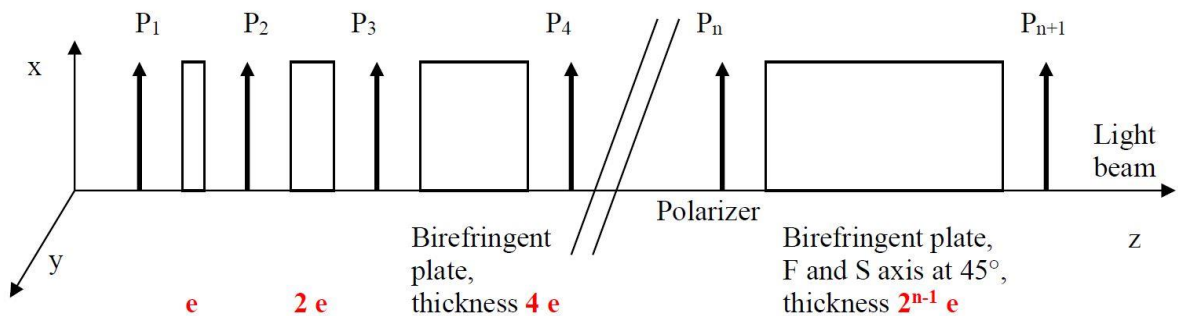

*Figure 43 : principle of the n-stage birefringent Lyot filter. The thickness of the quartz or calcite blocks varies in geometric progression of reason 2. $P_n$ parts are linear polarizers. OP Credit.*

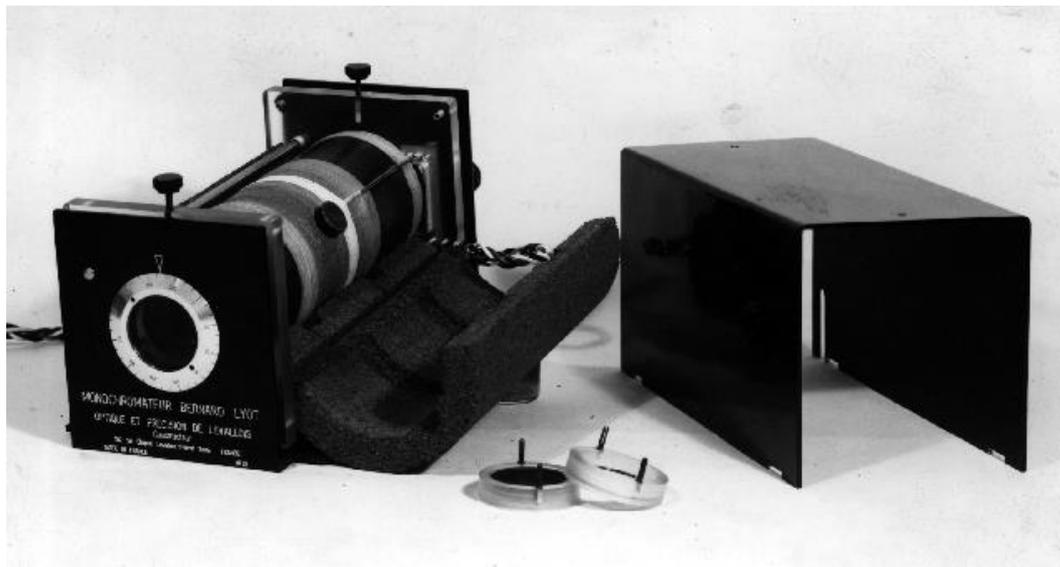
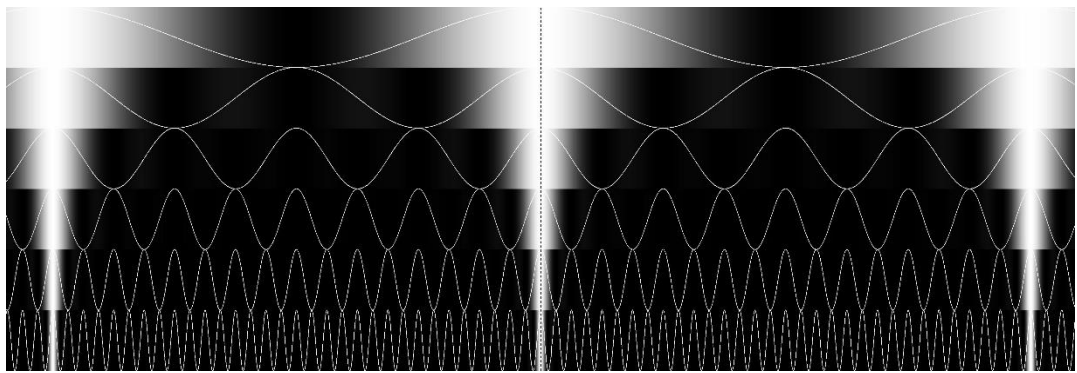

*Figure 44 : Lyot's monochromatic filter produced by the firm Optique de Précision Levallois (OPL) to equip many observatories, and its typical grooved spectrum with transmission peaks. Credit OP.*

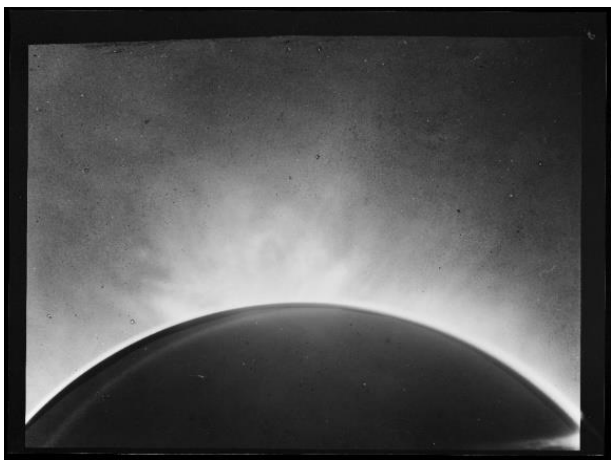

*Figure 45*

*Monochromatic image, in the green line of Fe XIV 13 times ionized at 530.3 nm. The hot magnetic loops and structures of the lower solar corona are observed with a Lyot filter installed on the coronagraph of the Pic du Midi in 1941 (about one Angström bandpass). The solar disk is hidden by an occulting cone, one of the main pieces (with the non-scattering optics) of the instrument.*

*Credit OP*

     The new Lyot filter had exceptional performance, since it was possible to produce monochromatic and instantaneous images of regions of the Sun much faster than the spectroheliograph (because the latter slowly scans the surface of the Sun with a slit, but in return has a better spectral selectivity). Indeed, at a time t, a filter provides an image (x, y) for a value of the wavelength λ, while a spectrograph gives a spectrum (λ, x); it is then necessary to scan the Sun along the y-axis with the input slit. Lyot was able to exploit this advantage by starting in 1937, at the Pic du Midi, a program of cinematography of prominences in the red Hα line of Hydrogen. He associated the professional filmmaker Joseph Leclerc with his work. The cinematographic possibilities of Lyot filters were put forward for the International Geophysical Year (IGY) of 1957. The aim was to film the flares and ejecta on the disk in the vicinity of the solar maximum. Lyot unfortunately died in 1952, but Henri Grenat and Gérard Laborde developed in 1954 in Meudon a monochromatic heliograph (figure

46) based on the Lyot filter (figure 44) and operating in the Hα line. It was precisely thermally controlled and recorded the images on 45 m long and 35 mm wide film reels (Grenat & Laborde, 1954).

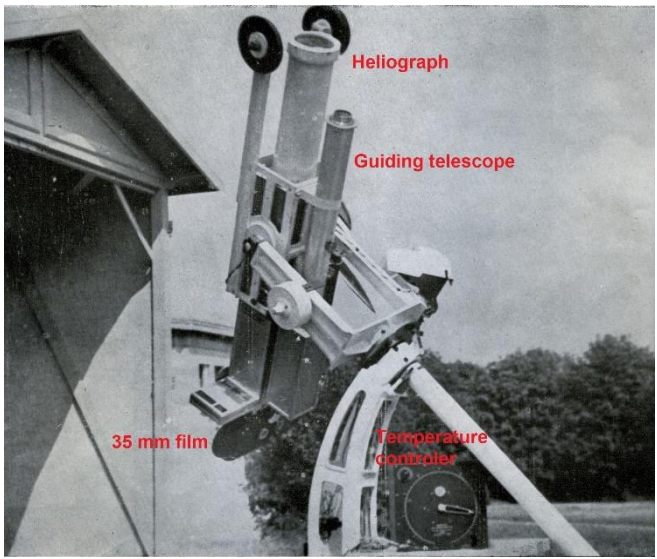

*Figure 46*

*Monochromatic heliograph made by Grenat and Laborde (1954) equipped with a Lyot filter to film solar flares in the Hα line for the International Geophysical Year of 1957. A large success, as the instrument worked during 50 years ! Thousands of 45 m reels were subsequently exposed until 2004, representing 130 km of 35 mm film and 6 million images. A CCD camera was implemented in 1999. Technical details are given by Malherbe (2023). Credit OP.*

It was decided, as part of the monitoring program of eruptions within the International Geophysical Year of 1957, to duplicate the Meudon Lyot filter heliograph to equip many observatories around the world. This task was given to the companies SECASI (Bordeaux), which made 6 refractors and mounts, and OPL (Levallois), which made more than 10 monochromatic filters for the Hα line. One of these instruments (Figure 47) was placed at the Observatoire de Haute Provence (OHP) where it operated from 1958 to 1994. The original instrument from Meudon was renovated several times, with increasingly sophisticated filters, and ran until 2004. In particular, wavelength-adjustable filters to explore not only the cores of the lines, but also their wings, were used (Figure 48), in order to estimate mass motions by the Doppler effect (0.5 Angström bandpass for the best filter). A total of 7 million images were taken in Meudon and at OHP, of which 6 million were on film (3000 reels of 45 m) and 1 million with a CCD digital detector (Figure 49).

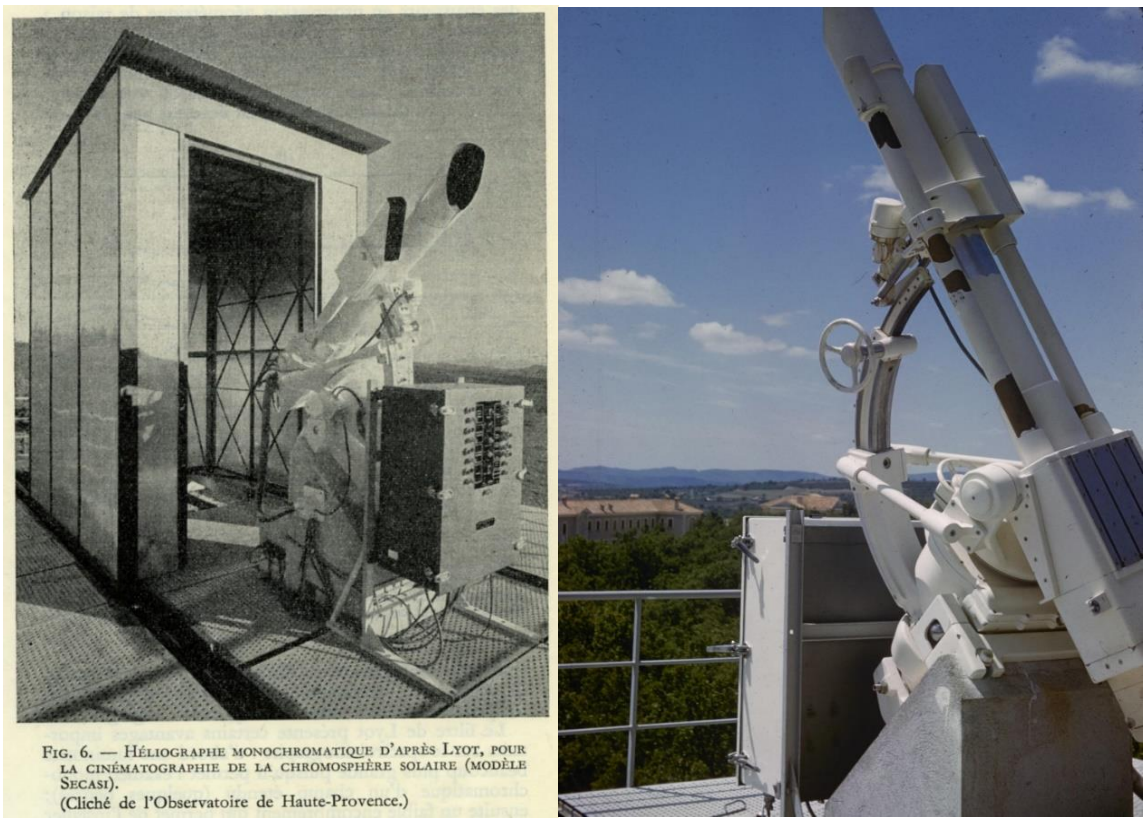

*Figure 47 : Hα heliograph from SECASI/OPL at the Observatoire de Haute Provence (credit OP).*

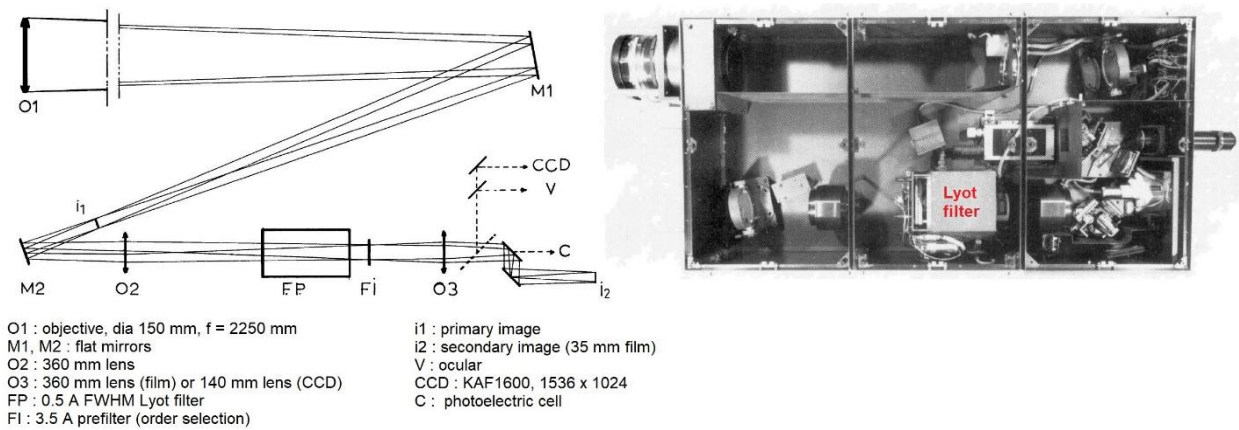

*Figure 48 : heliograph with wavelength-adjustable filter at Meudon from 1985 to 2004 (credit OP).*

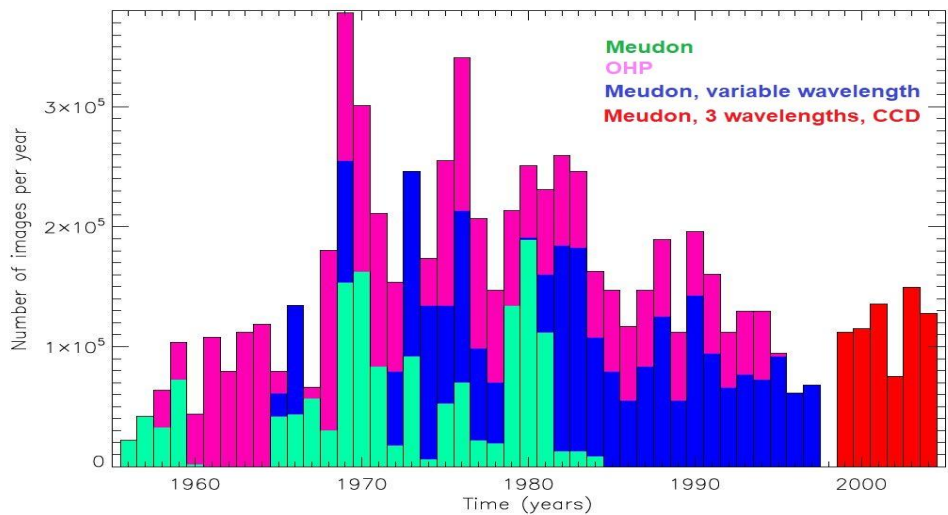

*Figure 49 : annual number of images taken by heliographs between 1956 and 2004 (credit OP).*

### 7 – Some developments in polarimetry at Meudon observatory

Lyot, who died prematurely in 1952, left much of his knowledge in polarimetry to his student Audouin Dollfus (1924-2010), who continued the coronameter and later made a selective solar polarizing filter (FPSS, figure 50). It was first installed in Meudon at the focus of the refractor fixed on the mount of the 1 m telescope, then at the turret dome refractor (50 cm aperture) of the Pic du Midi observatory, giving better images.

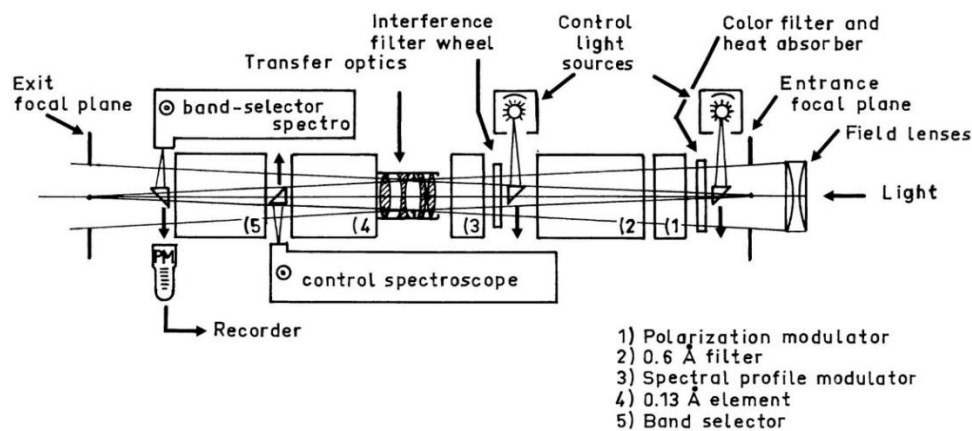

*Figure 50 : highly selective FPSS monochromatic filter (0.13 Angström) by Audouin Dollfus (credit OP).*

Pierre Charvin (1931-1990) developed the monochromatic and photoelectric coronameter (a coronal polarimeter) left by Lyot to Dollfus (Figure 51), with the aim of measuring the polarization of light in a coronal line (the green line of Fe XIV) isolated by a monochromatic Lyot filter (Charvin, 1965). Polarization gives access to the direction of the magnetic fields in the corona. Charvin was later director of the National Institute of Astronomy and Geophysics of the CNRS (INAG, which became later INSU), then president of the Paris Observatory from 1984 to 1990. Charvin installed the coronameter at the Pic du Midi to benefit from a clear sky; his work was taken up by Jean Arnaud (1945-2010).

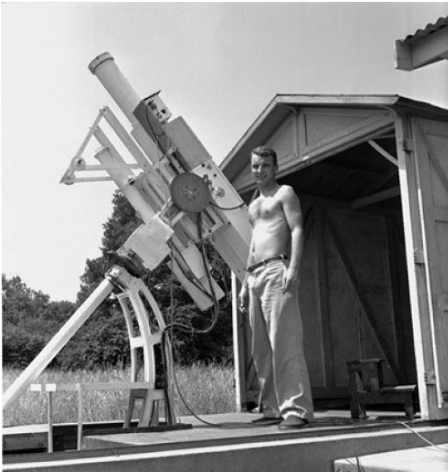

*Figure 51*

*Pierre Charvin and the photoelectric coronameter with a narrow monochromatic filter. Credit OP.*

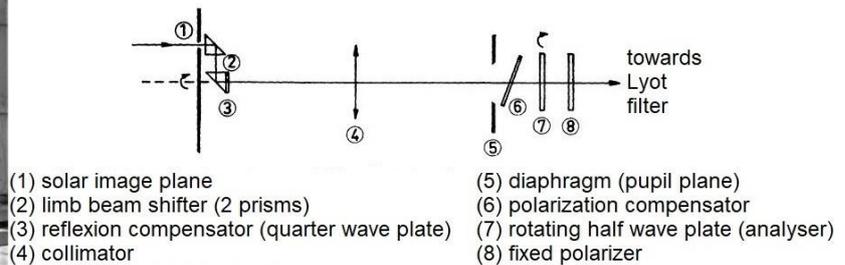

(1) solar image plane
(2) limb beam shifter (2 prisms)
(3) reflexion compensator (quarter wave plate)
(4) collimator
(5) diaphragm (pupil plane)
(6) polarization compensator
(7) rotating half wave plate (analyser)
(8) fixed polarizer

Jean-Louis Leroy (1936-2024) developed a photospheric polarimeter (Figure 52) to measure the polarization of light over a wide spectral range using colored filters, which gives access to transverse magnetic fields in the dark spots (Leroy, 1962). It is not possible to measure by this method the longitudinal magnetic fields that require a high-resolution spectrograph to discern the Zeeman effect along the resolved profiles of the lines (the mean value of Stokes V is null over unresolved lines). Leroy continued his career at the Pic du Midi where he studied, among other subjects, the polarization of the light of prominences, using the photo electric polarimeter developed by Ratier (1975), and analyzed the magnetic fields of prominences by interpretation of the Hanle effect, according to the theory developed by Bommier and Sahal in Meudon.

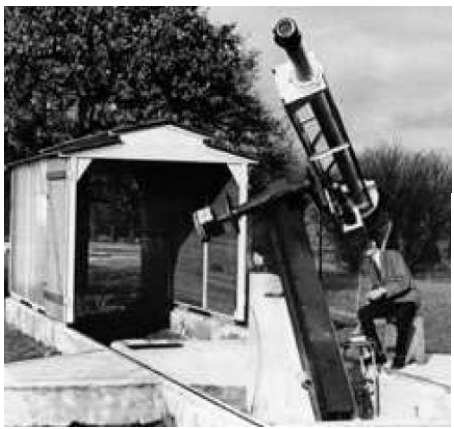

*Figure 52*

*Jean-Louis Leroy and the photoelectric photospheric polarimeter with colored filters in Meudon. Credit OP.*

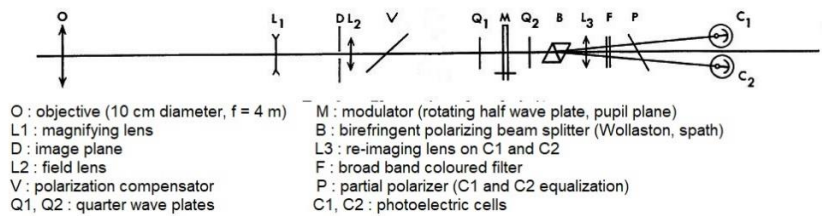

O : objective (10 cm diameter, f = 4 m)
L1 : magnifying lens
D : image plane
L2 : field lens
V : polarization compensator
Q1, Q2 : quarter wave plates
M : modulator (rotating half wave plate, pupil plane)
B : birefringent polarizing beam splitter (Wollaston, spath)
L3 : re-imaging lens on C1 and C2
F : broad band coloured filter
P : partial polarizer (C1 and C2 equalization)
C1, C2 : photoelectric cells

## 8 – Felenbok and the coronography at the Saint Véran station

In the early 1970s, INAG launched a site survey for a 3.60 m large night telescope project, that was finally completed in Hawaii in collaboration with Canada (Canada France Hawaï Telescope). Several places were prospected in terms of weather and image quality, including the Pic de Château Renard at 2930 m above the village of Saint Véran in the Queyras massif (Hautes Alpes, France). If this site was not suitable for a giant night telescope because of its difficult access, on the other hand it was a promising place for the installation of a small solar station. At that time, there were no digital detectors of the CCD or CMOS type, the most sensitive one being André Lallemand's electronic camera, consisting of a vacuum photocathode and an electronic optics focusing electrons on a photonuclear emulsion (Lallemand, 1960). Paul Felenbok (1936-2020) had the idea of applying this camera to the detection of faint structures in the solar corona or features located at a great distance from the limb (a solar radius). The project of a coronagraph with an electronic camera was born (Figure 53, Malherbe & Bellenger, 2025) and accompanied by a small spectrograph (Figure 54); this project was impossible to achieve at the Pic du Midi due to the lack of an available dome.

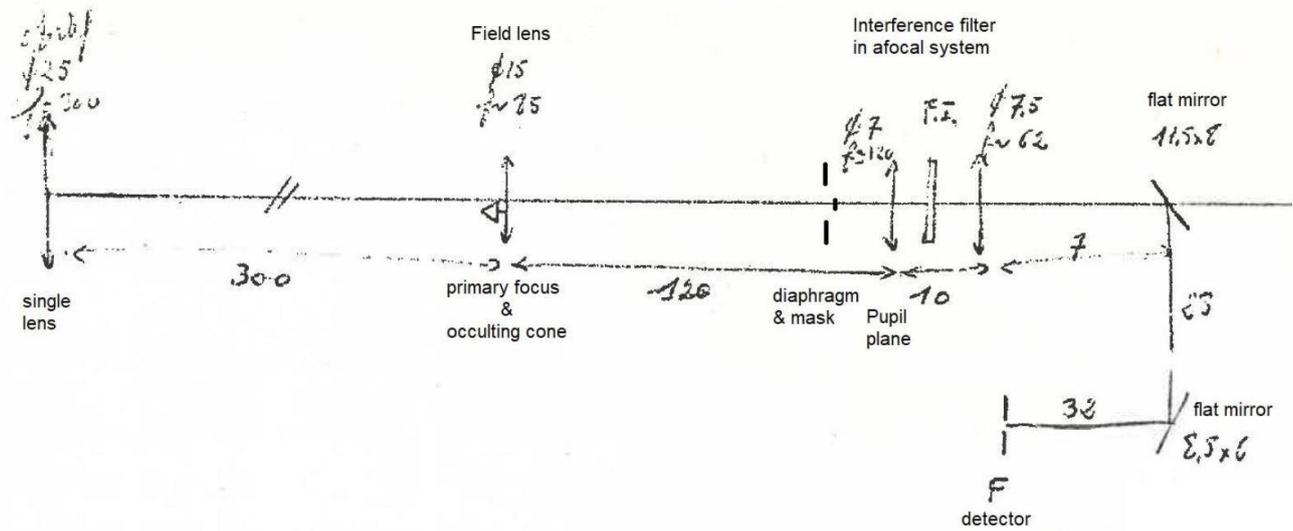

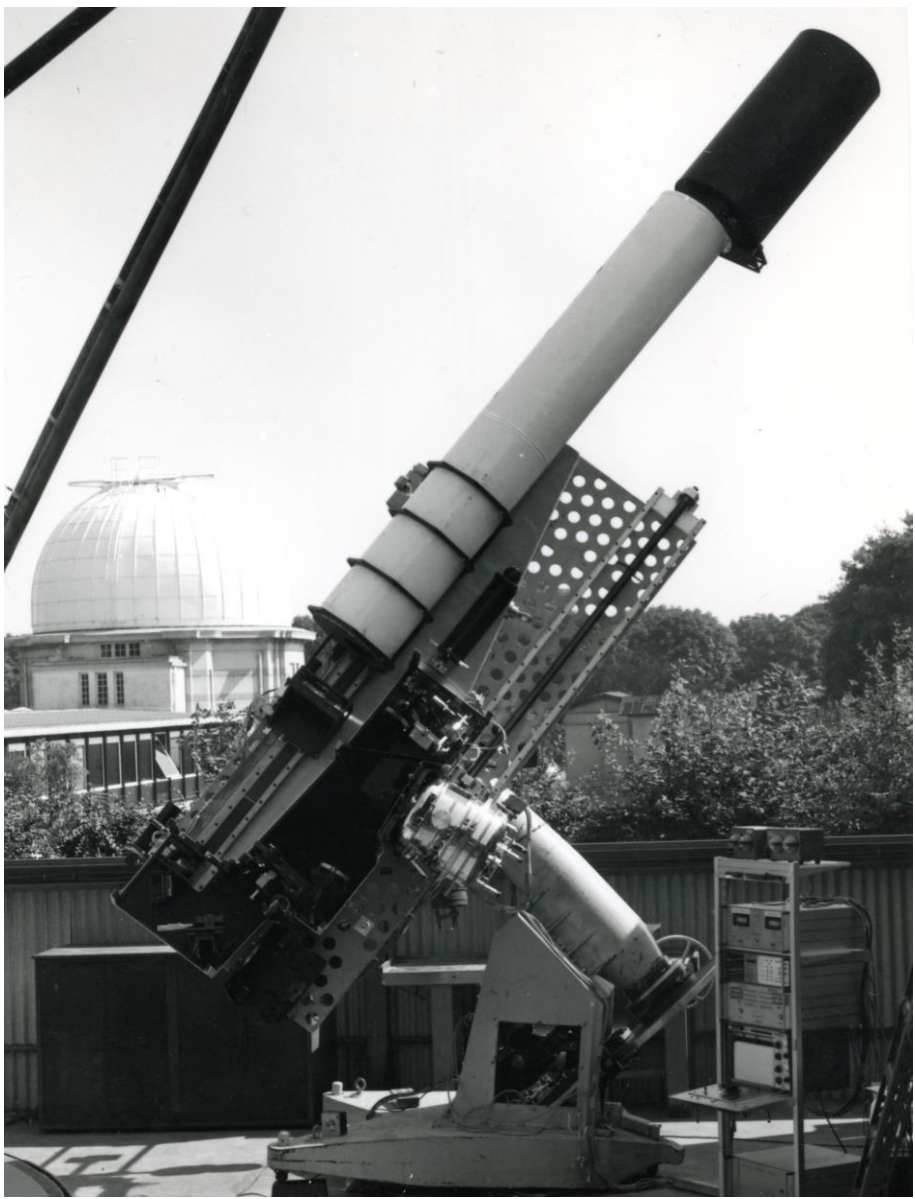

*Figure 53*

*Top: optical diagram of the Saint Véran coronagraph, a classic Lyot coronagraph with an aperture of 25 cm with two exits, one for imaging through a narrow interference filter, and the other for spectroscopy. The coronagraph had a focal length of 3 m, which was reduced to 1.5 m at the output by the two-lens afocal system located on either side of the filter.*

*Left: the coronagraph made and assembled in Meudon before transport to Saint Véran in 1976. The electronic camera is placed on this photo at the spectroscopic focus. It was a valve camera allowing a rapid plate change, which is essential in coronal physics because the structures can be very dynamic, with plasma ejections.*

*Credit OP.*

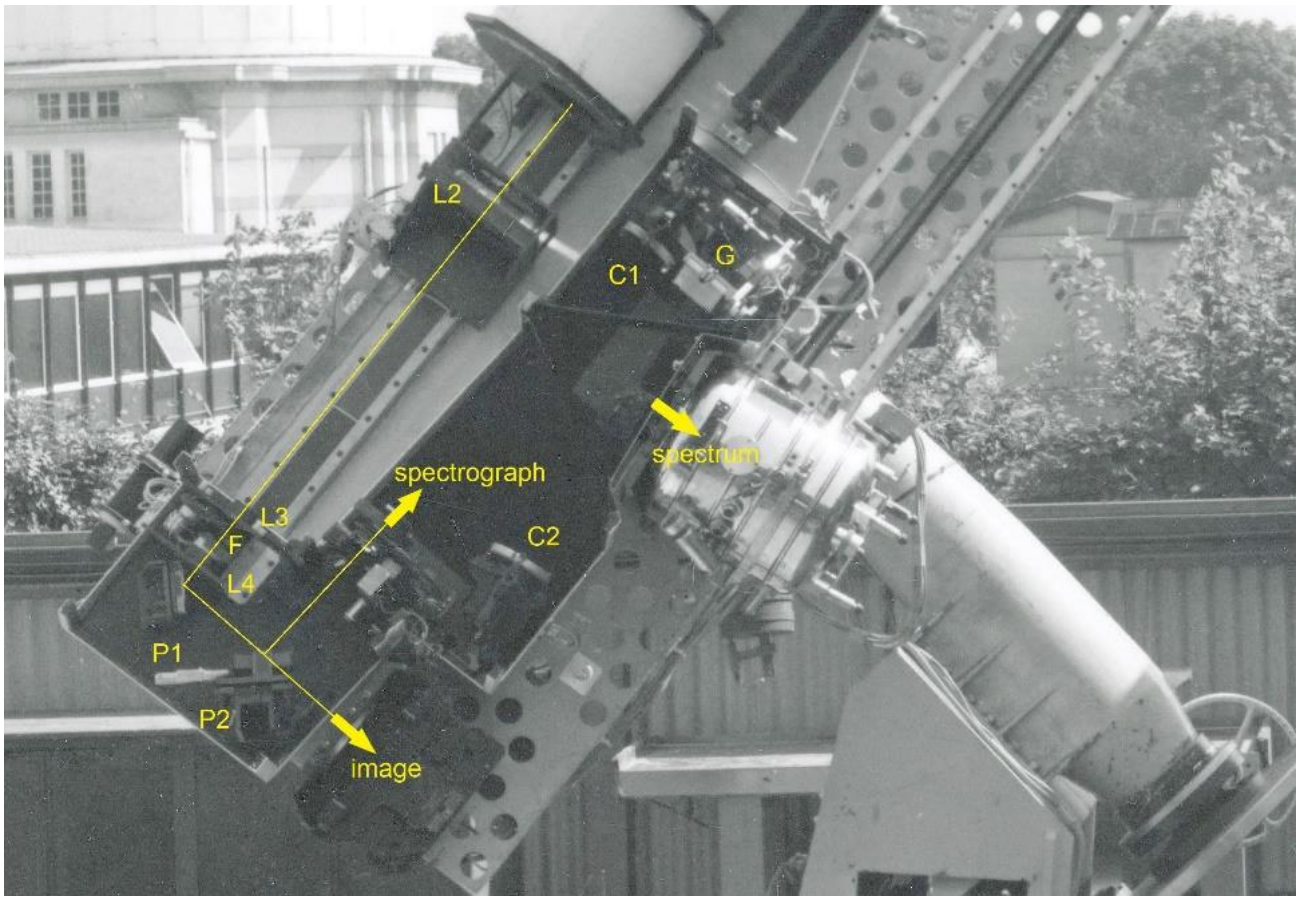

*Figure 54 : Detail of Figure 53 and the two outputs, imaging and spectroscopy, with the electron camera mounted in the spectrum. The focal length of the spectro is 1 m. C1 and C2 are the chamber mirror and the collimator. The grating is in G, it operates in order 2. Credit OP.*

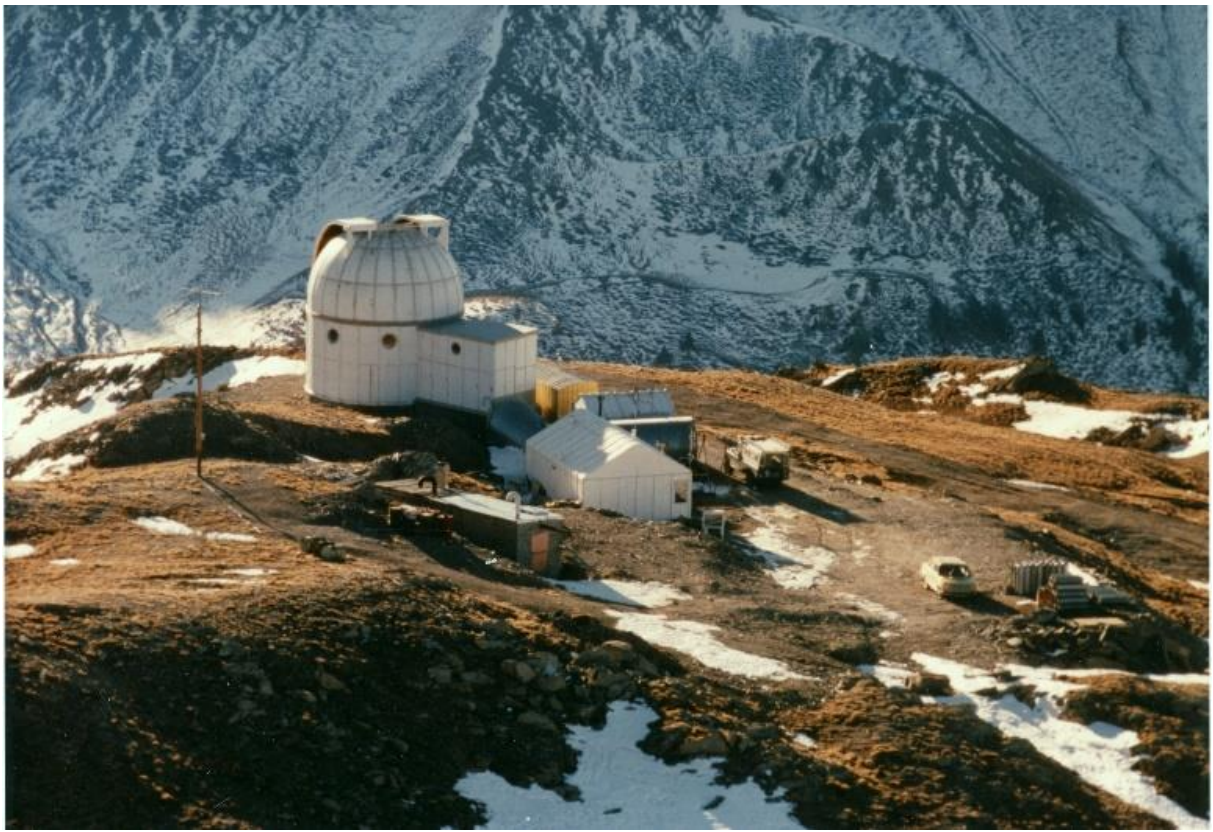

*Figure 55 : the station in 1976. The dome and its annex come from the Paris Observatory. OP credit.*

In 1974, the 7 m metal dome and its two-level annex of 5 x 5 m², located on the East Tower of the Paris Observatory, the Perrault building, without use, were dismantled, transported to Saint Véran and then reassembled during the summers of 1974 and 1975 (Figure 55). A living base was built, but the station was only available in the summer, because the access track could not be cleared of snow. Electricity was provided by a fuel generator, and the heating by gas bottles. At the same time, the coronagraph was made in Meudon and tested before its transport and installation in the dome during the summer of 1976 (figure 56).

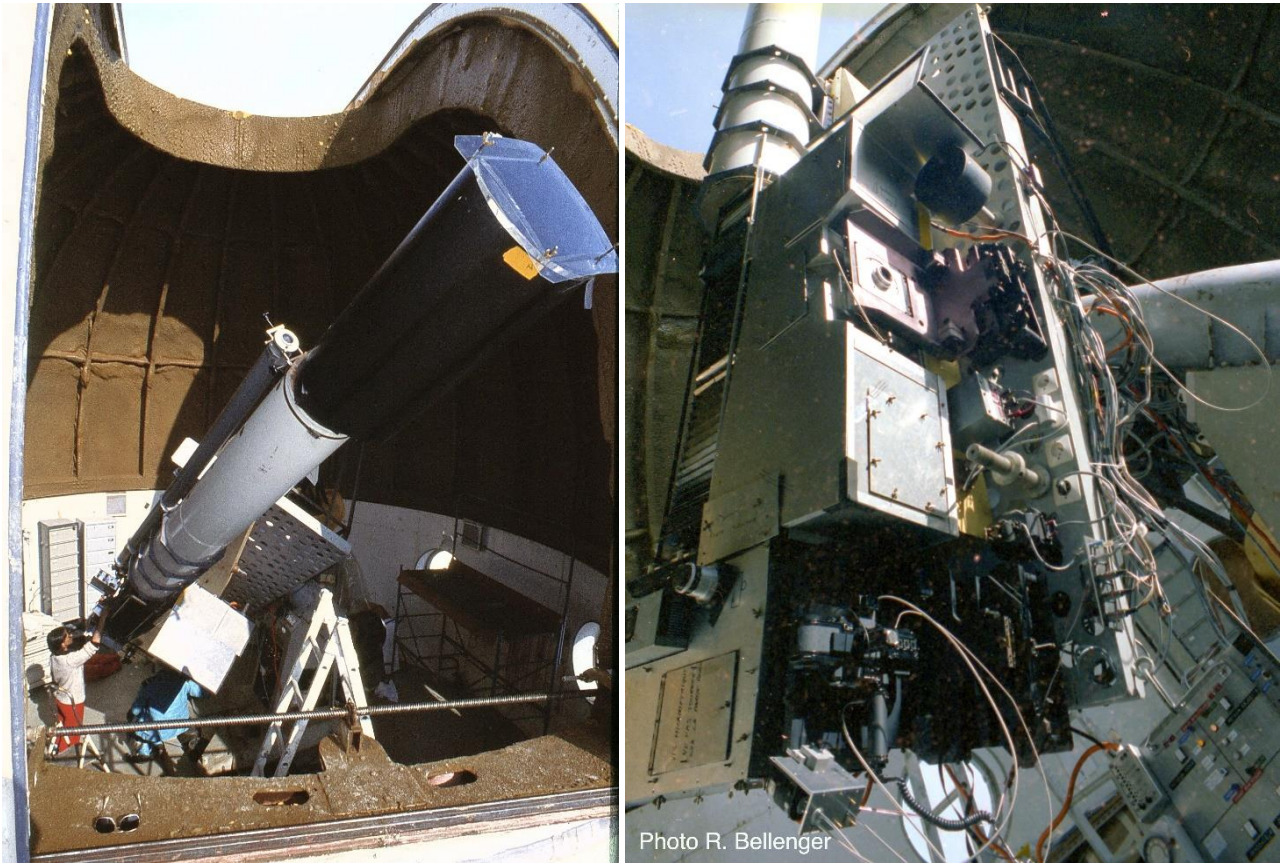

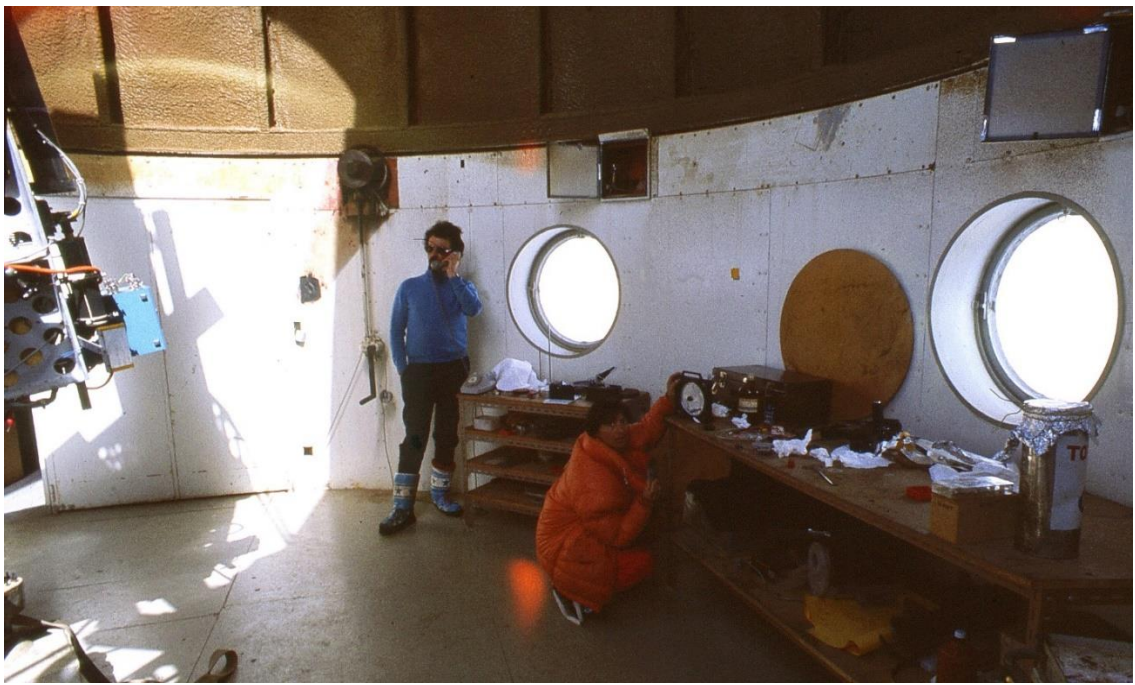

*Figure 56 : the cupola and the coronagraph in 1976. At the top left, Mireille Dantel makes adjustments. At the top right, the imaging exit with a Nikon camera, the spectrograph (closed box) and the two mounts for electronic cameras. Below, Jean-Pierre Picat and Mireille Dantel in the dome. Credit Rémy Bellenger.*

Unfortunately, the complicated logistics of the site overcame the great efforts made to operate the electronic camera, a real novelty of the operation, and the project was abandoned in 1980, also threatened by the emerging competition from NASA's Solar Maximum Mission satellite launched that year. The Felenbok team was also busy starting observations with the CFHT's valve camera, of the same type as in Saint Véran, but greatly improved, and which proved to be a great success (the valve camera allows the plates to be changed without breaking the vacuum). The observatory was taken over by the amateur AstroQueyras association, to which the Paris Observatory left the dome, which today houses a 60 cm night instrument lent by OHP.

**9 – the solar balloon observations by Dollfus, the beginnings of the space age**

In parallel with ground-based observations, the appearance of balloons has encouraged astronomers to try new and innovative experiments. The balloon can rise to very high altitudes, above most of the Earth's atmosphere, making it possible to capture radiation that does not reach the ground or to improve the quality of the images by eliminating turbulence from the air (although it can be complicated by the vibrations of the balloon). The first flight undertaken by an astronomer dates back to December 2, 1870. Paris being besieged by the Prussians, Jules Janssen obtained from the government the provision of a balloon to escape enemy lines and observe a total eclipse of the Sun in Oran, Algeria (Figure 57). This was the only scientific expedition of the 70 balloons, mounted during the siege of Paris.

*Figure 57*

*Below, the Gare d'Orléans in Paris was transformed into a balloon assembly site during the siege of Paris in 1870.*

*On the right, the stele of the statue of Janssen on the first terrace of Meudon, located below that of the observatory. Janssen took off aboard the balloon "le Volta" with his instruments to observe a total eclipse in Oran.*

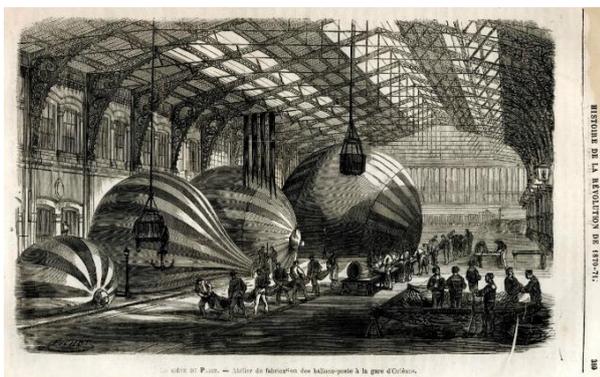
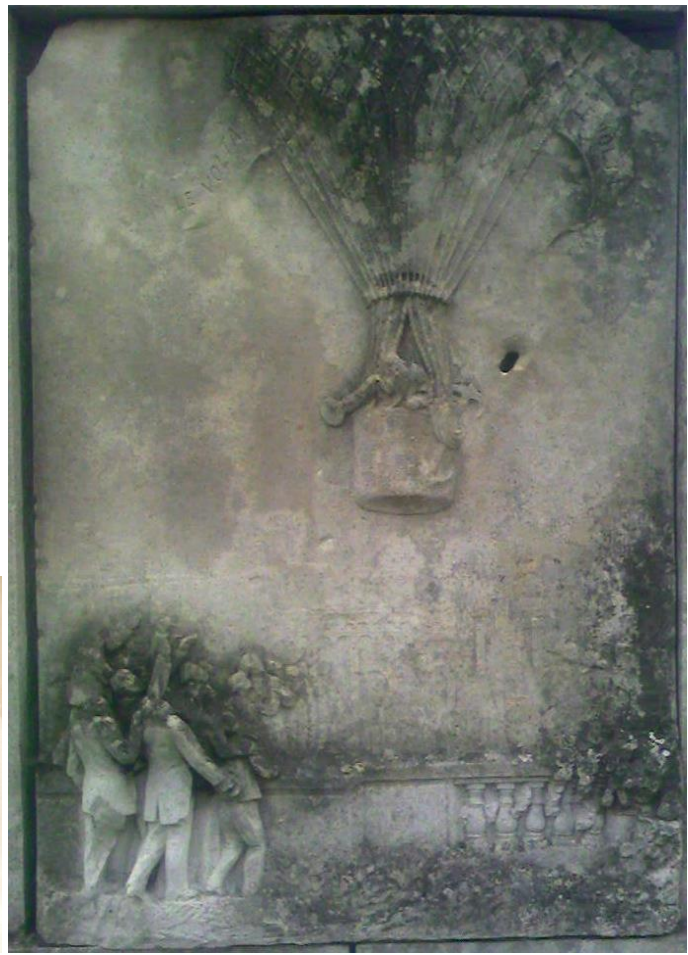

The balloons cannot be directed and are subject to the winds. Janssen flew away at night, discreetly, accompanied by a military captain and his scientific equipment. He flew safely, pushed by the wind blowing toward the Atlantic Ocean, and sometimes drove himself the balloon. He landed his balloon near Saint Nazaire. He later wrote about his expedition: "A special train took me to Nantes and from there I went to Tours where I arrived at 11 o'clock in the evening. I had left Paris at 6 a.m. From Tours I went by train to Bordeaux and Marseille where I embarked for Oran. There the rumour soon spread that a great French **gastronome** had just arrived in Algeria! ». Unfortunately, this extraordinary adventure remained fruitless due to the cloud cover on the day of the eclipse... The balloon "Le Volta" was then kept in Meudon, and donated by Mrs. Janssen to the Army Museum after the death of her husband in 1907.

A few years later, in 1899, Aymar de la Baume Pluvinel, a rich and experienced amateur astronomer, assiduous collaborator of Jules Janssen, instrumented a balloon without any passenger (figure 58) which reached the altitude of 9000 m and was recovered on the descent. The aim was to deepen the research carried out at Mont Blanc on the oxygen lines of the solar spectrum. The balloon embarked an automatic spectroscope with a photo plate that corroborated the observations of Mont Blanc, namely the decrease in intensity of the molecular lines, which could now be attributed with certainty to the Earth's atmosphere.

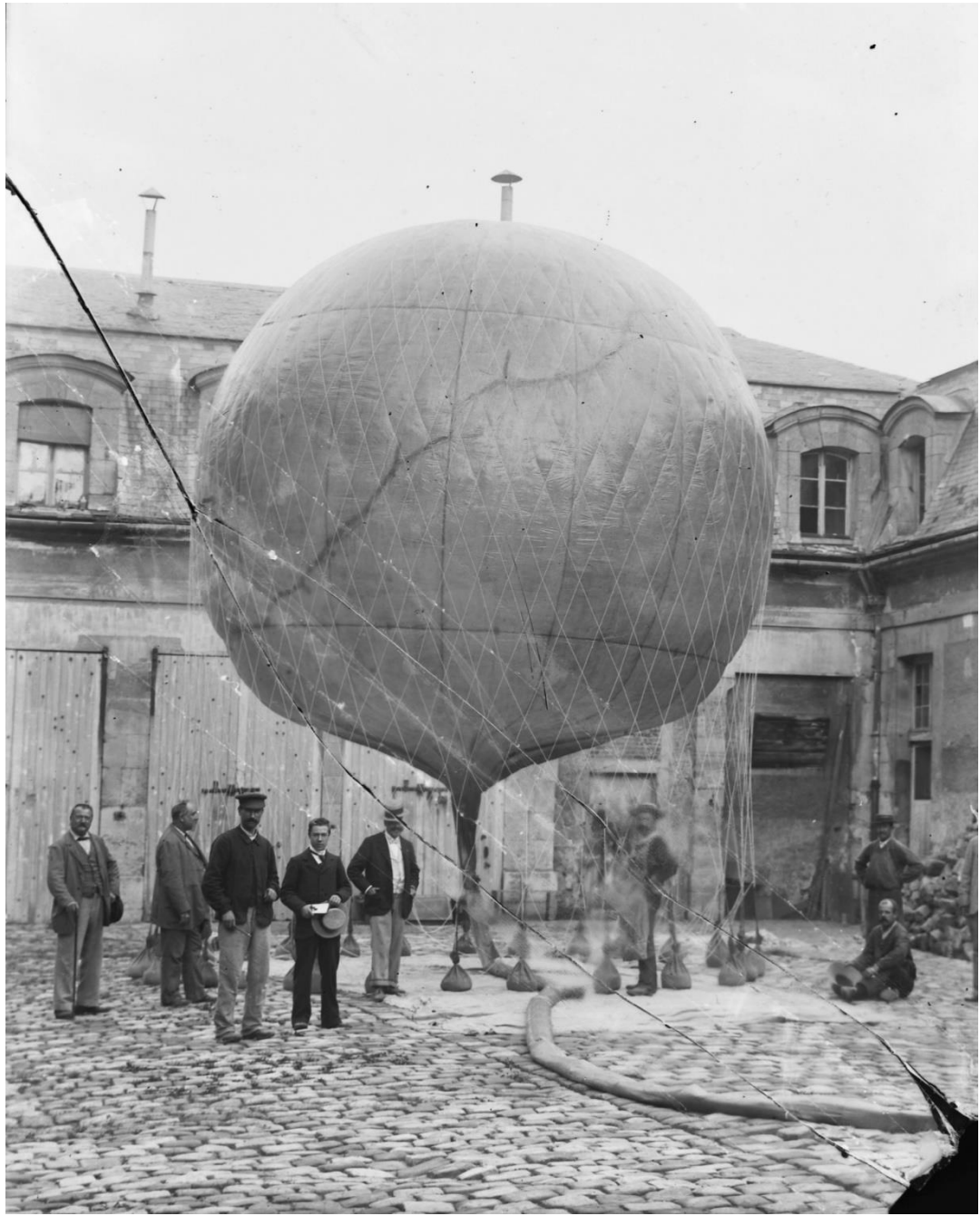

*Figure 58 : the instrumented balloon of Aymar de la Baume Pluvinel, in the courtyard of the outbuildings of the old Meudon castle, in the observatory, reached an altitude of 9000 m in 1899 (credit OP).*

The following adventures were the work of Audouin Dollfus, a famous aeronaut, as well as his father Charles Dofffus. In 1956 and 1957, Dollfus undertook the first observation of the Sun's surface using a large refractor attached to the balloon's basket (Dollfus, 1983), which climbed to an altitude of 6000 m, with its passenger (Figures 59 and 60). Why this attempt ? On the ground, astronomers could not do much better than the (already beautiful) Janssen's granulation shots, because the Earth's atmosphere is turbulent and blurs astronomical images. In order to improve the quality of the images, the only way was to go to high mountains, or better use the balloon, to reduce as best as possible the column of air above the observer.

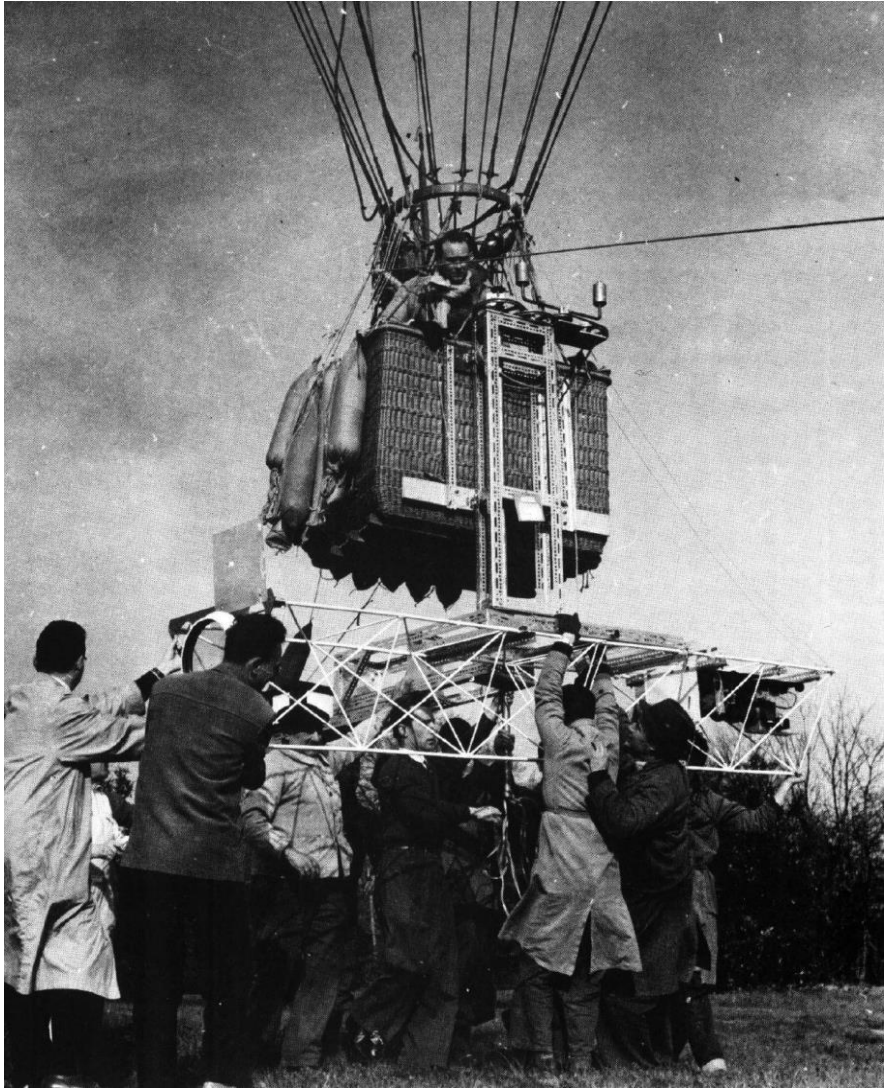

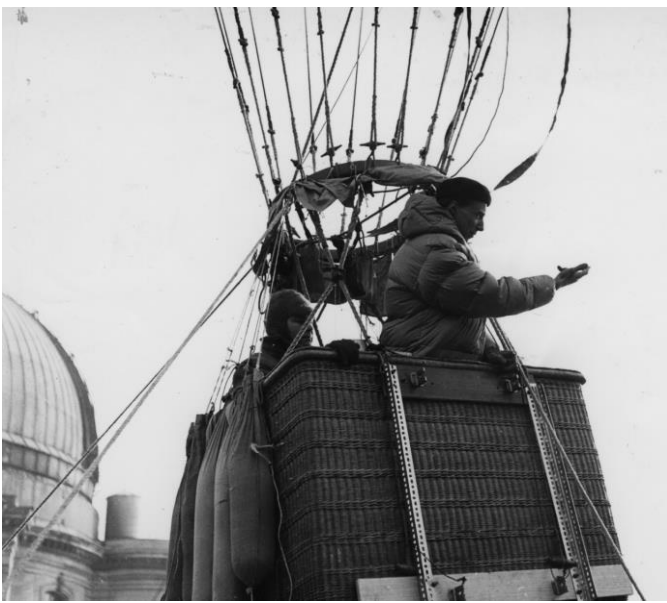
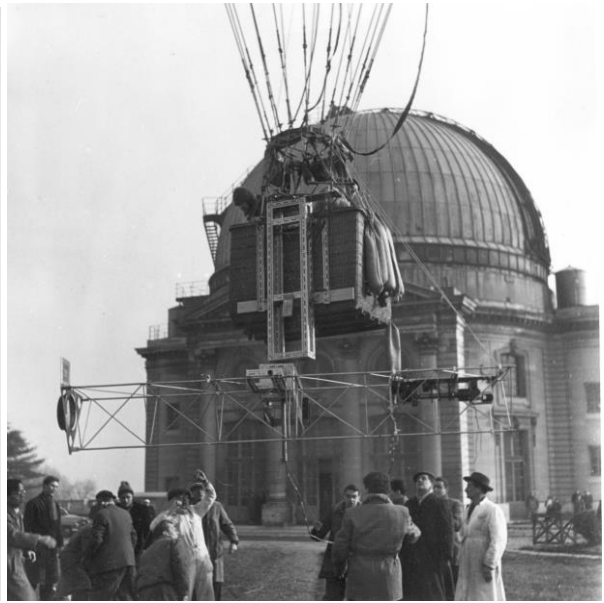

*Figure 59 : Audouin Dofffus' flight from Meudon on November 22, 1956 with a solar imaging refractor suspended below the ballon basket. Credit OP.*

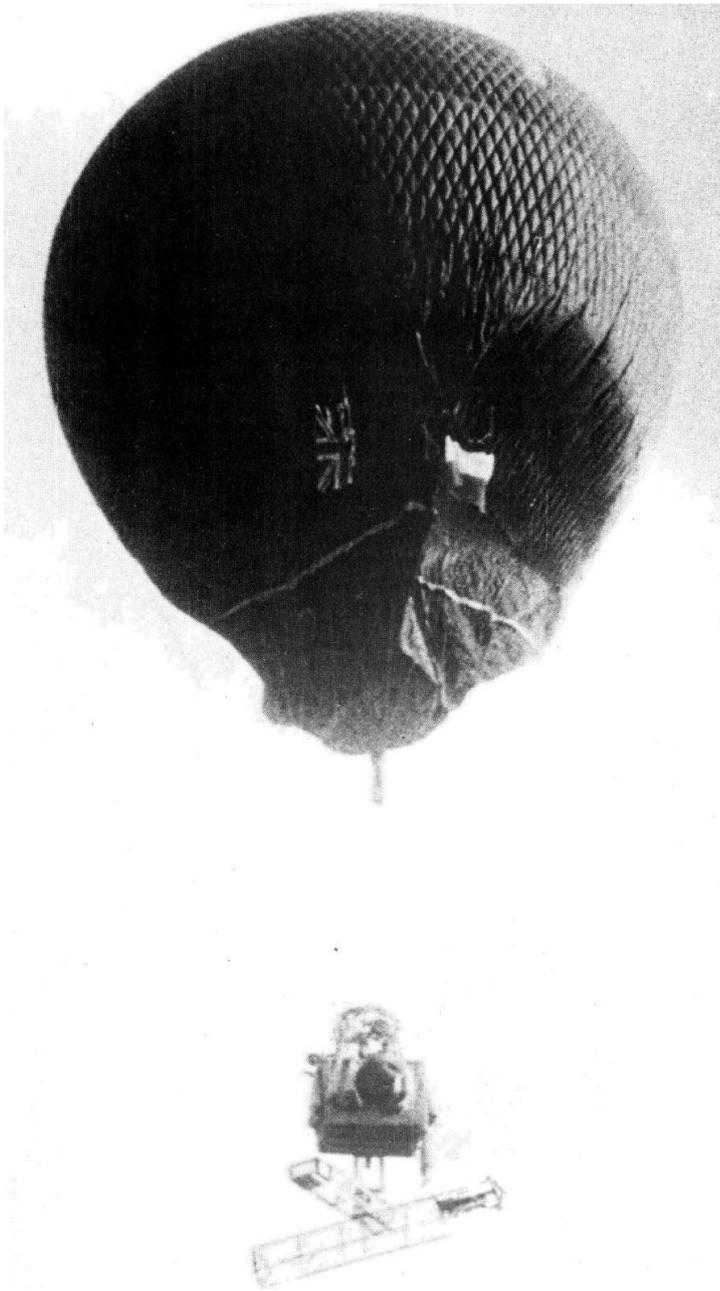

*Figure 60*

*Left: Audouin Dollfus observes the solar granulation at an altitude of 6000 m. A 30 cm diameter refractor with a focal length of 3 m is supported by the balloon basket and the instrument is piloted by the aeronaut.*

*Below: this experiment allowed Dollfus to bring back outstanding photographs (April 1, 1957) whose finesse was unequalled, thanks to the reduction of the disturbances caused by the turbulence of the atmospheric air. A worldwide exploit!*

*Credit OP.*

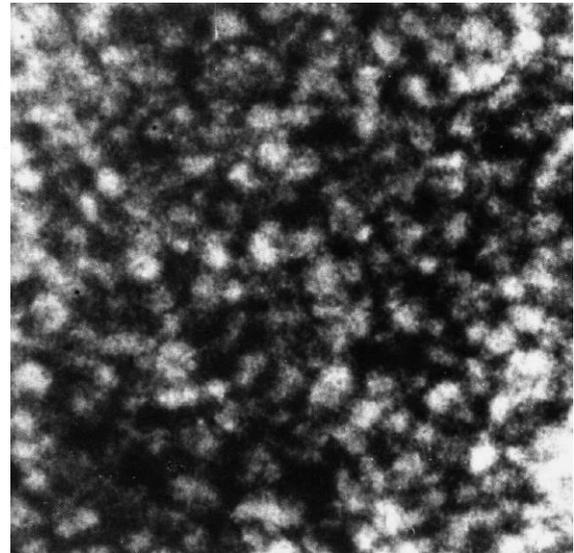

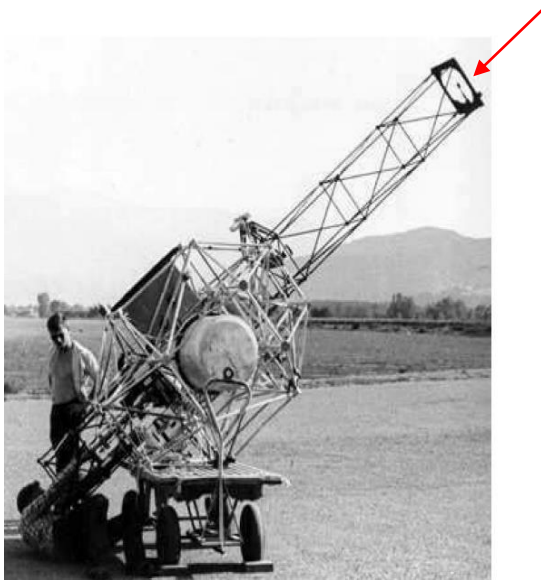
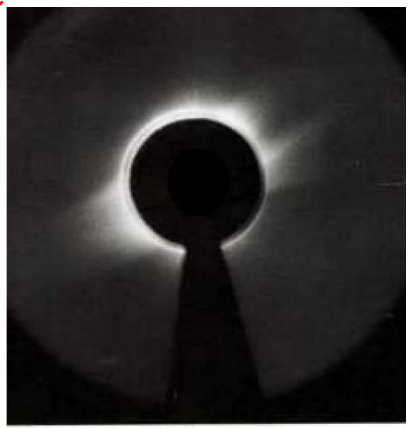

*Figure 61*

*Audouin Dollfus' external occultation coronagraph launched from Aire sur Adour (south of France) in 1973 with CNES. A masking disc (red arrow) is placed in front of the instrument. The corona and its jets are revealed at great distance from the limb. Credit OP.*

The observation at 6000 m altitude proved to be a great success and offered an unequalled spatial resolution on the granulation (Figure 60). As a passionate aeronaut, Dollfus dreamed of an experiment at an altitude of 25000 m with an instrumented telescope controlled from a pressurized capsule housing the operator. Although this project concerned nocturnal astronomy, we mention it because it is in the approach of the solar observation project at 6000 m. For the very high altitude, Dollfus turned to multi-balloon technology with a long cluster of 100 small balloons. He also tested expandable balloons under the Y hangar at Chalais-Meudon (a former airship hangar that has recently been restored). It took off in 1959 from Villacoublay and reached an altitude of 14000 m with the capsule and the telescope. Dollfus can thus be considered as a pioneer of space astronomy in France. The adventure continued with CNES after its creation in 1961. An automated external occultation coronagraph was able to observe the corona during several automatic flights at the altitude of 32000 m (Figure 61). Subsequently, coronal observation became essentially the prerogative of space telescopes, such as the coronagraphs carried by the international missions such as SOHO (1996) and STEREO (2006), or the EUV (Extreme Ultra Violet) imagers of SDO (2010) and SOLAR ORBITER (2020).

## 10 – Magnetography with spectro-polarimeters

Magnetography is the technique allowing the measurement of solar magnetic fields by interpreting the Zeeman effect. It is a high-spectral resolution polarimetric and spectroscopic technique that generates line spectra in the form of Stokes profiles I($\lambda$), Q($\lambda$), U($\lambda$), V($\lambda$). The circular polarization rate profile V/I($\lambda$) indicates the longitudinal field (projected along the line of sight) and those of the linear polarization rates Q/I($\lambda$) and V/I($\lambda$) provide the transverse field (projected on the sky) and its orientation. In the direction of the progression of light, we will therefore encounter a polarimeter before injecting the light into the spectrograph. The first polarimeter used at Meudon from 1962 consisted of a Hale grid (Figure 62) with alternating quarter- and three-quarter waveplates, followed by a linear polarizer, so that the alternating measurements of I+V($\lambda$) and I-V($\lambda$) were not cospatial. The first magnetograph operated by Michard & Rayrole (1965) consisted of a 40 cm, 7 m focal length telescope fed by the Foucault's large siderostat (Figure 63); it used elements of the 7 m spectrograph of d'Azambuja and Deslandres, which was no longer in use (Malherbe, 2025).

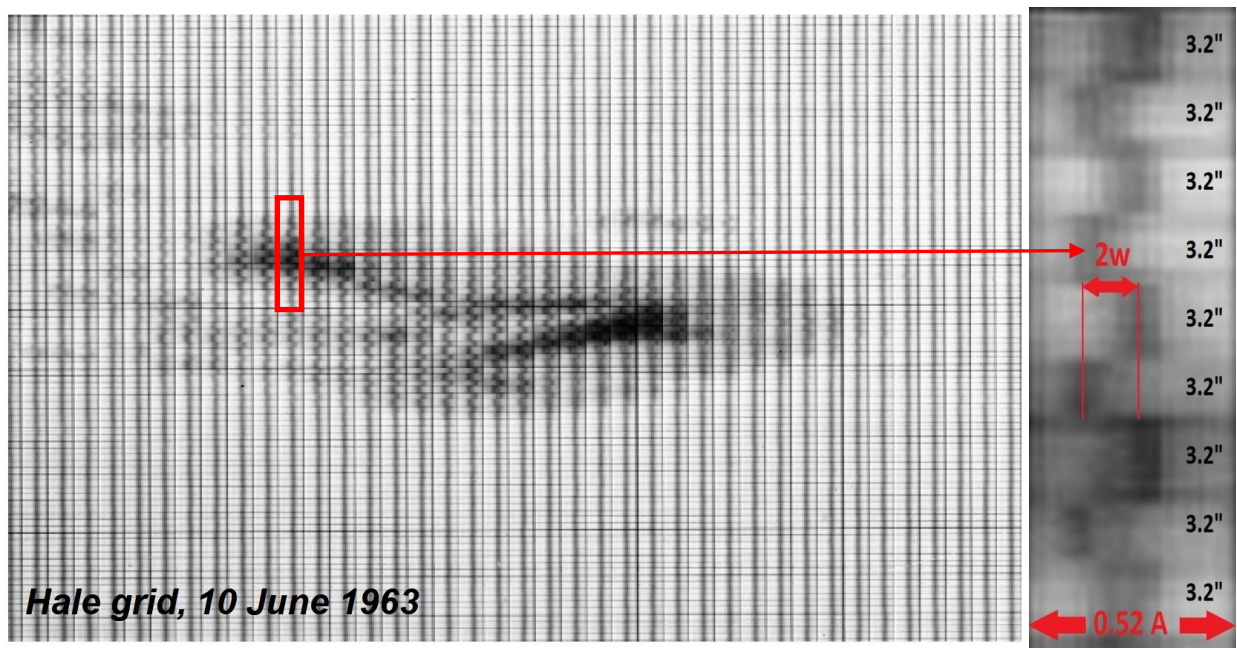

*Figure 62 : polarimetric observation of a sunspot with the Hale grid in an iron line sensitive to the Zeeman effect. The horizontal bands transmit I+V and I-V alternately, the difference in wavelength 2w between the two Zeeman components is proportional to the longitudinal magnetic field (strong in the spot). The separation of the bands is 3.2'' (detail at right). The disadvantage of this method is the lack of cospatiality between the measurements of I+V and I-V, so that the spatial resolution is limited and the determination of the magnetic field remains approximate. Credit: OP.*

The second version of the magnetograph is described by Rayrole (1967); he replaced the Hale grid with a thermally controlled quarter-wave plate in association with a parasitic polarization compensator (as the beam does not have any cylindrical symmetry) and a rotating analysing polarizer (Figure 64).

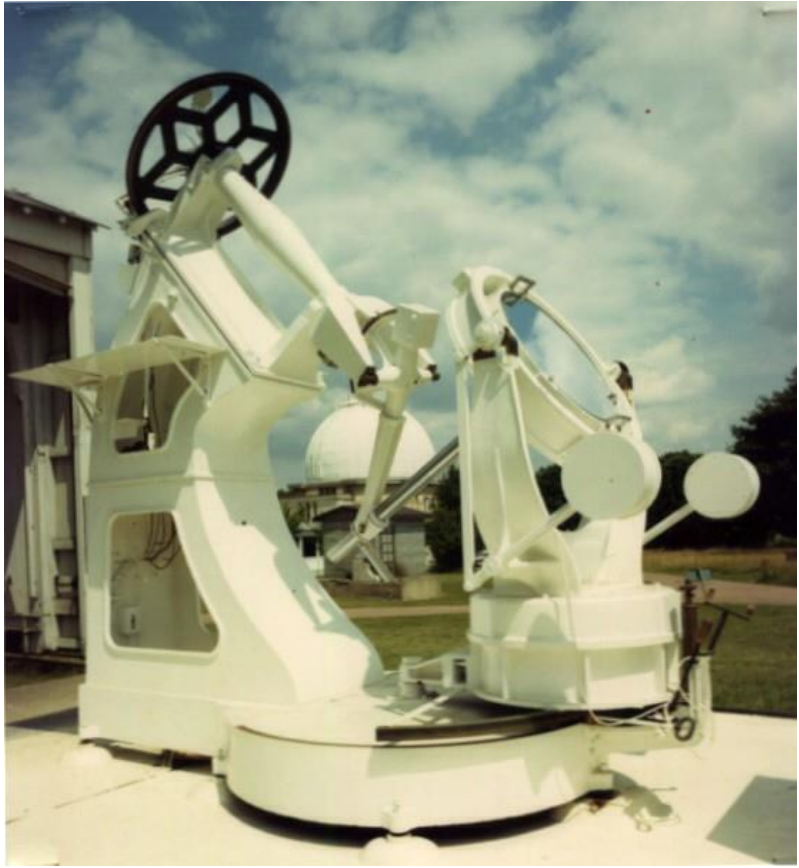

Figure 63 : Large siderostat (75 cm) of Foucault in Meudon. It fed the magnetograph. Credit: OP.

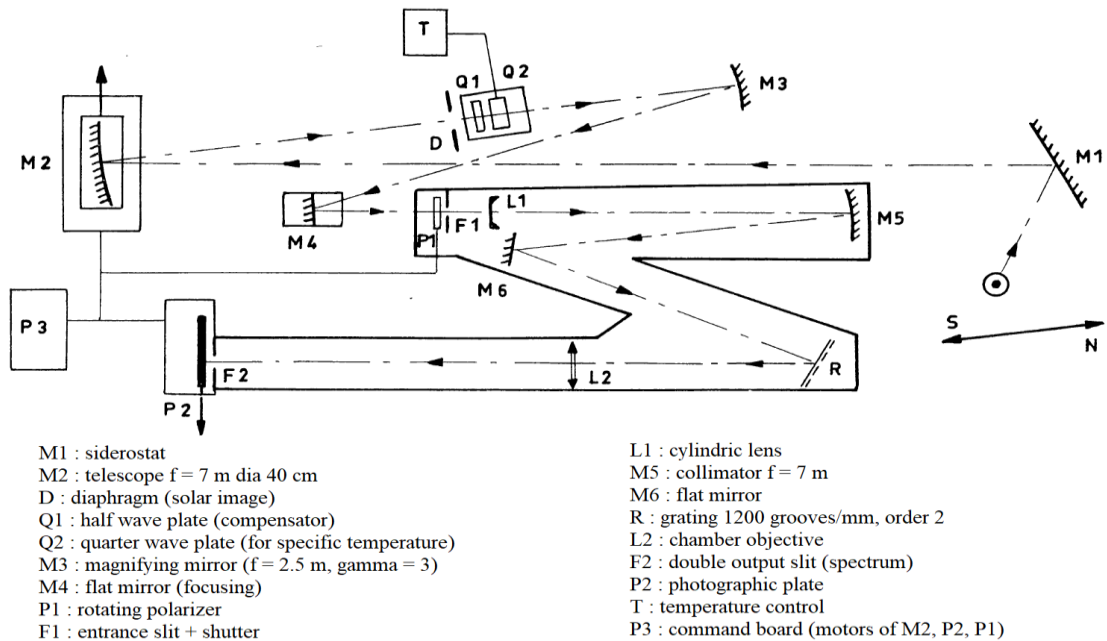

Figure 64 : second magnetograph of Meudon (Rayrole, 1967). The focal length equivalent to M2/M3 is 21 m. It no longer used the Hale grid but a Q1 quarter-wave plate and a motorized P1 analysing polarizer. A double slit in the spectrum selected the two Zeeman components. The photographic plate P2 was in synchronous translation with the M2 mirror, in order to scan the solar surface. The spectrograph had a focal length of 7 m, with the M5 concave mirror collimator and the L2 chamber objective. Credit OP.

The third version (1980) of Rayrole's magnetograph (Figure 65) was more sophisticated, as it allowed the observation of several lines formed at different altitudes; the acquisition became digital thanks to linear arrays of 256 RETICON photodiodes controlled by a 16-bit Texas Instruments microcomputer using 8-inch floppy disks of 512 KB capacity of storage. Numerous observing campaigns, coordinated with NASA's Solar Maximum Mission satellite (SMM) and other ground based instruments, such as the MSDP at Meudon or Pic du Midi, took place. The processing was numerical. For the output of magnetograms, a COMTAL vision one/20 image processor, acquired in 1980, replaced the paper listings encoded by letters or numbers according to the value of the radial magnetic field. The data (256 x 256 maps), read from diskettes, appeared in modern form as false-colour maps with opposite polarities in blue/green and yellow/red (Figure 66).

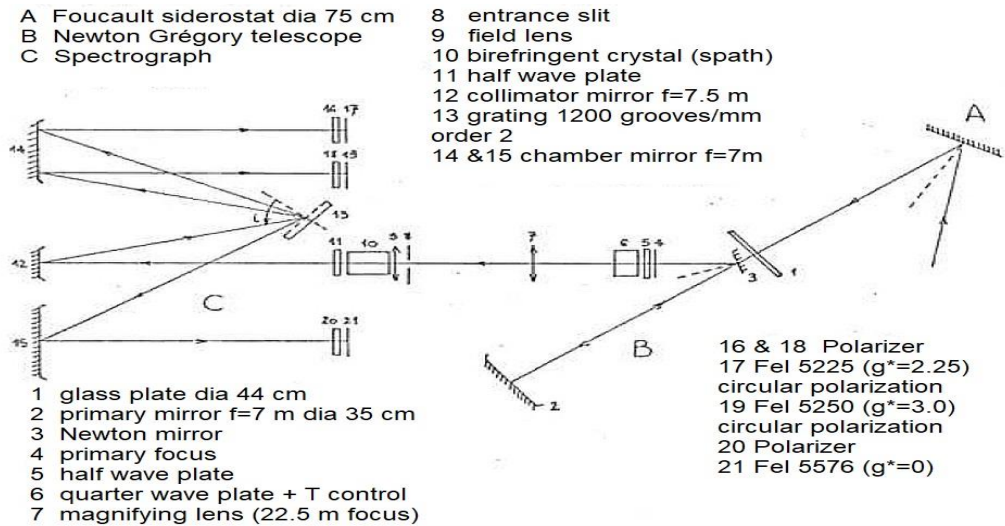

*Figure 65 : third version of the Meudon magnetograph (1980). The focal length equivalent to M2/M3 is 21 m. The photographic plate has been replaced by RETICON diode arrays. The grating spectrograph has a 7 m collimator and two 7 m mirror chambers for two simultaneous lines. OP Credit.*

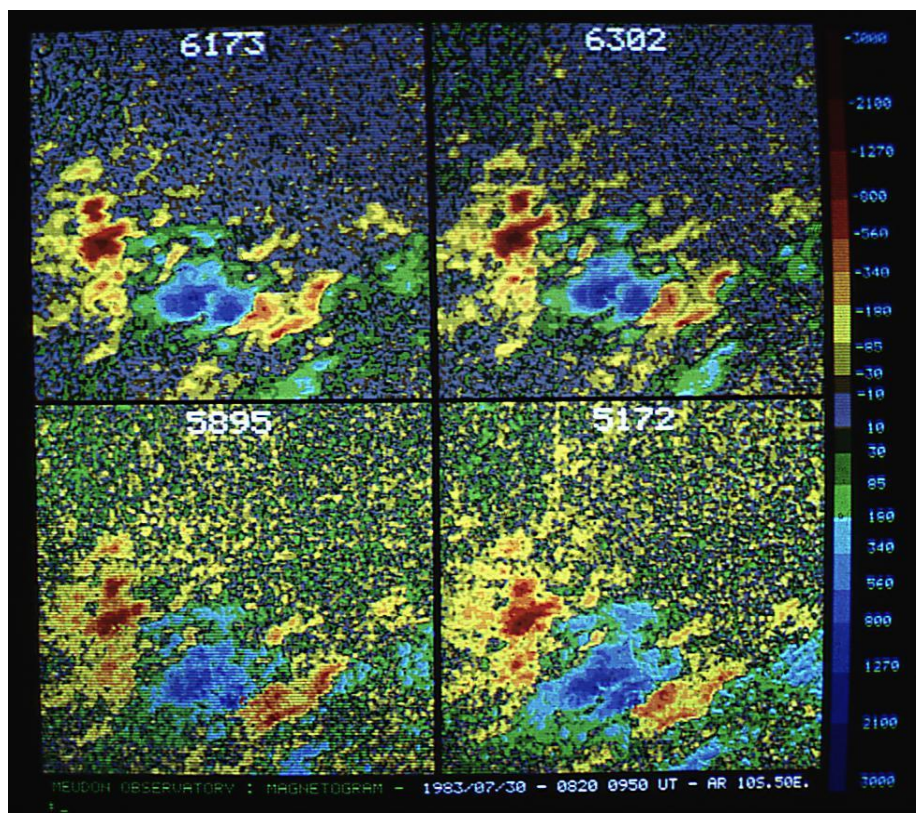

*Figure 66 : maps of the radial magnetic field in two lines of Iron (top), Sodium and Magnesium (bottom), at different atmospheric altitudes, obtained on July 30, 1983 and encoded by an image processor. Credit OP.*

The Meudon magnetograph, which was shut down after 1985, allowed Rayrole, with the help of other people such as Meir Semel and Pierre Mein, to design a large multi-line magnetograph, free of instrumental polarization, the THEMIS telescope (figures 67 and 68, Rayrole & Mein, 1993, Ceppatelli, 2004 ) which was opened in 1999 by INSU at the Teide Observatory (2370 m) at Tenerife (Canary Islands). The telescope, with an aperture of 90 cm, 57 m of equivalent focal length, was in azimuthal mount and guaranteed observations at high spatial resolution of the full Stokes vector (I, Q, U, V). The double spectrograph with a focal length of 8 m (Figure 69) was equipped with a predisperser and a disperser. The 20 THOMSON 384 x 288 CCD cameras allowed to observe 10 simultaneous lines in two polarization states, i.e. I+V($\lambda$) and I-V($\lambda$) simultaneously for each line, then by rotation of the polarimeter plates (Figure 70), I+Q($\lambda$) and I-Q($\lambda$), then finally I+U($\lambda$) and I-U($\lambda$). As THEMIS was built without any parasitic polarization, it could measure the magnetic field vector (radial component, transverse component and its direction to within 180°) by exploiting all the information provided by the Zeeman effect (which was impossible in Meudon with siderostats or coelostats).

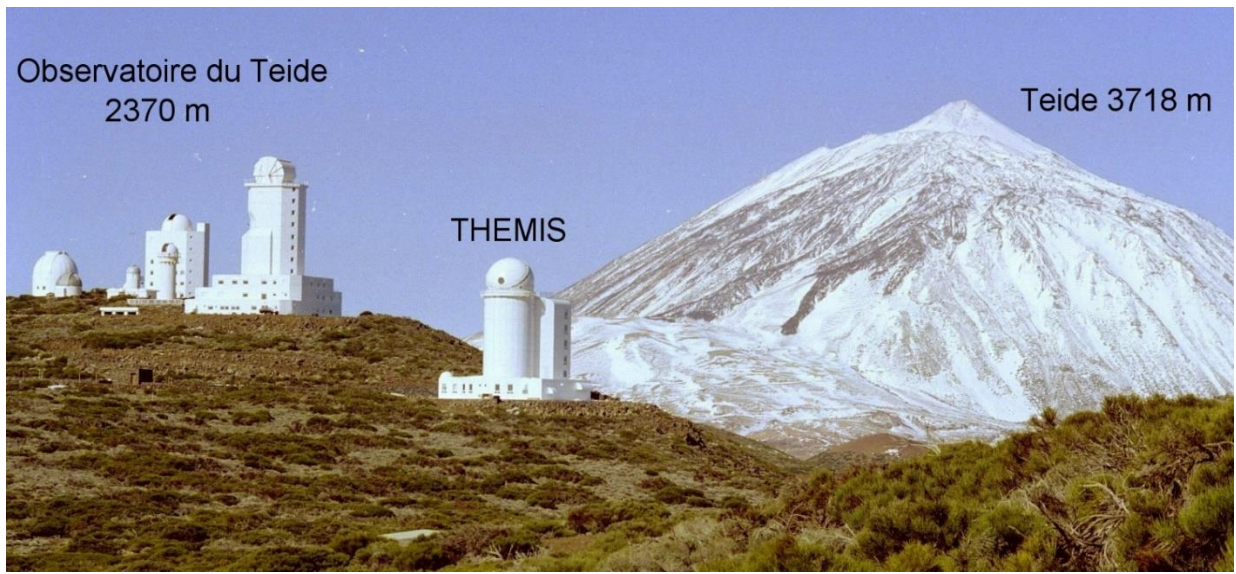

*Figure 67 : The Teide Observatory in Tenerife. Credit J.-M. Malherbe.*

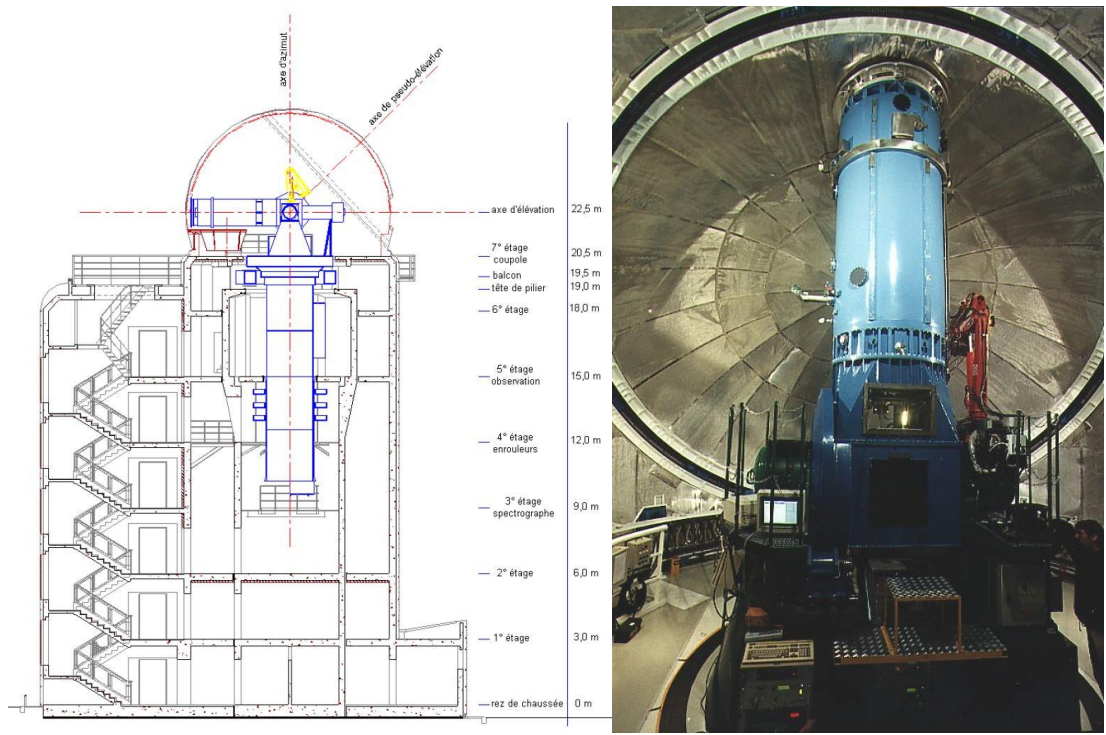

*Figure 68 : THEMIS building (left) and the telescope inside the dome (right). The 8 m spectrographs are vertical under the telescope and rotate with the azimuthal mount around the vertical. Credit INSU/OP.*

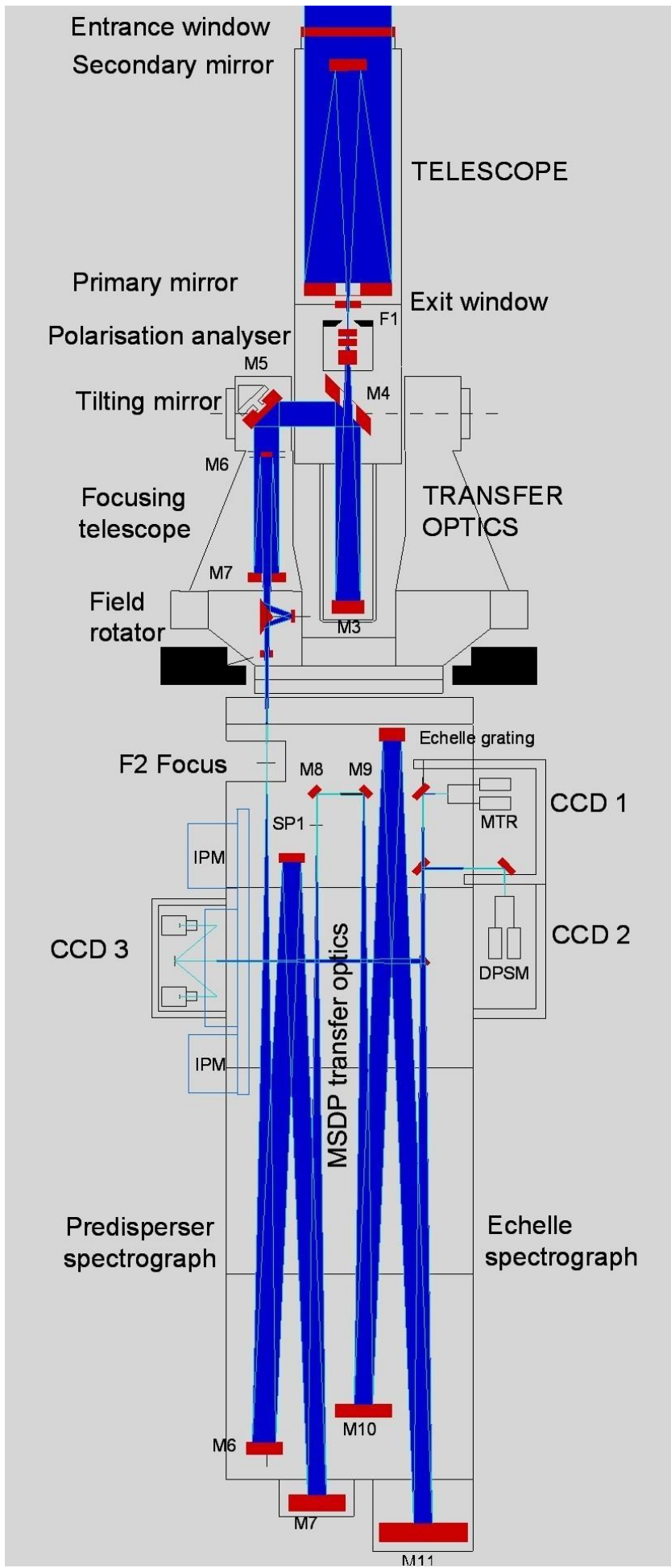

*Figure 69*

*The THEMIS telescope, as commissioned in 1999 by INSU/CNRS, consisted of:*

*1 – a closed telescope under Helium with an aperture of 90 cm and a focal length of 17 m in azimuthal mount*

*2 – a polarization analyser at the F1 focus of the telescope, on its axis. Two beams exit the polarimeter with two different polarization states. It delivered Stokes combinations I+V and I-V, or I+Q and I-Q, or I+U and I-U (V, Q, U were therefore measured in sequence by digital subtraction of the signals)*

*3 – a transfer optic comprising a small telescope with two concave mirrors M3 and M7 increasing the focal length to 57 m at the F2 focus. An image stabilizer, a high frequency tip tilt, was incorporated into the beam*

*4 – a field derotator to compensate the field rotation of the azimuthal mount*

*5 – a predisperser spectrograph with concave mirrors M6 and M7 serving as an order selector for the following disperser spectrograph, with a focal length of 8 m. The predisperser could also serve as a full disperser with an echelle grating.*

*6 - a dispersing spectrograph with M10 and M11 concave mirrors with an echelle grating, focal length 8 m*

*7 – multiple CCD outputs with up to 20 cameras in multi-line spectro polarimetry mode (called MTR)*

*8 – a wide-slit MSDP (Multichannel Subtractive Double Pass) spectro imaging mode (2D window in F2), in which the predisperser served as a first disperser, and the second disperser as a subtractive spectrograph, both with the same echelle grating*

*Credit INSU/OP.*

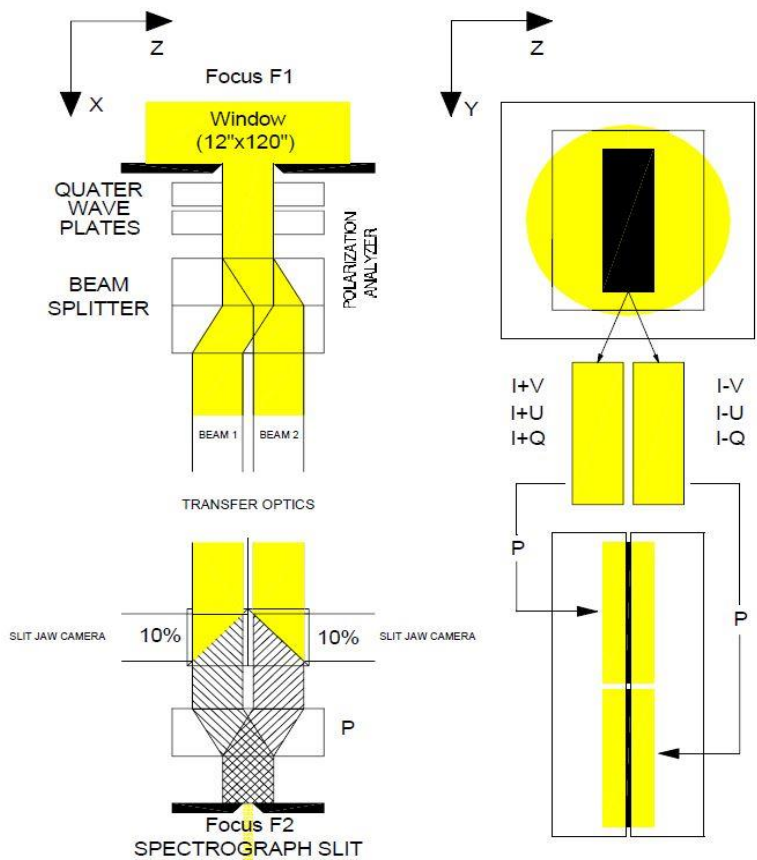

*Figure 70*

*THEMIS polarimeter in 1999 when it was commissioned. It is composed of two rotating and roughly achromatic quarter-wave plates (between 400 and 900 nm) followed by a beam splitter shifter analyser. It was made of two calcite crystals delivering two parallel and linearly polarized beams in two orthogonal directions. These two beams contained the pairs I+V and I-V, then by turning the quarter wave plates, I+Q and I-Q, and finally in another position of the plates, I+U and I-U. The two parallel beams were aligned before injection into the spectrograph by prism translator (P, right part of the figure). Credit OP.*

   Several modes of polarimetric observation, all with their own advantages and disadvantages, have been tested and then used according to the preferences of the observers or the precision requested:

- the 2-camera mode (Figure 71) in which the first camera observes the I+S($\lambda$,y) line profiles and the second I-S($\lambda$,y), where S = Q, U, V in sequence, and y is the ordinate along the slit (Lopez *et al*, 2000)

- the 1-camera mode and half spatial field (Figure 72) in which the signals I+S($\lambda$,y) and I-S($\lambda$,y) are placed on the same sensor (implying a spatial field reduced by 2 in the y direction), where S = Q, U, V in sequence

- the 1-camera and grid mode (Figure 73) in which the I+S($\lambda$,y) and I-S($\lambda$,y) signals are placed on the same detector, with a grid at the F1 focus implying a striped and globally halved spatial field, where S = Q, U, V in sequence. This polarimetric method, proposed by Semel (1932-2012), is explained in Figure 74 and described by Semel (1980). It guaranteed the best polarimetric accuracy but the data processing was complicated.

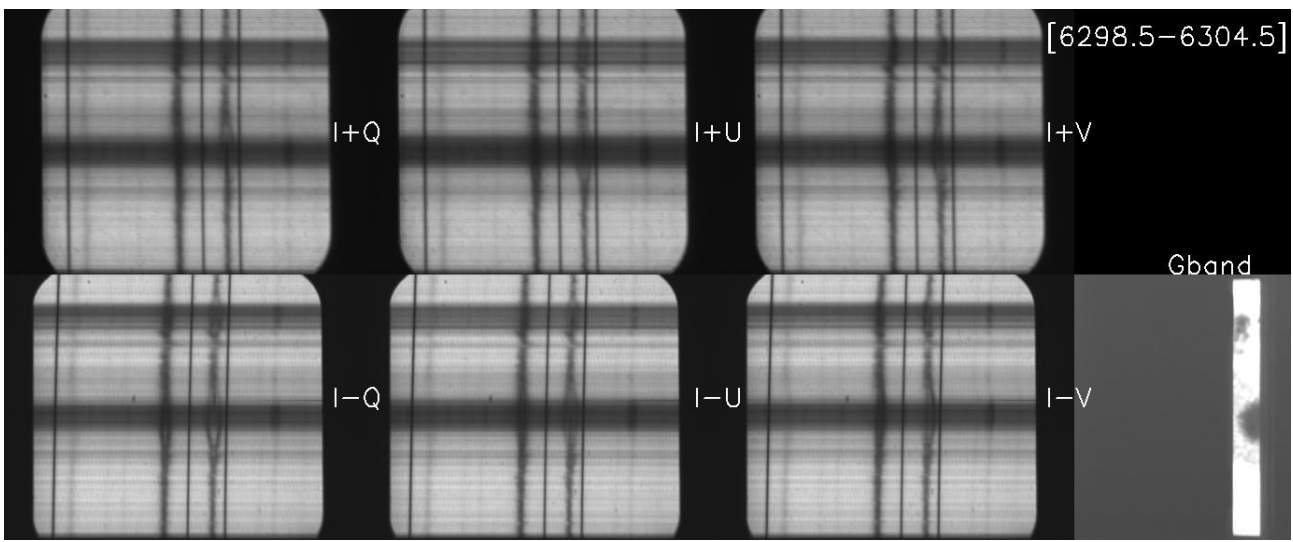

*Figure 71 : Polarimetric mode with two cameras, 2' vertical field, Q, U, V in sequence. OP credit.*

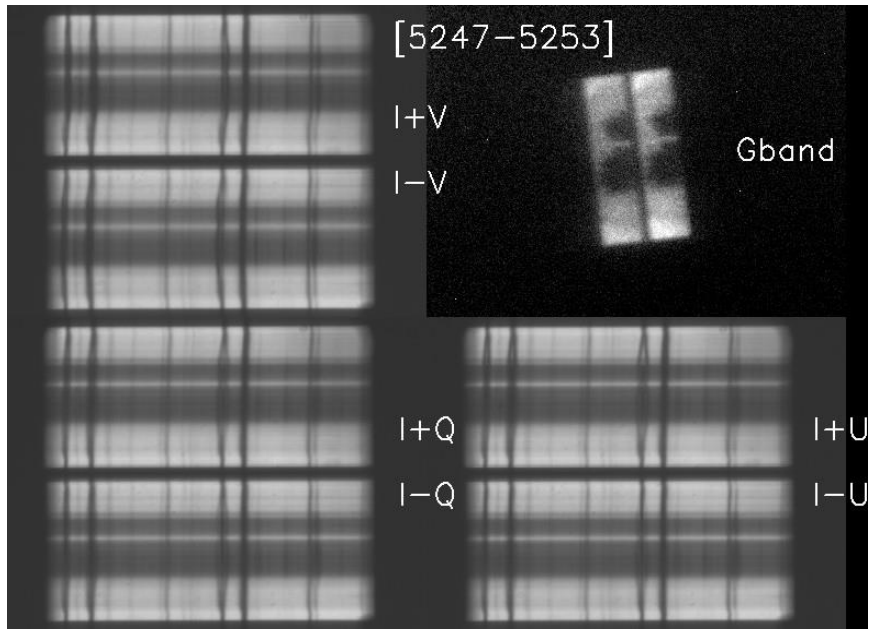

*Figure 72 : single-camera polarimetric mode, 1' vertical field, Q, U, V in sequence. OP credit.*

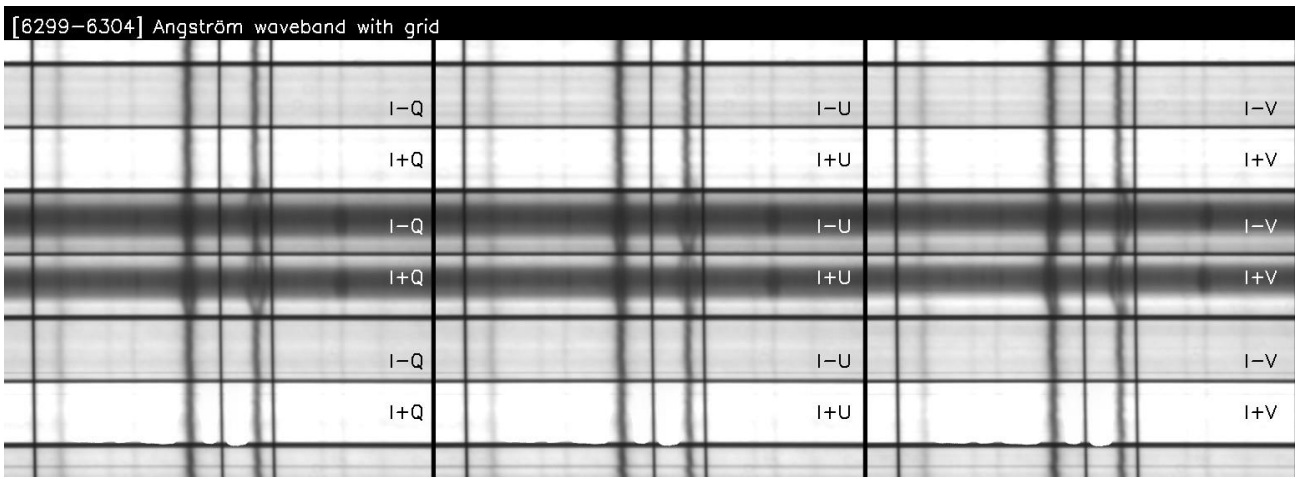

*Figure 73 : Polarimetric mode with Semel's grid, 15'' x 3 vertical field, Q, U, V in sequence. OP credit.*

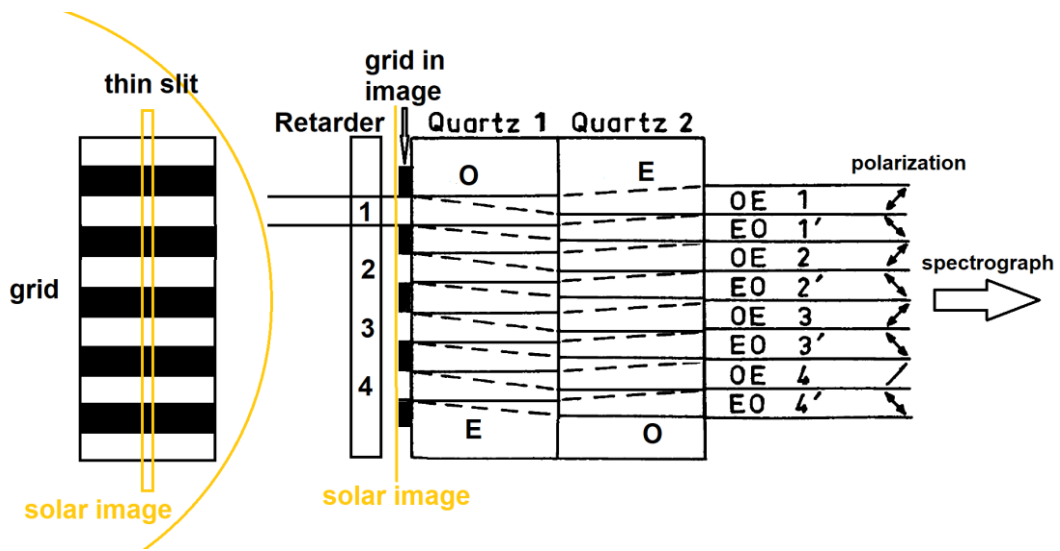

*Figure 74 : Semel's grid polarimetric method (1980), vertical field of 15'' x 5. A striped mask at the F1 focus, in front of the polarimeter, obscured half of the field, like a zebra. It was therefore necessary to translate the field by a telescope displacement to cover the total field of view by a second exposure. Credit: OP.*

For the measurement of weak polarizations, Semel *et al* (1993) proposed the "beam exchange" made of two successive polarimetric exposures, after rotating the polarimeter plates. It consisted in alternating, on the same spatial field of view and the same area of the CCD detector, I+S and I-S signals (S = Q, U, V in sequence). This method consequently implied 6 successive measurements to obtain the 6 Stokes combinations involving I, Q, U and V. The advantage of beam exchange is that it is independent of the flat field in the presence of a stabilized image, which allows to measure polarization rates as low as $10^{-5}$ after summing hundreds or thousands of spectra. This method was fruitful for the study of the second solar spectrum (or linear polarization spectrum of the limb), whose polarization rates are low.

As an example, Figure 75 shows that the Stokes profiles V/I(x,y,λ), Q/I(x,y,λ) and U/I(x,y,λ), obtained on an active region by slit scanning, allowed to deliver on a 2D field of view the intensity of the continuum, the intensity of the line, the rate of circular polarization V/I (the interpretation of which provides the radial magnetic field), the linear polarization rate $(Q^2+U^2)^{1/2}/I$ (whose interpretation gives the transverse magnetic field), and ½ arctan(U/Q), which leads to the direction of the transverse field to within 180°.

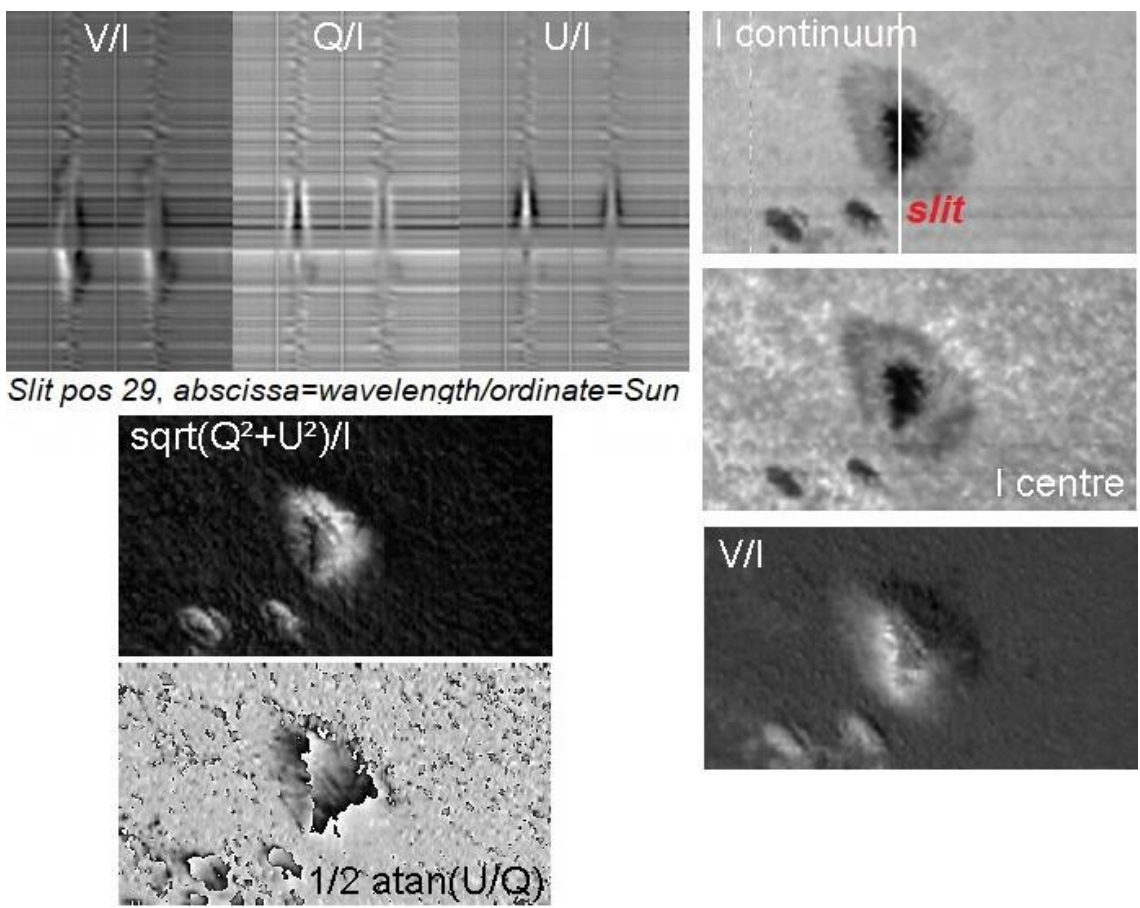

*Figure 75 : spectro-polarimetric scanning of a sunspot. The Stokes profiles V/I(λ,y), Q/I(λ,y) and U/I(λ,y) as a function of wavelength at the x = 29 over 60 position of the spectrograph slit have been plotted. OP credit.*

## 11 – The large 9 m spectrograph of the International Geophysical Year (IGY 1957)

For IGY 1957, Raymond Michard (1925-2015), who directed the Meudon solar department, defined an ambitious spectroscopic program in collaboration with Jean Rösch (1915-1999), director of the Pic du Midi observatory. It was the spectroscopy of solar flares. A small universal spectrograph with a focal length of 4 m was first installed at the Pic du Midi, fed by a 40 cm coelostat with two mirrors. A second and more powerful spectrograph, providing a dispersion of 5 mm/Angström, was built at Meudon (Laborde *et al*, 1959), at the small siderostat building, it was fed by a coelostat with two flat mirrors of 40 cm (Figure 76). This coelostat is now at the "Maison du Soleil" (an outreach house) in Saint Véran since 2015, where it feeds the 30 cm lens of the "Grande Lunette du Mont Blanc", which forms a white light image for educational purposes.

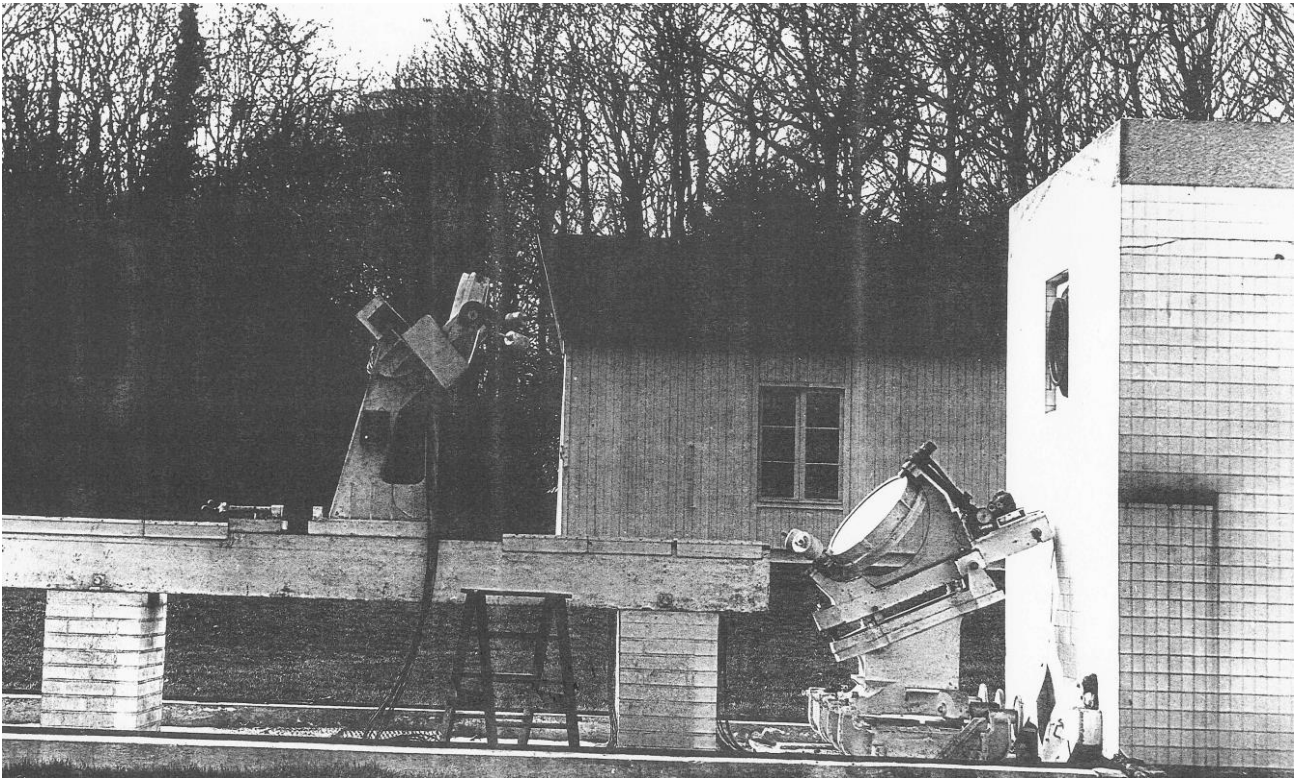

*Figure 76 : coelostat with two plane mirrors which fed the 9 m spectrograph before transport to the Pic du Midi. In the background, the solar tower being finished in 1970. OP Credit.*

      The project was, after testing campaigns in Meudon, to transport the 9 m spectrograph to the Pic du Midi to carry out spectroscopy at high spatial resolution, which was done in 1959. The spectrograph was fed by a telescope with an aperture of 50 cm and a focal length of 11 m (Figure 77); it was soon replaced by a 5 m focal length objective accompanied by an enlarging spherical mirror when the 50 cm telescope was needed to feed the small 4 m spectrograph located at the Pic du Midi. The Meudon spectrograph included a spherical collimator with a focal length of 7.4 m illuminating an echelle grating, and a chamber mirror with a focal length of 9 m. The whole system was finally transported to the Pic du Midi where it was operational after 1960. It should be noted that the 50 cm telescope has been donated to the "Maison du Soleil" in Saint Véran to feed the 7 m SHARMOR spectrograph, also on loan from the Paris Observatory for educational purposes.

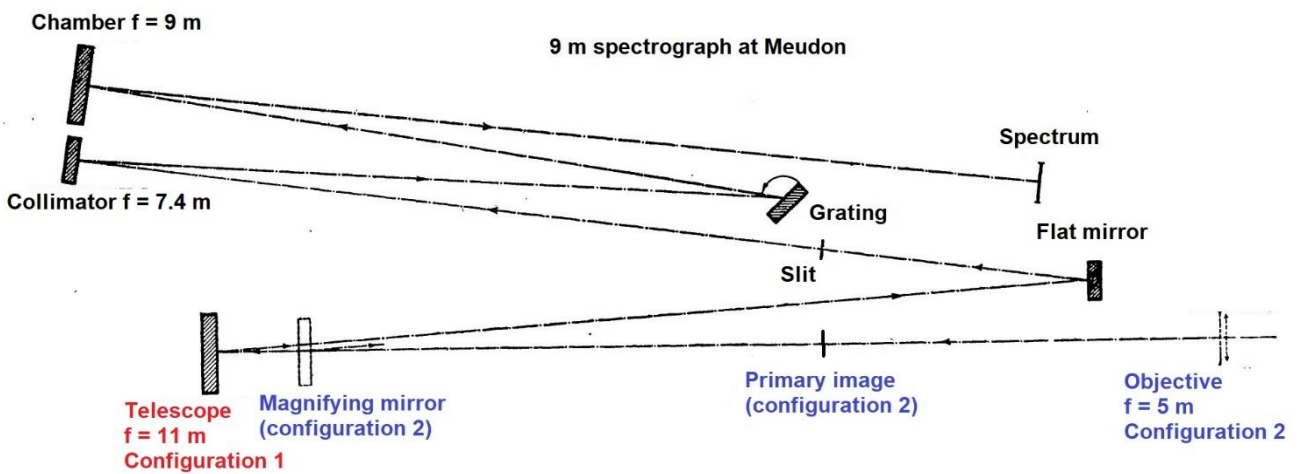

*Figure 77 : the 9 m spectrograph in Meudon. At first, it was illuminated by a 50 cm telescope with an 11 m focal length (configuration 1), then by a 5 m focal length objective with a concave secondary mirror extending the focal length by a factor of 2 (10 m, configuration 2). OP credit.*

Figures 78 to 81 show the schematic installation transported to the Pic du Midi, in operation after 1960.

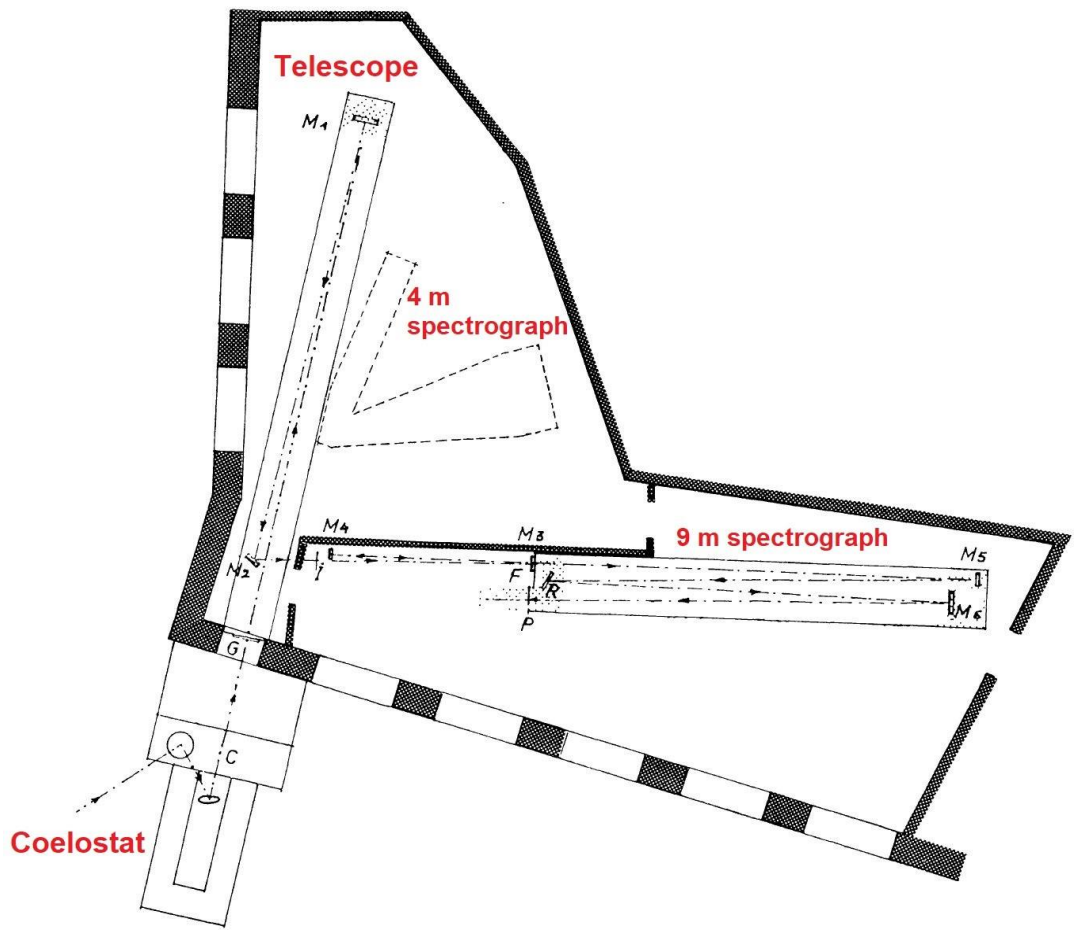

**C** : coelostat  
**G** : uviol glass  
**M1** : 11 m telescope  
**I** : primary image (dia 10 cm)  
**M2** : flat mirror  
**M3** : magnifier mirror (x 2)  
**M4** : flat mirror  
**F** : reflecting slit (20 cm solar diameter)  
**M5** : 9 m collimator  
**R** : 600 grooves/mm grating (20°/55° blaze)  
**M6** : 9 m chamber  
**P** : spectrum (20 cm solar image height)

*Figure 78 : the 9 m spectrograph at the Pic du Midi after 1960. Credit OP.*

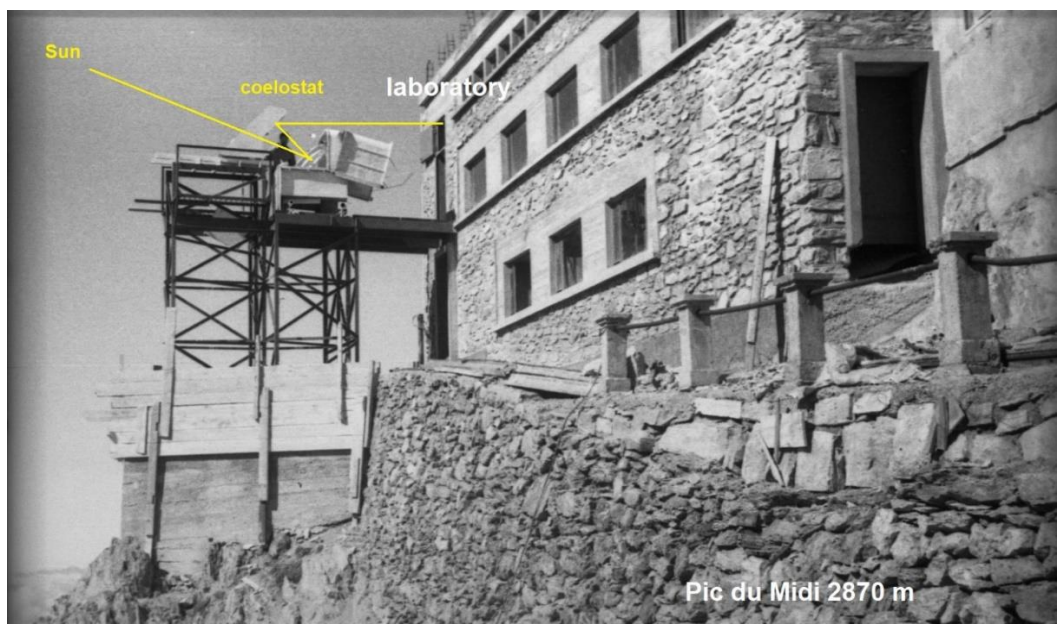

*Figure 79 : the first coelostat of the solar laboratory at Pic du Midi before 1960. OP Credit.*

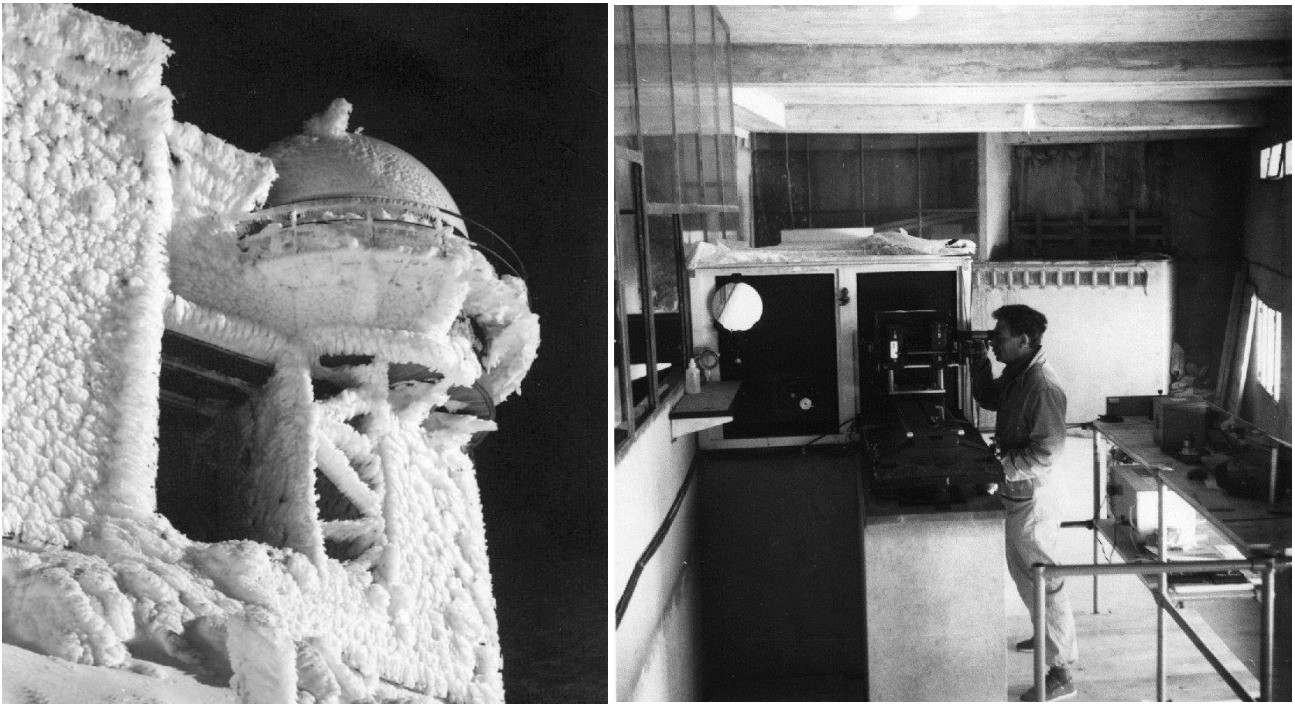

*Figure 80 : the second coelostat at Pic du Midi after 1960 with Roger Servajean at the 9 m spectrograph eyepiece. The M3 concave mirror doubles the focal length of the M1 telescope from 11 to 22 m. OP credit.*

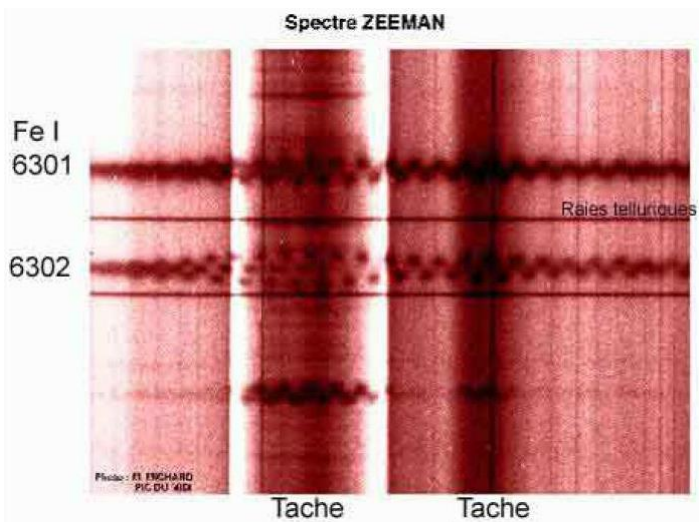

*Figure 81*

*An example of observations of the Zeeman effect on sunspots carried out by Michard et al (1961) in the early 60s using a Hale grid. It was placed at the entrance of the 9 m spectrograph. Here we see the lines of Iron at 6301.5 and 6302.5 Angström, they are very sensitive to the Zeeman effect (strong Landé factor), in circular polarization (I+V, I-V alternate on the field of view in the spatial direction x, while the wavelength λ is on the y-axis). Credit OP.*

**12 – The Turret Dome spectrograph**

The 9 m spectrograph was used by the Meudon team during the 60s. It appeared, however, that the image quality for obtaining high-resolution spectra was lower than expected, probably due to the location of the coelostat to the east of the Pic du Midi, with the Sun passing over the buildings. For this reason, a new refractor (called "turret dome" because its objective was rejected away from the dome, which remained closed by a system of moving curtains, figure 82) was installed to the west crest by Rösch (Roudier *et al*, 2021). Initially 38 cm wide, it quickly increased to 50 cm, providing very fine images of the granulation. Zadig Mouradian (1930-2020), a former user of the 9 m spectrograph, then started work in Meudon on a new 8 m spectrograph that he installed around 1980 on the equatorial mount of the Pic du Midi refractor (Figure 82). This spectrograph (Figure 83, Mouradian *et al*, 1980) was folded on itself twice to save space. It was equipped with a MSDP spectro-imaging system developed by Mein (1980). A CCD detector replaced the 35 mm and 70 mm films in the 90s and a liquid crystal polarimeter (Figure 84) was implanted by Malherbe in 2003, on the optical axis of the refractor, a location free of instrumental polarization (Malherbe, 2007).

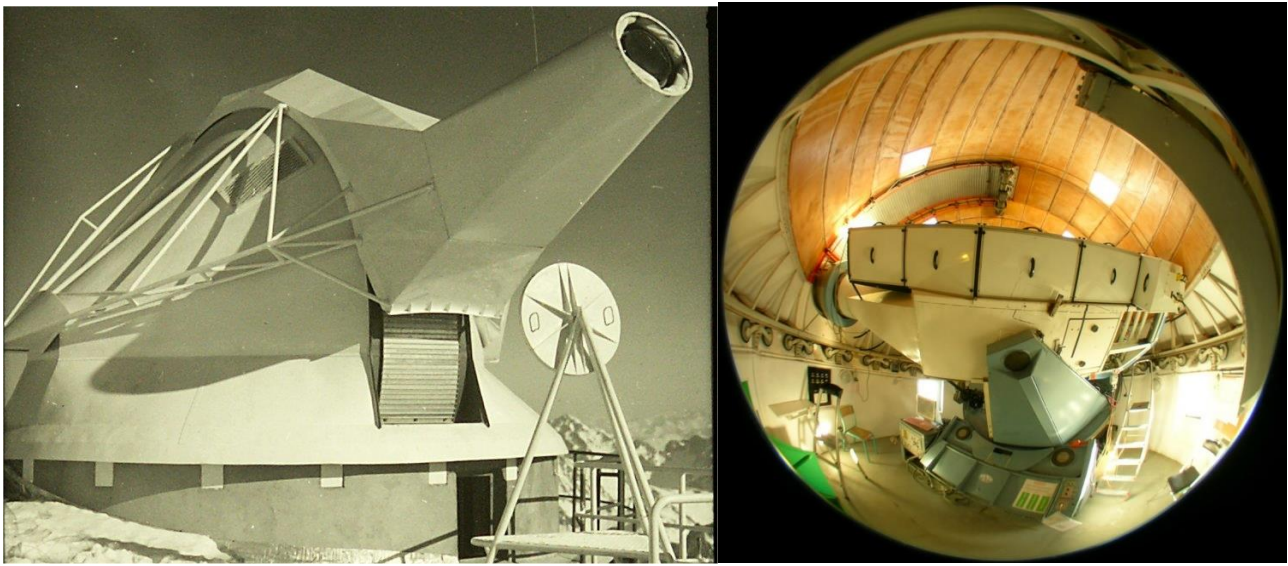

*Figure 82 : the turret dome and its spectrograph designed and built in Meudon. The focal length of the refractor was 6.5 m, increased to 30, 45 or 60 m with various magnifying objectives before injecting the beam into the thin-slit spectrograph or 2D MSDP window. Credit OMP and Sylvain Rondi.*

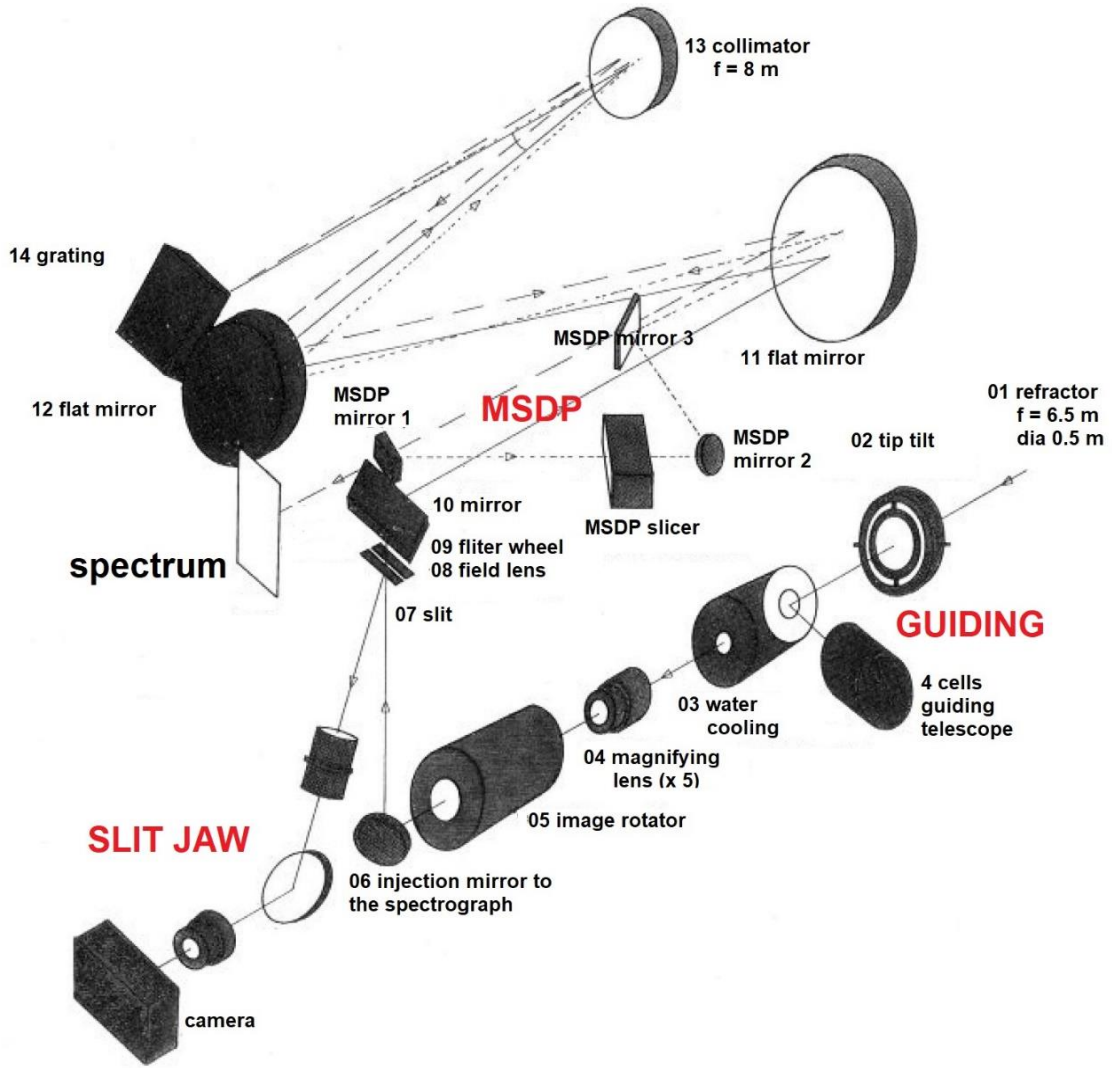

*Figure 83 : the 8 m spectrograph of the turret dome refractor, designed and built in Meudon by Mouradian. Its dispersion was 5 mm/Angström. The orders were isolated by interference filters. Credit OP.*

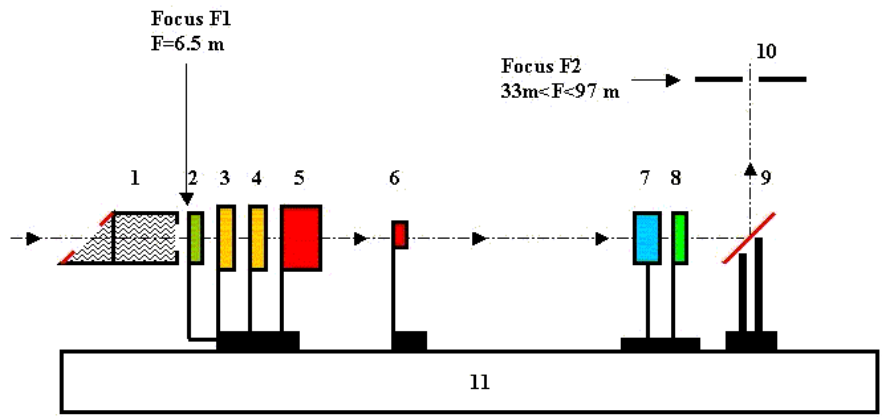

1: water cooling device (field stop)
2: UV/IR filter 390-700 nm
3: variable retarder 1
4: variable retarder 2
5: magnification lens (f=40/60/95 mm)
6: beam shifter and flat field lenses (MSDP only)
7: interference filter
8: precision dichroïc linear polarizer
9: flat mirror
10: spectrograph entrance slit
11: optical rail

*Figure 84 : the two-liquid crystal polarimeter designed in Meudon allowed to measure the Stokes combinations I+V, I-V, I+Q, I-Q, I+U, I-U in sequence. The liquid crystals used were perfectly quarter wave or half wave for any observed line by precise adjustment of the voltages at the electrodes. Credit OP.*

## 13 – The Multi-channel Subtractive Double Pass spectrographs (MSDP)

At the end of the 60s, the observatory decided to build the solar tower of Meudon, which was commissioned in the early 1970s. This 35 m high tower contains a vertical and fixed telescope with an aperture of 60 cm and a focal length of 45 m (Figure 85). The concave mirror of the telescope, at the bottom of the tower, is fed by a large coelostat with two flat 80 cm mirrors located on the upper terrace. The light rays are thus captured above the Meudon forest, which stabilizes the images. However, during the past 50 years, image quality has suffered from the excessive development of urban planning and the construction of a motorway cutting through the forest. At the foot of the tower, there is a large horizontal grating spectrograph, with a focal length of 14 m, providing an exceptional dispersion of 1 cm/Angström. It is equipped with a collimator and two chamber mirrors thus offering two outputs for spectra.

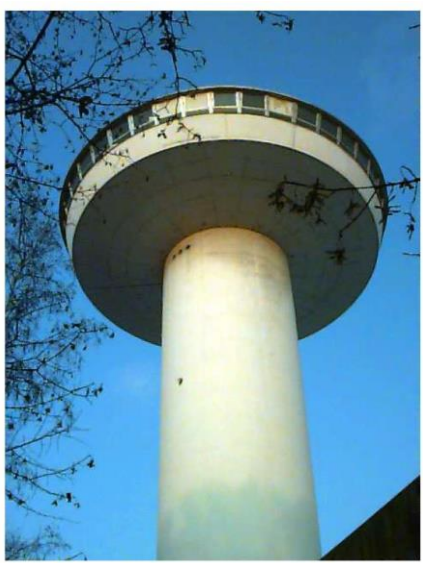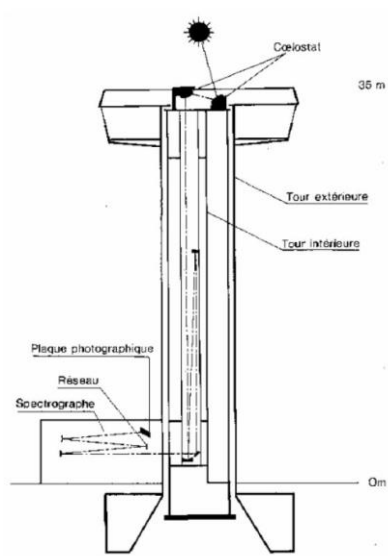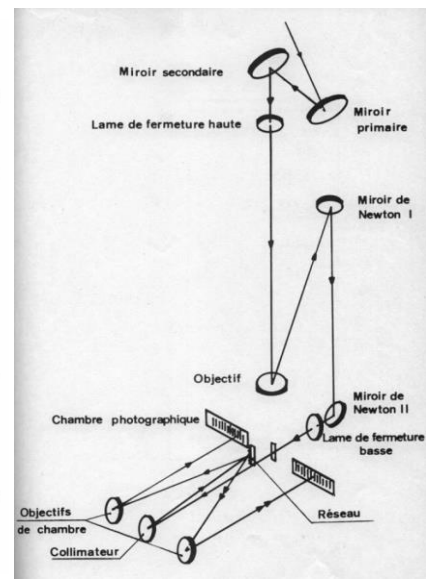

*Figure 85 : the Meudon solar tower and its large 14 m spectrograph; optical diagram. OP credit.*

The spectrograph was completed in 1977 by Pierre Mein with a detailed spectro imaging device explained by figures 86 and 87 (Mein, 1977). In this device, the thin slit is replaced by a rectangular window F1 with a field of view of the order of 30" x 4'. In the F2 spectrum, a slicer (a grid followed by prism translators, such as the various prototypes in Figure 87) is incorporated which forms N adjacent channels of the same spatial field of view and covering the wavelength of the chosen spectral line. As technology progressed, we had N = 7, then 9, 11 and 16 (Figure 87). Thus, with N channels, it is possible to reconstruct the line using N spectral points for any pixel (x, y) in the observed field. This system is called Multichannel Subtractive Double Pass (MSDP), because the N beams coming out of the slicer are redirected to the grating for a second pass that subtracts the dispersion. In F3 we obtain N contiguous spectral images, in which there is a multiplexing of the x and λ coordinates on the x-axis (but not on the y-axis). For this reason, raw images, such as the one in Figure 86 showing a prominence at the limb, are difficult to interpret when the diagram in Figure 88 is not in mind. It gives the transmission functions of the N channels in wavelength as a function of the x-axis within each channel. Each transmission is an affine function of x, and they form a set of N parallel and equidistant straight lines. The distance between two successive lines is the spectral resolution (of the order of 80 to 300 milli Angström depending on the slicer). It can be seen that the sampling of the line profiles depends on the x-axis location of the pixel, but not on the y-coordinate. The advantage of the MSDP is the great speed in covering a 2D area on the Sun, since the telescope can scan a region in a few distant steps, compared to a few hundred tight steps in conventional spectroscopy; the spatial resolution is higher; on the other hand, the spectral resolution is lower than in thin slit spectroscopy where it is usually close to 20 milli Angström. The MSDP is therefore the most suitable for the observation of wide chromospheric or photospheric lines.

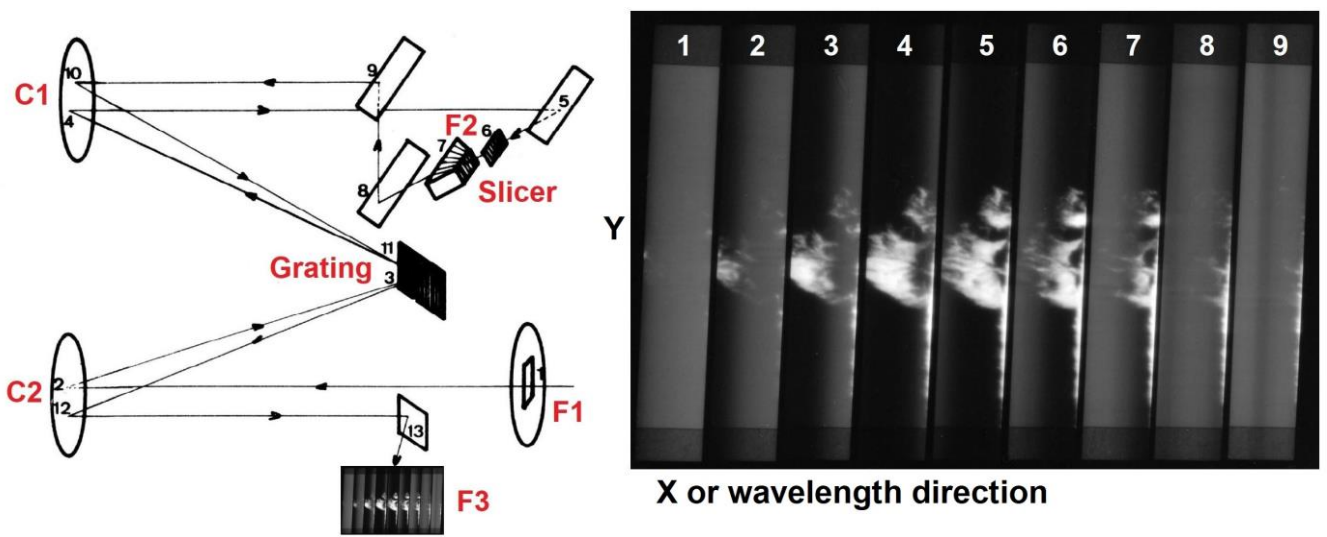

*Figure 86 : principle of the MSDP spectro imaging technique. Credit OP.*

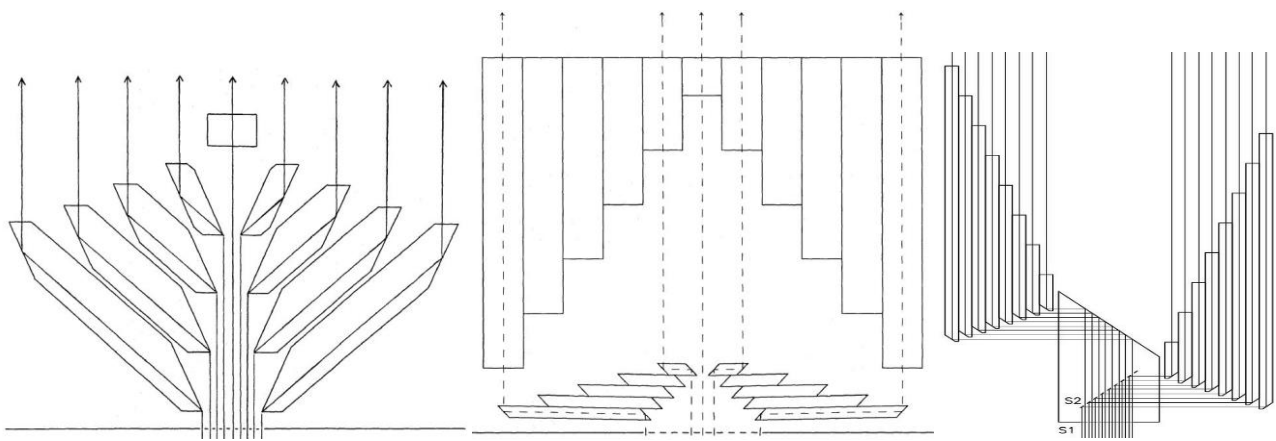

*Figure 87 : three generations of slicers used with the MSDP, on the left N = 9 channels, in the centre N = 11 and on the right N = 2 x 8 = 16 interleaved channels. A grid with N rectangular slits is placed in the spectrum, in front of the shifting prisms. There is a prism behind each 2D slit of the grid. Credit OP.*

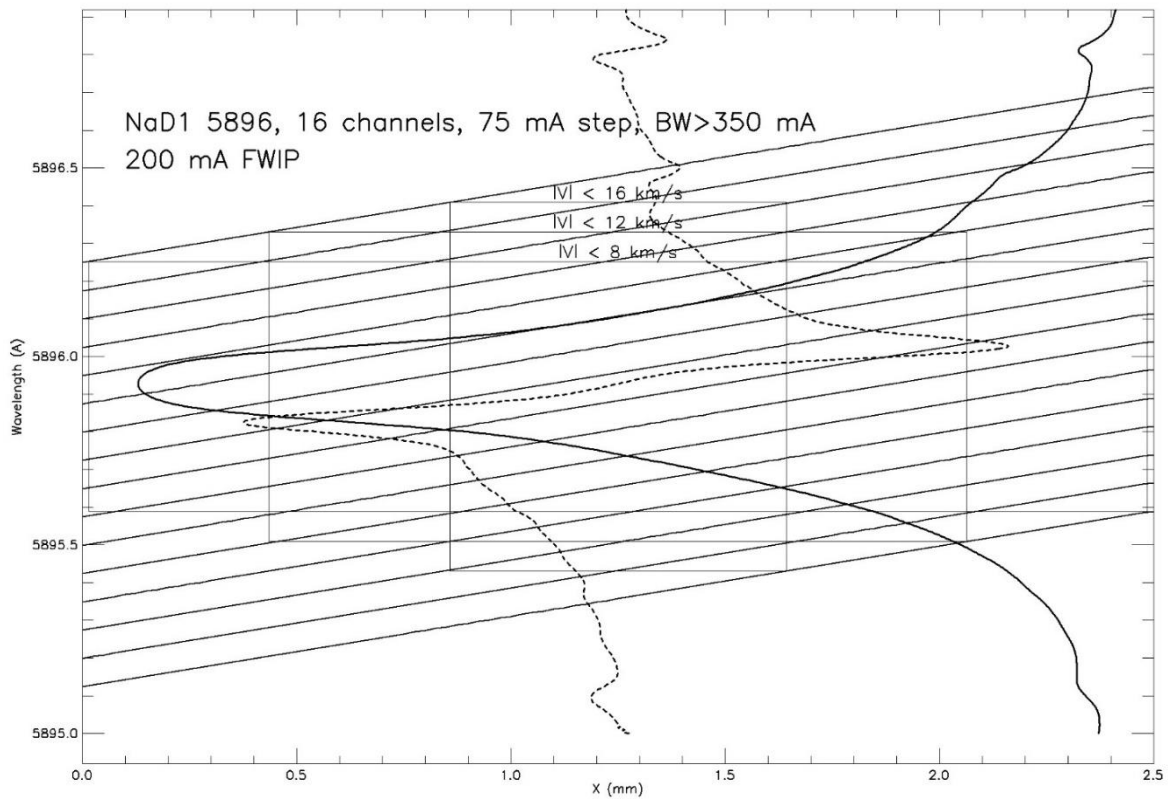

*Figure 88* : transmission functions of the N channels of the MSDP slicer. They provide the wavelength for the N channels as a function of the x-axis position. Here is the case of the NaD1 5896 Angström line for the 16-channel slicer of the THEMIS telescope. 75 milli Angström resolution. In dotted lines, the derivative of the line profile, the peaks identify the inflection points (best sensitivity for Doppler measurements). The rectangles delimit the width of the field observable in the x-direction to have a band of 0.35 Angström on the line. For a radial velocity of less than 8 km/s, the full width of the field of view is available. It decreases at higher velocities due to the Doppler displacement of the line. Credit OP.

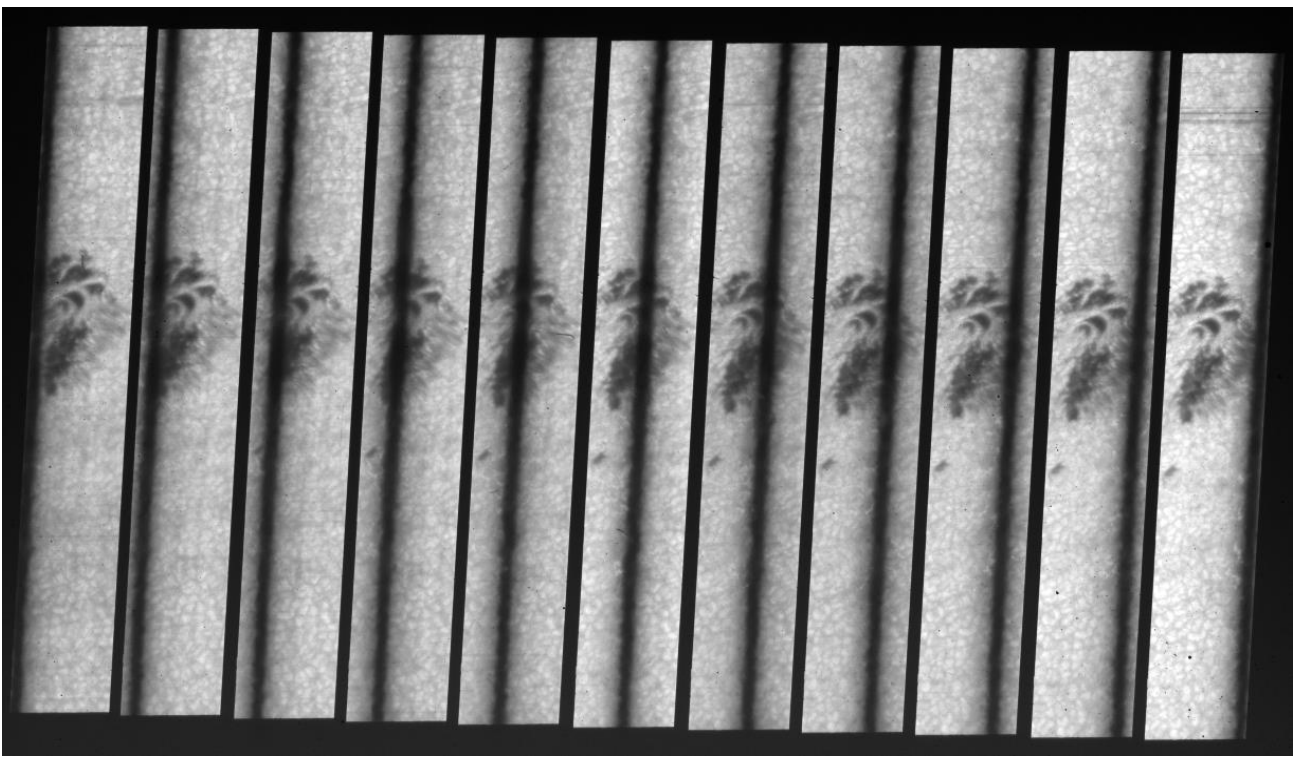

*Figure 89* : high spatial resolution with the 11-channel MSDP of the Pic du Midi. NaD1 line. Credit OP.

The second MSDP (Mein, 1980) was incorporated in the early 80s into the turret dome spectrograph of the Pic du Midi for observations at high spatial resolution (Figures 89 and 90) with 11-channel slicers. Initially photographic, the spectrograph benefited from a CCD camera at the end of the 90s. The dynamics of the granulation and its magnetism could be well studied in spectro polarimetry, as shown in Figure 90.

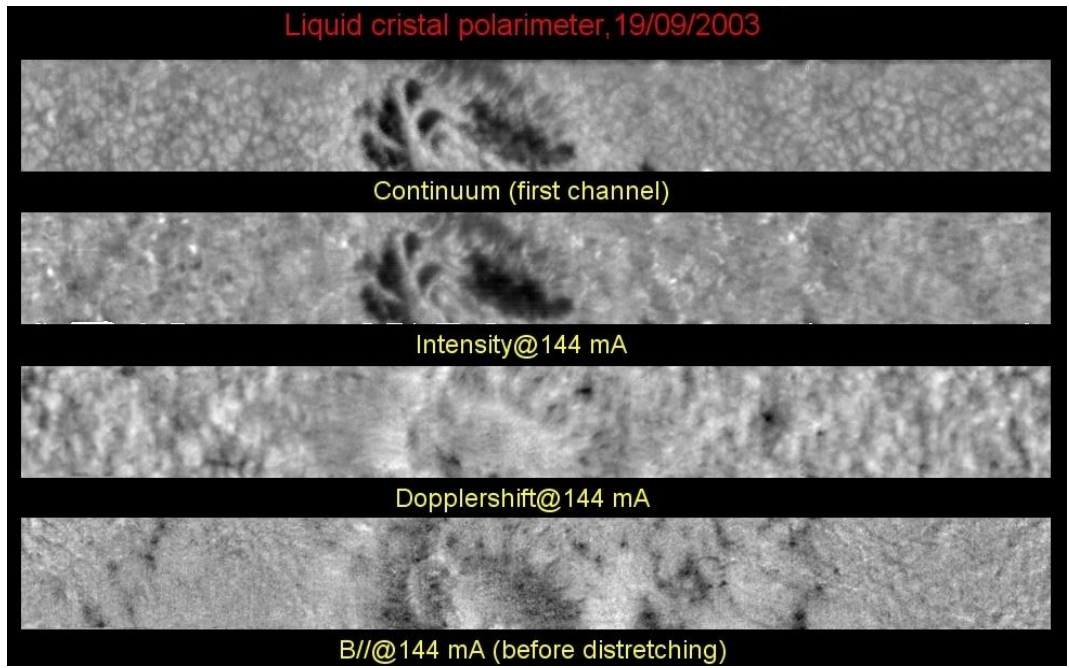

*Figure 90 : the granulation around a sunspot observed at the 11-channel MSDP of the Pic du Midi. NaD1 5896 Angström line. From top to bottom: intensity of the continuum, intensity in the wings, upward and downward Doppler velocities, radial magnetic field and its polarities. September 19, 2003. OP credit.*

The third MSDP (Figure 91 and Mein, 1991) was integrated into the large 15 m vertical spectrograph of the German Vacuum Tower Telescope (VTT) of the Teide Observatory (Canary Islands) in 1990. The VTT (optical characteristics comparable to the Meudon solar tower) gave excellent images thanks to its 60 cm vacuum telescope and its good site. Two simultaneous lines could be observed at the MSDP; initially photographic, it soon benefited from two Thomson CCDs made available by colleagues of Freiburg.

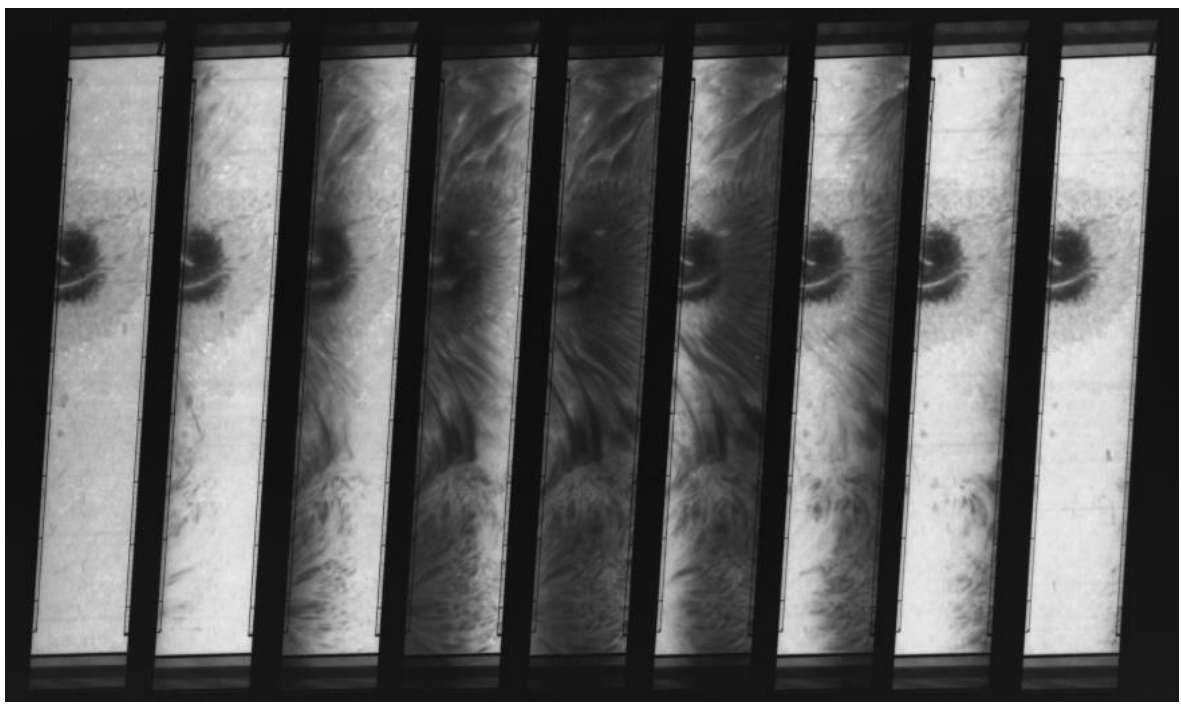

*Figure 91 : MSDP with high spatial resolution and 9 channels at the VTT in Tenerife. Hα line. Credit OP.*

The fourth MSDP was incorporated into the spectrograph of the THEMIS telescope in 1999 and was operational in spectro polarimetric mode (Mein, 2002). It used the Semel (1980) grid as shown in Figure 92 and had 2 blocks of 8 interleaved channels, i.e. 16 channels in total giving 80 milli Angström spectral resolution. It was able to record simultaneously I+V and I-V, then in sequence I+Q and I-Q, and finally I+U and I-U. It made it possible, with an unequalled speed, to map magnetic fields by interpreting the Zeeman effect on some relatively wide lines such as NaD1 5896 or CaI 6103 Angström (Figure 93).

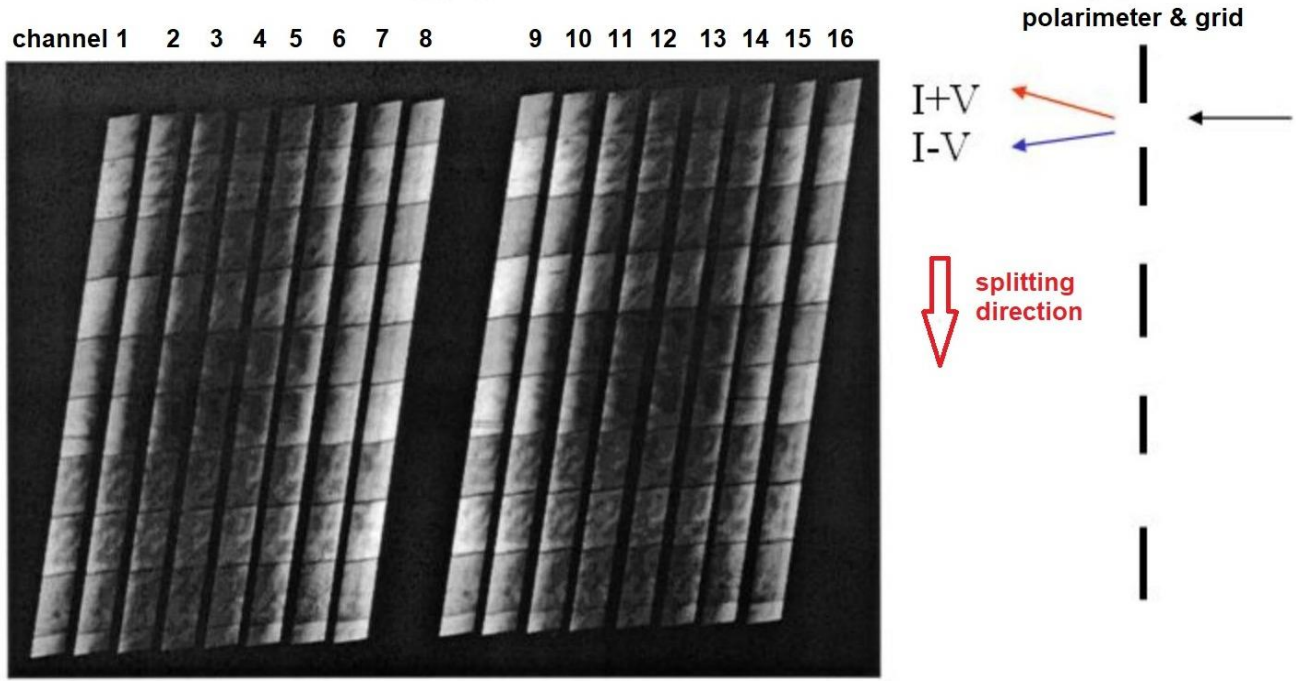

*Figure 92 : 16-channel polarimetric MSDP with Semel's grid at THEMIS in Tenerife. Hα line. OP credit.*

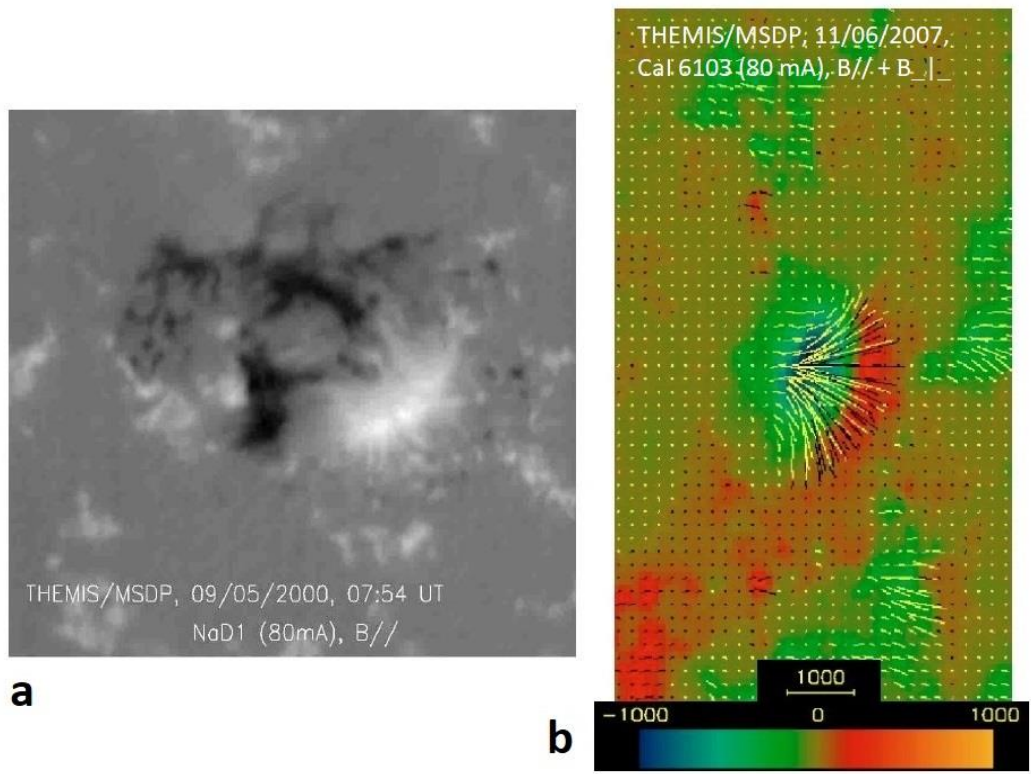

*Figure 93 : an example of results with the MSDP THEMIS. Left: radial magnetic field in NaD1 5896 Angström. Right: radial magnetic field (in colour) and transverse magnetic field (dashes represent the vector) in CaI 6103 Angström. The direction of the transverse field is within 180°. Credit OP.*

Finally, a new generation of slicers was developed and tested in Meudon by Sayède and Mein in the 2010s; in order to gain in spectral resolution (30 milli Angström) and increase the luminosity and the number N of channels, the multi-grid placed in the spectrum and the N adjustable shifting prisms were replaced, respectively, by a multifaceted one-piece micro-mirror and N small adjustable mirrors (Figure 94).

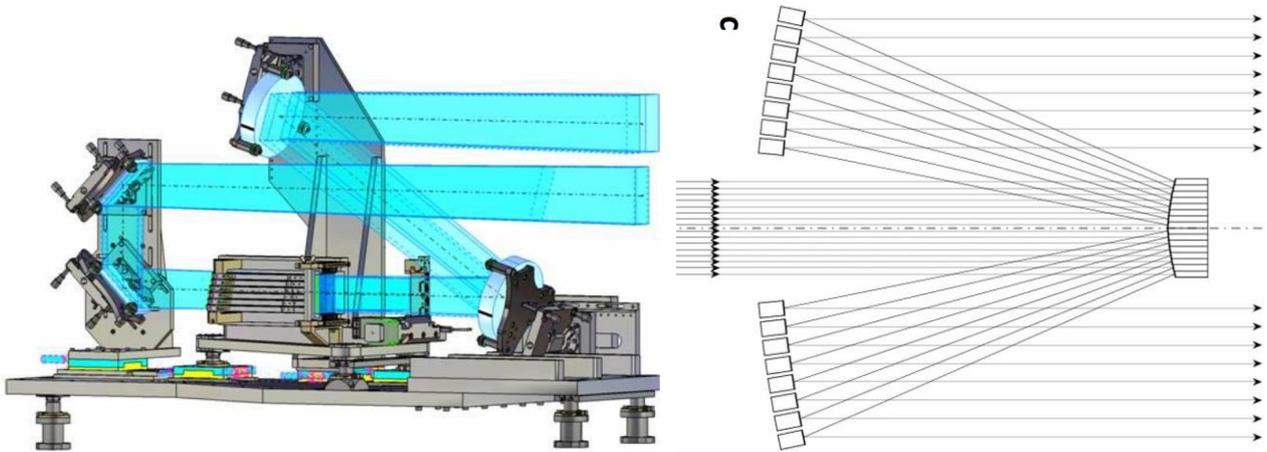

*Figure 94 : The new slicer of the Meudon solar tower, composed of a one-piece micro-mirror with 18 facets and 18 small shifting mirrors, a technique that favours a greater number of channels. OP credit.*

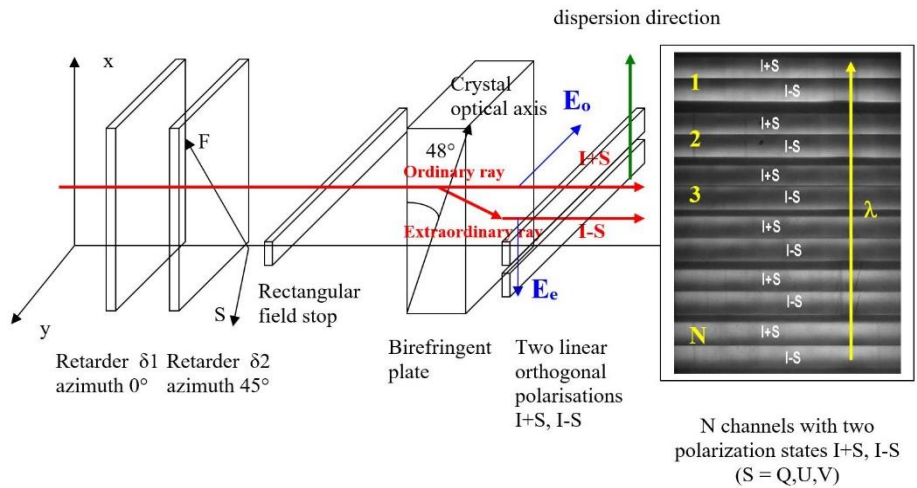

*Figure 95 : the new liquid crystal polarimeter of the Meudon solar tower divides each channel into two polarimetric subchannels by splitting the beam in the direction of dispersion. OP Credit.*

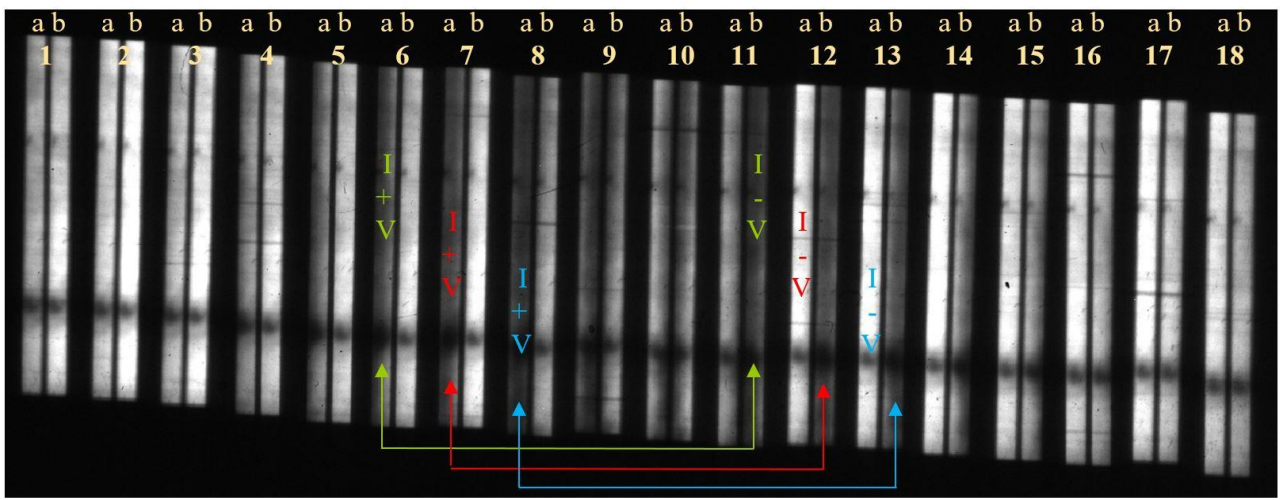

*Figure 96 : Polarimetric MSDP of the Meudon solar tower using the 18-channel slicer. The new polarimeter divides each channel into two sub-channels delivering I+V and I-V. Mgb1 5173 Angström line. Credit OP.*

A new method of polarimetric analysis was developed by Malherbe in 2015, for the simultaneous measurement of I+V and I-V by avoiding the use of a grid, which has the disadvantage of chopping the field of view. To do this, the birefringent calcite crystal was rotated by 90° compared to what was done at THEMIS with the grid, in order to split the field of view no longer in the y direction, but in the x or λ direction (Figure 95); the spectrograph entrance window was thus reduced by a factor of 2 in the x direction (Figure 96) which allows each of the 18 channels to be divided into 2 sub-channels (a, b), one giving I+V and the other I-V. The new wavelength transmission functions are then given by the diagram in Figure 97, the left part corresponding to the 18 sub-channels (a), and the right part to the 18 sub-channels (b).

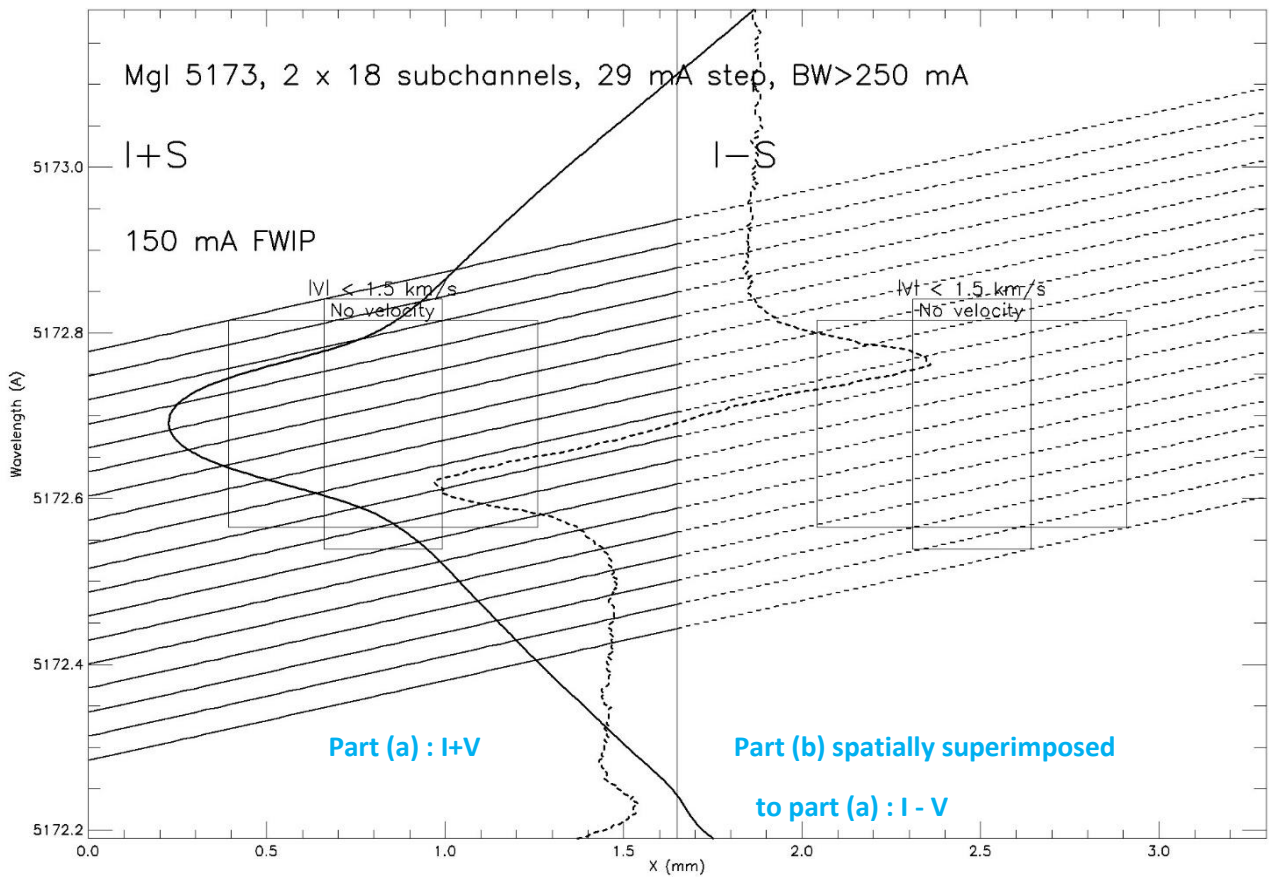

*Figure 97 : transmission functions of the 18 channels of the new slicer of the Meudon tower in polarimetric mode. They provide the wavelength for the 18 channels as a function of the x-axis. Here is the case of the Mgb1 line at 5173 Angström. 29 milli Angström resolution. In dotted lines, the derivative of the line profile, the peaks identify approximately the position of Stokes V peaks. The rectangles delimit the width of the field observable in the x-direction to have a band of 0.25 Angström on the line. The left (a) and right (b) parts are spatially superimposed, but (a) transmits I+V and (b) transmits I-V. OP credit.*

## 14 – The birth of solar radio astronomy

Let us now come to radio astronomy, a part that will be less detailed, as the author is not a specialist of this topic. The first radio signal of solar origin was identified in 1942 by James Hey by analysing the signals recorded by British military radars during the Second World War. The German army has built and disseminated 1500 Würzburg-Riese radars of 7.50 m in diameter during the conflict; a few examples were recovered by Yves Rocard after 1945 for scientific purposes. One of the radars was installed in Meudon, on the edge of the "Bel Air" lake (Figure 98). These radars operated on the 55 cm wavelength band. The one in Meudon was transformed into a solar radio antenna by Laffineur (1949, 1951). The Nançay radio astronomy station was created shortly afterwards in 1953 under the impetus of Yves Rocard at the "Ecole Normale Supérieure", who supported Jean-François Denisse, Jean-Louis Steinberg and Jacques-Emile Blum in this innovative and fundamental task because it opened a new window on the universe. Two antennas were placed at Nançay on rails (figures 99 and 100) allowing their respective distance to be varied. The Meudon antenna was later transferred to the Bordeaux observatory. One antenna remained in Nançay, the second was given to the radar museum of Douvres la Délivrande in Calvados, near Caen (France). Radio astronomy allows to capture plasma emissions, from the chromosphere in GHz range (Nobeyama, Japan) to the upper corona at frequencies around 100 MHz (Nançay). These frequencies correspond to increasing altitudes (because the plasma frequency is proportional to the square root of the density). Solar flares also emit in radio wavelengths because they accelerate particles (or slow down by impact on the chromosphere); indeed, any charged particle in motion emits radiation according to its acceleration.

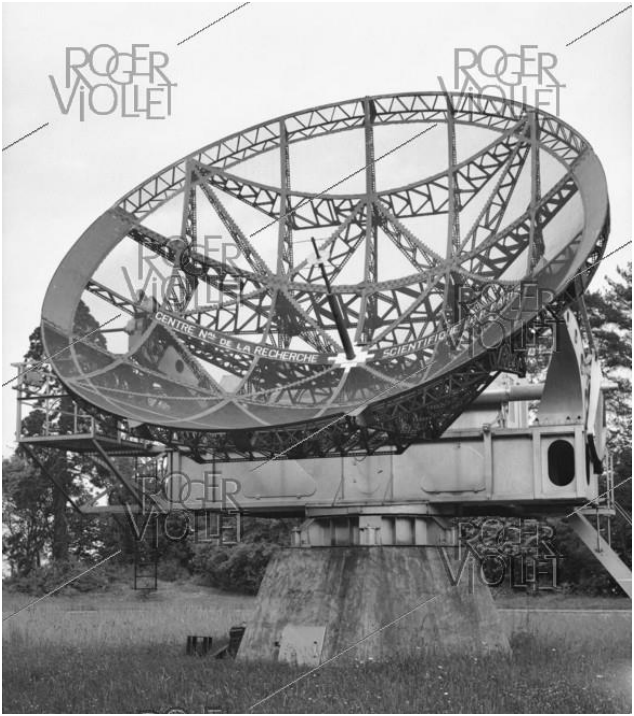

*Figure 98*

*Würzburg radar recovered from the 1939-1945 war, manufactured by the Germans, then installed in Meudon in 1949 to start the development of radio astronomy in France. The radar operated at 55 cm wavelength (545 MHz, the low corona).*

*Credit Roger-Viollet.*

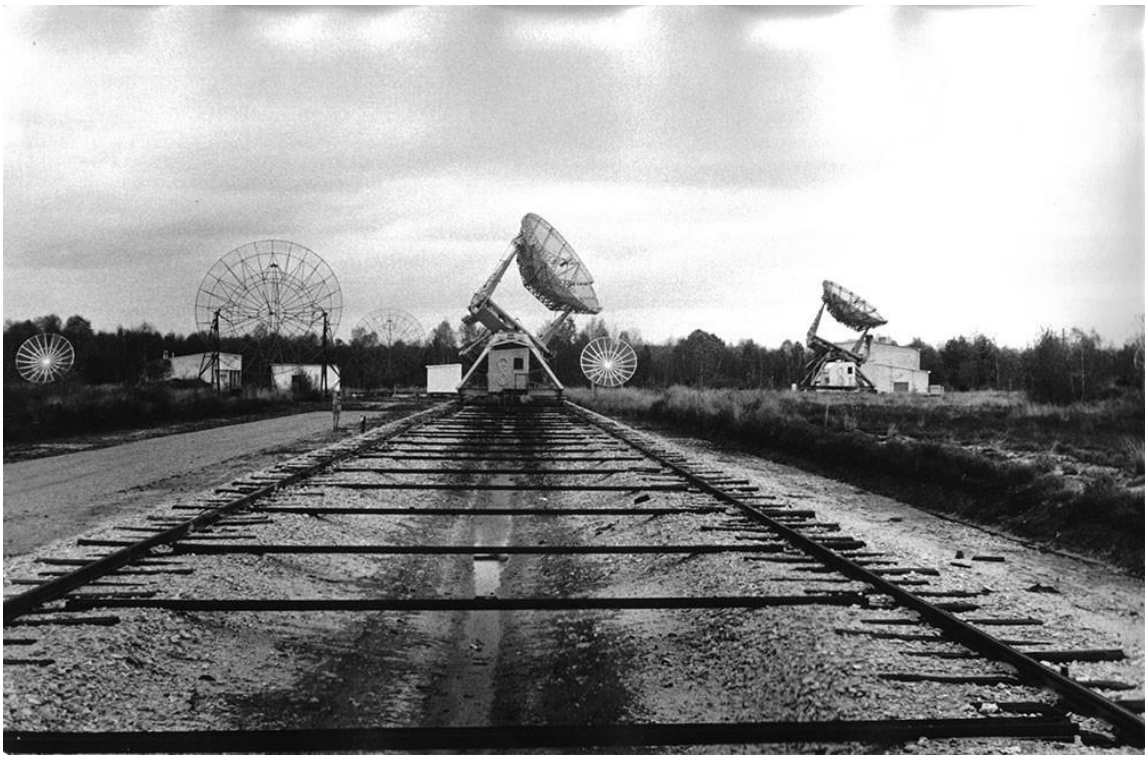

*Figure 99 : the two Würzburg radars installed in Nançay in 1953 on rails, allowing to change their respective distance and develop interferometry. OP Credit.*

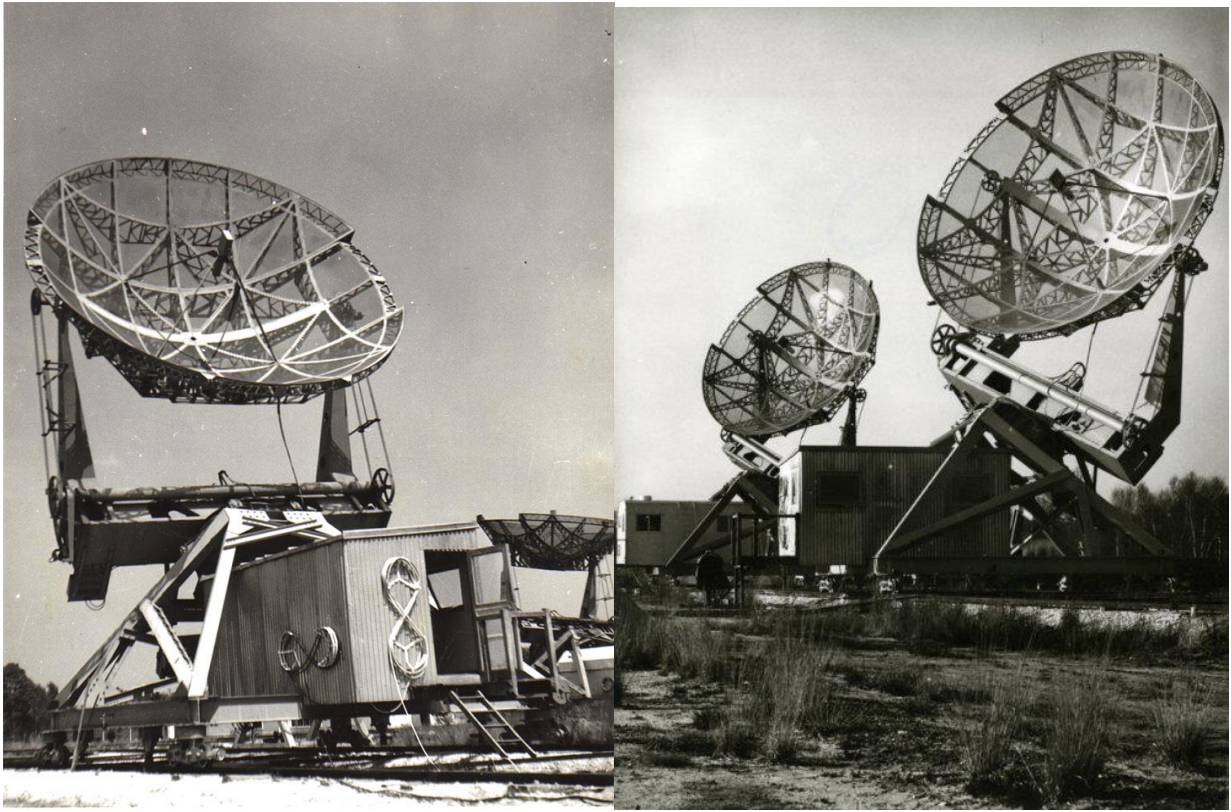

*Figure 100 : two the Würzburg radars in Nançay. Credit OP.*

Subsequently, an interferometer (Figure 101) operating on the 3 cm wavelength (10 GHz) was set up and allowed Monique Pick (1933-2023) to study the solar radio emissions of the chromosphere. A network of non-directional antennas in the decametric wavelength range (Figure 102) was also created, not only for the study of solar bursts, but also for the study of the magnetospheres of planets such as Jupiter. Pick then concretized the project of a large radioheliograph (figures 103 to 105) with two kilometric North-South and East-West branches, composed of about fifty antennas; the aim was to produce by interferometry 2D images of the Sun, with a spatial resolution of 1', in frequencies ranging from 150 to 450 MHz; the decrease of the electron plasma frequency with altitude allows to explore the corona in 3 dimensions.

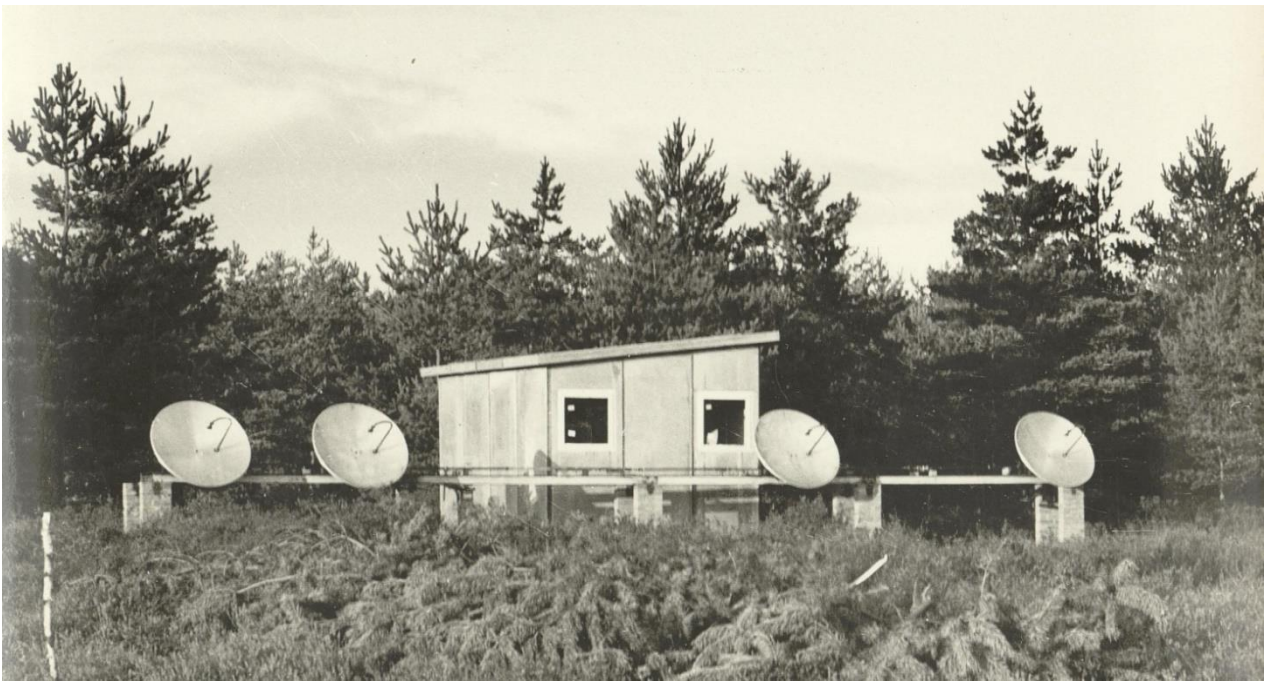

*Figure 101 : solar radio interferometer at 3 cm in Nançay. OP credit.*

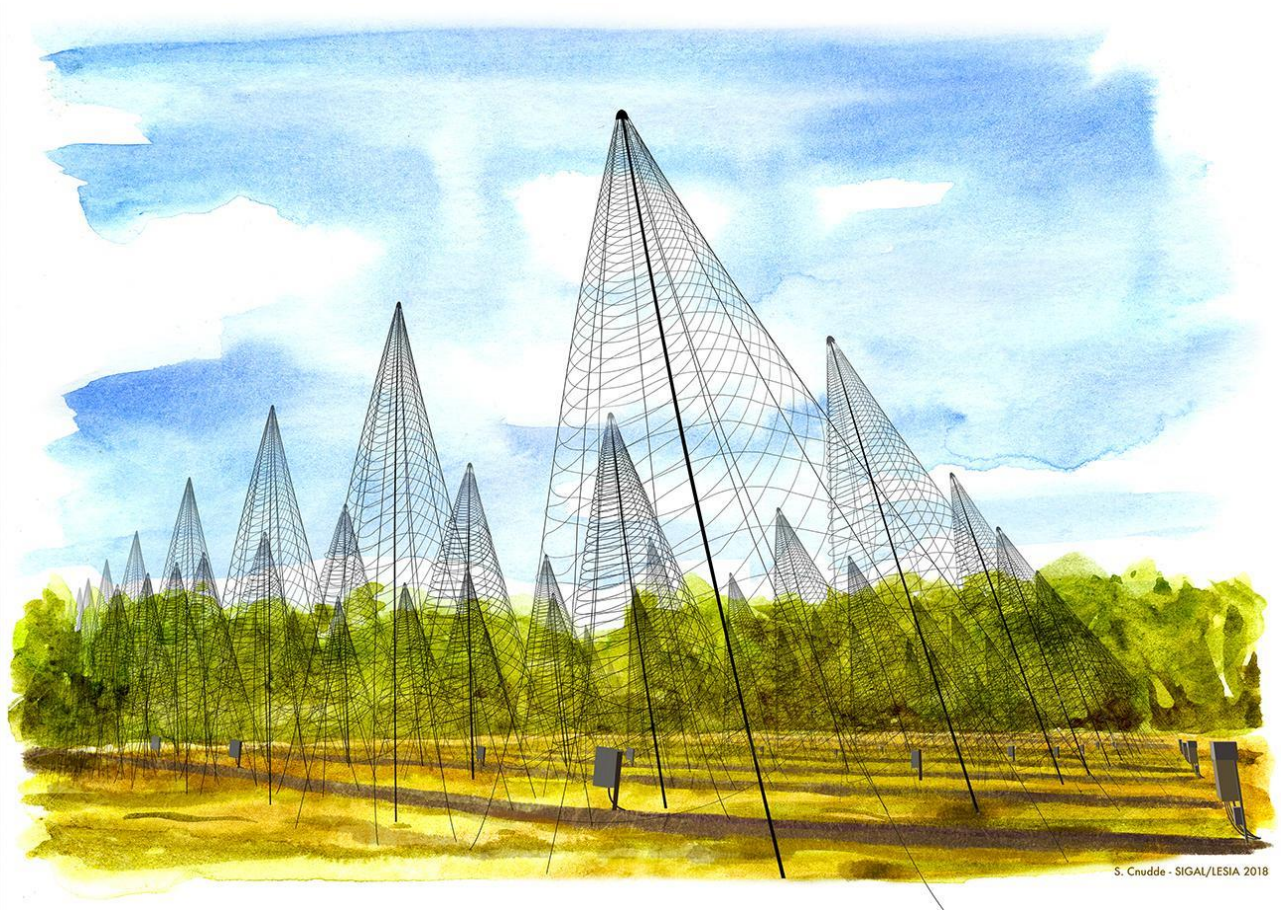

*Figure 102 : Nançay decametric antenna array . Artist's impression. Credit Sylvain Cnudde (OP).*

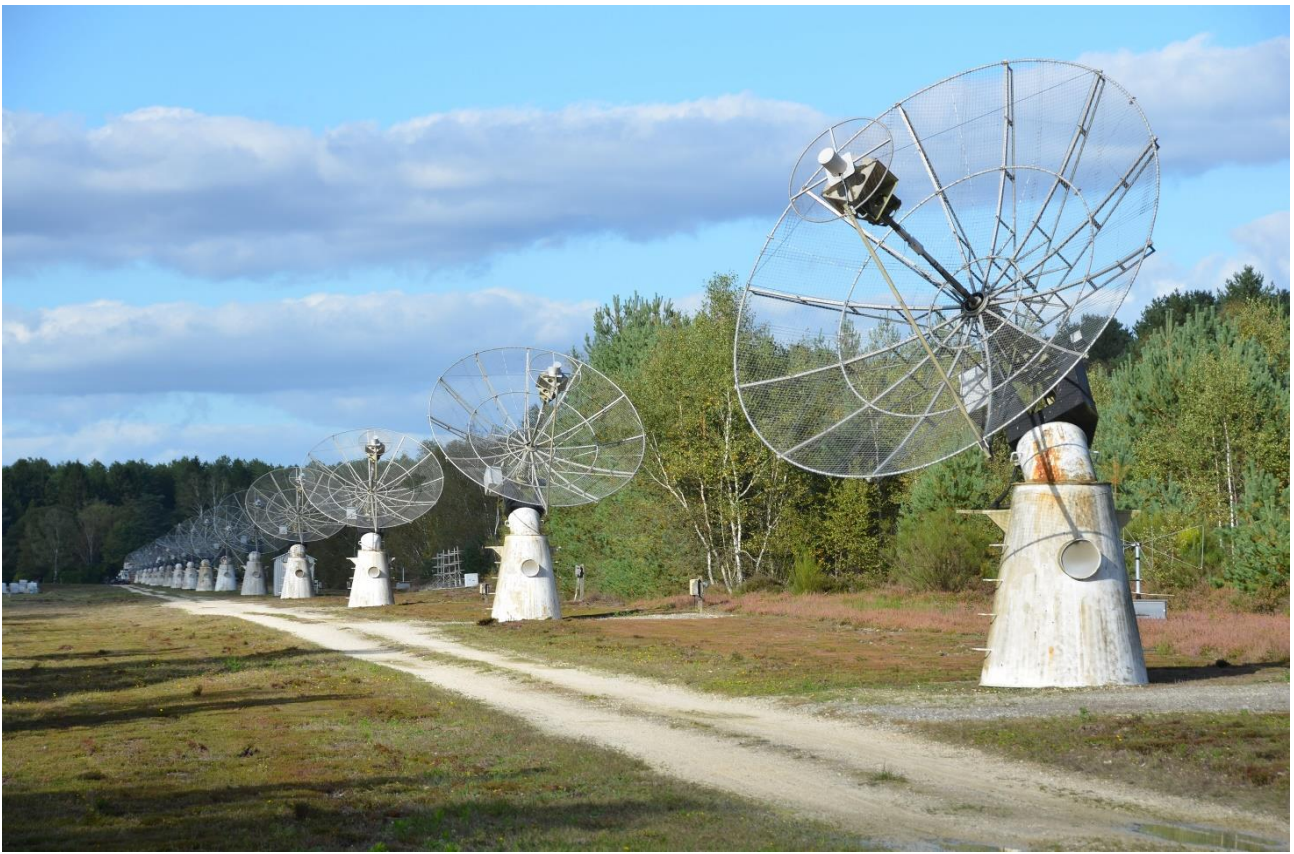

*Figure 103 : metric radioheliograph (150 to 450 MHz) of Nançay, with two orthogonal kilometre-length baselines totalling about 50 antennas. Credit OP.*

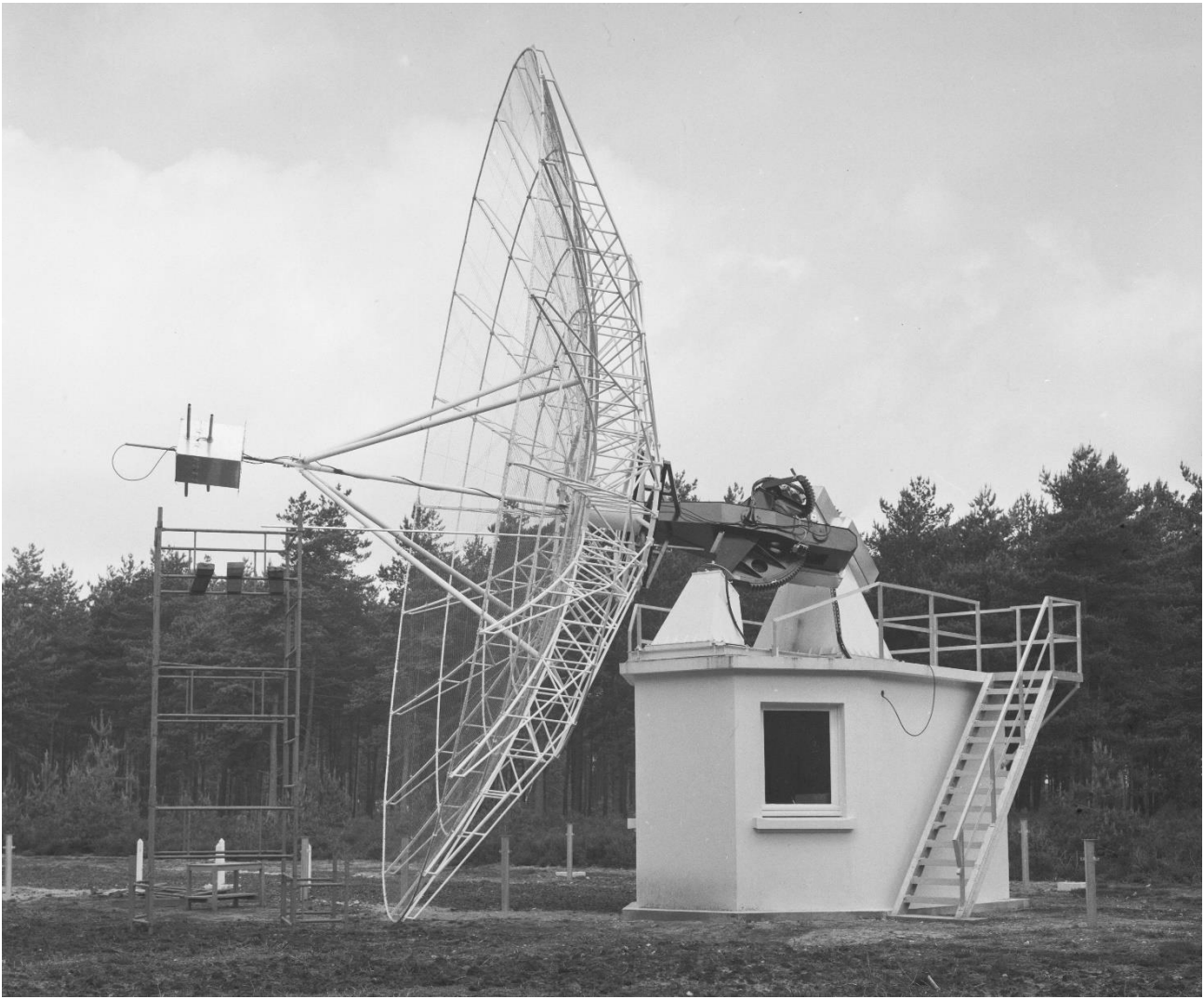

*Figure 104 : additional large antenna of 10 m from the Nançay radioheliograph. OP credit.*

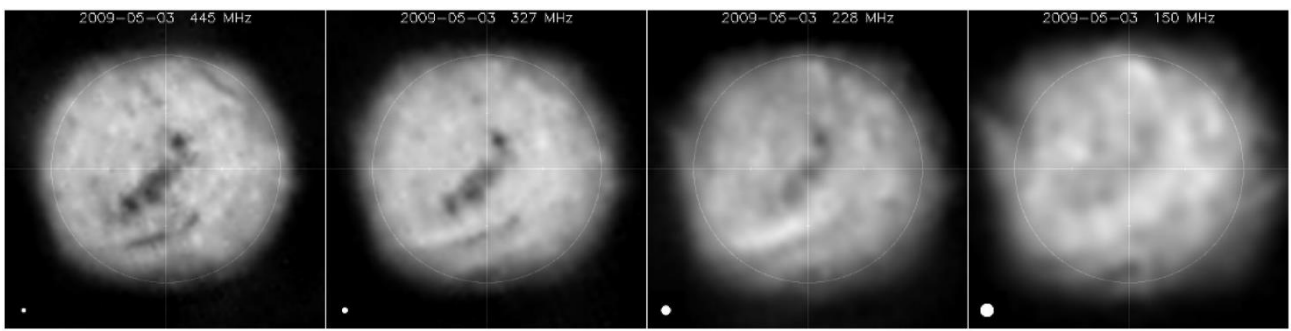

*Figure 105 : images of the Sun at 445, 327, 226 and 150 MHz from the Nançay radioheliograph. OP credit.*

**Conclusion**

The year 2026 marks the 150$^{th}$ birthday of the Meudon Observatory. We have reported it only from the point of view of ground-based solar instrumentation, because it was the purpose of the foundation of the observatory by Jules Janssen, a specialist of the Sun, but not only. Under his impulse, the establishment was equipped with large nocturnal instruments, such as the large double refractor (67 cm and 83 cm objectives) or the 1 m telescope. Over time, Meudon grew considerably and diversified, to the point that the part of activity related to the Sun became the minority, with the development of other sciences, such as planetology, stellar, galactic and extragalactic astronomy, requiring the international pooling of major equipment, both on the ground (giant telescopes) and in space. We deliberately did not address space solar astronomy, which developed since 1980 within strong international collaborations, as France obviously has neither the means nor the vocation to act alone in this topic. Ground-based instruments allowed to carry out coordinated observation campaigns with space missions, as early as 1980 with NASA's SMM satellite (Malherbe, 2025), and even early with SkyLab (1973). After the beginnings of balloon exploration, launched by Dollfus in 1960, the foundation of the DESPA department, transformed into LESIA and then LIRA, enabled the observatory to participate in major developments in space solar physics, culminating in the launch of SOLAR ORBITER in 2020. But that's another story to tell.

### About the author

I am an engineer by initial training (Ecole Centrale Paris, 1979); I then pursued a more specialized training in physics at the University of Paris Diderot : Diploma of Advanced Studies in Astrophysics in 1980 ("DEA"), postgraduate doctorate in 1983 ("doctorat de 3ème cycle") and State Doctorate in Physical Sciences in 1987 ("Doctorat d'Etat"). As an astronomer since 1983, I spent my entire career at the Paris Observatory – PSL University, then I was named emeritus astronomer in 2023. My career has been devoted to the physics of the Sun, in many aspects, observational, instrumental and data processing, MHD modelling and MHD numerical simulation. At the same time, I taught physics at the PSL University and at the Paris Observatory. I was also responsible for instrumental developments (such as the spectroheliograph of the Meudon Observatory, MeteoSpace telescopes of the Calern plateau, polarimetry at the Pic du Midi and Meudon). I am much interested in the history of solar instrumentation.